\documentclass[iop]{emulateapj}
\newcommand{\cha}{\textit{Chandra }}
\def\xmm{{XMM-{\it Newton\/}}}

\def\lu{{ erg s$^{-1}$}}
\def\flu{{ erg s$^{-1}$} cm$^{-2}$}

\def\leg{{\em COSMOS-Legacy}\/}

\shorttitle{The \cha \leg\ survey: Source X-ray spectral properties}
\shortauthors{Marchesi et al.}

\usepackage{natbib}
\usepackage{float}
\usepackage{color}
\usepackage{graphicx}
\usepackage{gensymb}
\usepackage{enumitem}
\usepackage{longtable}
\begin{document}

\slugcomment{Accepted to the Astrophysical Journal on August 9, 2016}
\title{The \cha \leg\ survey: source X-ray spectral properties}

\author{S. Marchesi\altaffilmark{1,2}, G. Lanzuisi\altaffilmark{2,3}, F. Civano\altaffilmark{4}, K. Iwasawa\altaffilmark{5}, H. Suh\altaffilmark{6}, A. Comastri\altaffilmark{2}, G. Zamorani\altaffilmark{2}, V. Allevato\altaffilmark{7}, R. Griffiths\altaffilmark{8}, T. Miyaji\altaffilmark{9}, P. Ranalli\altaffilmark{10}, M. Salvato\altaffilmark{11}, K. Schawinski\altaffilmark{12}, J. Silverman\altaffilmark{13}, E. Treister\altaffilmark{14,15}, C.M. Urry\altaffilmark{16}, C. Vignali\altaffilmark{2,3}} 

\altaffiltext{1}{Department of Physics \& Astronomy, Clemson University, Clemson, SC 29634, USA}
\altaffiltext{2}{INAF--Osservatorio Astronomico di Bologna, via Ranzani 1, 40127 Bologna, Italy}
\altaffiltext{3}{Dipartimento di Fisica e Astronomia, Universit\`a di Bologna, viale Berti Pichat 6/2, 40127 Bologna, Italy}
\altaffiltext{4}{Harvard-Smithsonian Center for Astrophysics, 60 Garden Street, Cambridge, MA 02138, USA}
\altaffiltext{5}{ICREA and Institut de Ci\`encies del Cosmos (ICC), Universitat de Barcelona (IEEC-UB), Mart\'{\i} y Franqu\`es 1, 08028 Barcelona, Spain}
\altaffiltext{6}{Institute for Astronomy, 2680 Woodlawn Drive, University of Hawaii, Honolulu, HI 96822, USA}
\altaffiltext{7}{Department of Physics, University of Helsinki, Gustaf H\"allstr\"omin katu 2a, FI-00014 Helsinki, Finland}
\altaffiltext{8}{Physics \& Astronomy Dept., Natural Sciences Division, University of Hawaii at Hilo, 200 W. Kawili St., Hilo, HI 96720, USA}
\altaffiltext{9}{Instituto de Astronom\'ia sede Ensenada, Universidad Nacional Aut\'onoma de M\'exico, Km. 103, Carret. Tijunana-Ensenada, Ensenada, BC, Mexico}
\altaffiltext{10}{Lund Observatory, P.O. Box 43, 22100 Lund, Sweden}
\altaffiltext{11}{Max-Planck-Institut f{\"u}r extraterrestrische Physik, Giessenbachstrasse 1, D-85748 Garching bei M{\"u}nchen, Germany}
\altaffiltext{12}{Institute for Astronomy, Department of Physics, ETH Zurich, Wolfgang-Pauli-Strasse 27, CH-8093 Zurich, Switzerland}
\altaffiltext{13}{Kavli Institute for the Physics and Mathematics of the Universe (WPI), The University of Tokyo Institutes for Advanced Study, The University of Tokyo, Kashiwa, Chiba 277-8583, Japan}
\altaffiltext{14}{Universidad de Concepci\'{o}n, Departamento de Astronom\'{\i}a, Casilla 160-C, Concepci\'{o}n, Chile}
\altaffiltext{15}{Pontificia Universidad Cat\'{o}lica de Chile, Instituto de Astrofisica, Casilla 306, Santiago 22, Chile}
\altaffiltext{16}{Yale Center for Astronomy and Astrophysics, 260 Whitney Avenue, New Haven, CT 06520, USA}

\begin{abstract}
We present the X-ray spectral analysis of the 1855 extragalactic sources in the \cha \leg\ survey catalog having more than 30 net counts in the 0.5-7 keV band. 38\% of the sources are optically classified Type 1 active galactic nuclei (AGN), 60\% are Type 2 AGN and 2\% are passive, low-redshift galaxies. We study the distribution of AGN photon index $\Gamma$ and of the intrinsic absorption $N_{\rm H,z}$ based on the sources optical classification: Type 1 have a slightly steeper mean photon index $\Gamma$ than Type 2 AGN, which on the other hand have average $N_{\rm H,z}$ $\sim$3 times higher than Type 1 AGN. We find that $\sim$15\% of Type 1 AGN have $N_{\rm H,z}$$>$10$^{22}$ cm$^{-2}$, i.e., are obscured according to the X-ray spectral fitting; the vast majority of these sources have L$_{\rm 2-10keV}>$10$^{44}$ \lu. The existence of these objects suggests that optical and X-ray obscuration can be caused by different phenomena, the X-ray obscuration being for example caused by dust-free material surrounding the inner part of the nuclei.
$\sim$18\% of Type 2 AGN have $N_{\rm H,z}$$<$10$^{22}$ cm$^{-2}$, and most of these sources have low X-ray luminosities (L$_{\rm 2-10keV}<$10$^{43}$ \lu). We expect a part of these sources to be low-accretion, unobscured AGN lacking of broad emission lines.
Finally, we also find a direct proportional trend between $N_{\rm H,z}$ and host galaxy mass and star formation rate, although part of this trend is due to a redshift selection effect.
\end{abstract}


\section{Introduction}
A proper understanding of the properties of the supermassive black holes (SMBHs) in the center of galaxies, and of their evolution across cosmic time, requires unbiased samples of active galactic nuclei (AGN), both obscured and unobscured (i.e., sources with hydrogen column density $N_{\rm H,z}$ below and above the 10$^{22}$ cm$^{-2}$ threshold conventionally adopted to separate unobscured and obscured sources, respectively), over a wide range of redshifts and luminosities. Multiwavelength data are also required to avoid selection effects. Mapping the typical AGN population, i.e, those moderate luminosity sources that produce a significant fraction of the X-ray background emission \citep[see, e.g.,][]{gilli07,treister09}, is possible only with surveys that combine depth, to detect AGN up to z$\sim$6, and area, to find statistically significant numbers of sources at any redshift.

X-ray data are strategic in the AGN selection process, for several reasons. First, at X-ray energies the contamination from non-nuclear emission, mainly due to star-formation processes, is far less significant than in optical and infrared wavelengths \citep{donley08,lehmer12,stern12}. Moreover, \cha and \xmm\ can select both unobscured and obscured AGN, and can also detect a fraction of Compton thick AGN, i.e., sources with hydrogen column densities, $N_H\geq$10$^{24}$ cm$^{-2}$, up to redshift z$\sim$2--3 \citep{comastri11,iwasawa12,georgantopoulos13,buchner15,lanzuisi15}. Therefore, combining X-ray and optical/NIR observations of AGN allows one to study simultaneously the properties of the accreting SMBHs and their host galaxies.

The proper characterization of AGN X-ray spectra requires observations with high signal-to-noise ratio (S/N) 
to properly model many spectral features, such as the so-called ``soft excess'', warm absorbers, emission and absorption lines different from the Iron K$\alpha$ line at 6.4 keV, and a reflection component \citep[see, e.g.,][for a review of these features]{risaliti04}. However, AGN spectra with low S/N can be modelled in the 0.5--10 keV band with an absorbed power-law, where the intrinsic absorption is caused by the gas and dust surrounding the SMBH, or by the host galaxy itself. The Iron K $\alpha$ line at 6.4 keV can also be properly modelled in low S/N AGN spectra. Therefore, the X-ray spectroscopy of large numbers of AGN, combined with extended multiwavelength coverage, makes possible to study the distribution of parameters such as the intrinsic absorption ($N_{\rm H,z}$), the Iron K$\alpha$ equivalent width and the power-law photon index ($\Gamma$), and to look for trends between these quantities and redshift or intrinsic X-ray luminosity.

The primary power-law component of the AGN X-ray spectra is caused by inverse Compton scattering emissions of UV photons. These photons are first emitted by the disc and then up-scattered by the hot corona electrons that surround the disc \citep{haardt91,siemiginowska07}. The power-law produced by this process has typical photon index $\Gamma$1.9, with dispersion $\sigma$=0.2 \citep{nandra94,piconcelli05,tozzi06,mainieri07,lanzuisi13}. Moreover, $\Gamma$ seems to be independent of different SMBH physical parameters, such as BH mass and spin, while a directly proportional trend between $\Gamma$ and the SMBH Eddington ratio ($\lambda_{edd}$=L$_{\rm bol}$/L$_{\rm edd}$, with L$_{\rm edd}$=1.2 $\times$10$^{38}$ (M$_{\rm BH}$/M$_\odot$) \lu) has been reported in several works \citep[e.g., ][]{wang04,shemmer08,risaliti09,jin12}. 

A proper analysis of the different X-ray spectral parameters requires large datasets, with both good X-ray statistics and complete multiwavelength characterization. Since developing these types of datasets is not trivial, even the dependences of the photon index and the intrinsinc absorption from other quantities (e.g., redshift, X-ray or bolometric luminosity) are still debated, as are the X-ray spectral properties of optically classified Type 1 and Type 2 AGN \citep[see, e.g., ][]{mateos05,page05,shemmer06,young09,sobolewska09,lusso12,lanzuisi13,fotopoulou16}.

The \cha \leg\ survey \citep{civano16}, with its relatively deep average coverage of $\sim$160\,ks over 2.15 deg$^2$, for a total of 4.6\,Ms, provides an unprecedented dataset to study the X-ray properties of AGN over a wide range of redshifts and luminosities. Moreover, the COSMOS field \citep{scoville07} has been covered with extended multiwavelength photometric \citep{capak07,koekemoer07,sanders07,schinnerer07,taniguchi07,zamojski07,ilbert09,mccracken10,laigle16} and spectroscopic \citep{lilly07,lilly09,trump07} observations, thus enabling to identify and characterize $\sim$97\% of the X-ray sources \citep{marchesi16a}. Therefore, a complete analysis of the X-ray spectral parameters for different classes of optical sources is possible in \cha \leg. In this work, we present the X-ray spectral analysis of the 1855 extragalactic sources with more than 30 net counts in the 0.5--7 keV band in \cha \leg. In Section \ref{sec:survey} we describe the \cha \leg\ survey, the X-ray catalog and the optical/IR properties of the X-ray sources. In Section \ref{sec:spec_analysis} we present the spectral extraction procedure, while in Section \ref{sec:fit} we describe the different fitting models we used, and the results of the fitting. In Section \ref{sec:fit_distr} we discuss the fit parameters distribution. Finally, in Section \ref{sec:high-z} we discuss the properties of the z$>$3 subsample, and in Section \ref{sec:concl} we summarize the results of this work. Throughout the paper, we assume a cosmology with H$_0$ = 71~km~s$^{-1}$~Mpc$^{-1}$, $\Omega_M$ = 0.3 and $\Omega_{\Lambda}$= 0.7. Errors are at 90\% confidence if not otherwise stated.

\section{The \cha \leg\ survey}\label{sec:survey}
\subsection{The X-ray catalog}
\begin{figure}[H]
\centering
\includegraphics[width=0.5\textwidth]{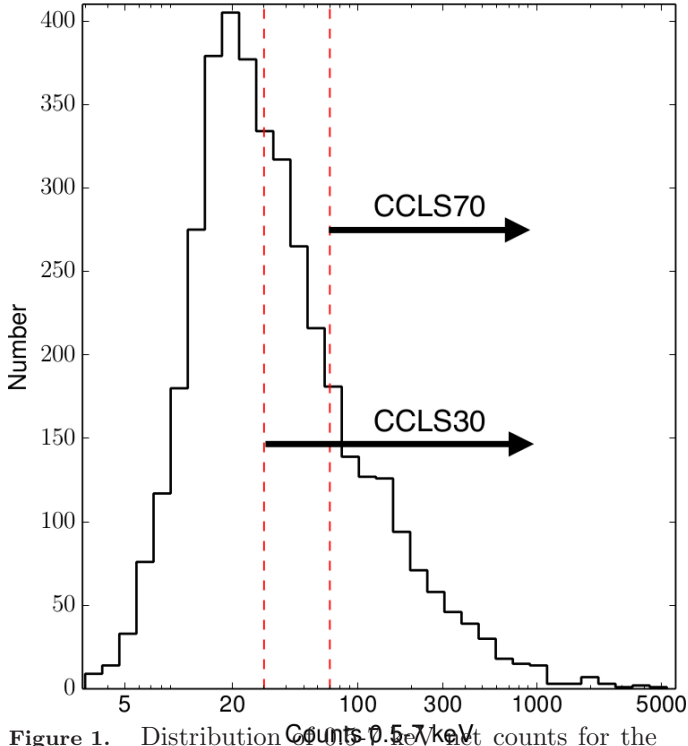}
\caption{{\normalsize Distribution of 0.5-7 keV net counts for the whole \cha \leg\ survey. Red dashed lines mark the two different thresholds adopted in the X-ray spectral analysis, i.e., 30 and 70 net counts.}}\label{fig:histo_counts}
\end{figure}

The \cha \leg\ X-ray catalog is described in \citet{civano16}. The catalog is the result of 4.6 Ms of observations with \cha on the 2.2 deg$^2$ of the COSMOS field. The final catalog includes 4016 sources, detected in at least one of the following three bands: full (F; 0.5--7 keV), soft (S; 0.5--2 keV) and hard (H; 2--7 keV). Each source was detected in at least one band with a maximum likelihood detection value DET\_ML$>$10.8, i.e., with probability of being a spurious detection P$<$2 $\times$ 10$^{-5}$. The survey flux limit in each of the three bands is f =1.2 $\times$ 10$^{-15}$ \flu\ in the 0.5--10 keV band, f = 2.8 $\times$ 10$^{-16}$ \flu\ in the 0.5--2 keV band and f =1.9 $\times$ 10$^{-15}$ \flu\ in the 2--10 keV band. Fluxes have been computed assuming as a model a power-law with no intrinsic absorption and photon index $\Gamma$=1.4\footnote{This is the slope of the cosmic X-ray background \citep[see, e.g., ][]{hickox06} and well represents a distribution of both obscured and unobscured AGN at the fluxes covered by \cha \leg.}. Fluxes in the full (hard) band are computed over the 0.5--10 keV (2--10 keV) energy range, instead than over the 0.5--7 keV (2--7 keV) one, for an easy comparison with other works in the literature. In Figure \ref{fig:histo_counts} we show the 0.5-7 keV net counts distribution for the 4016 \cha \leg\ sources. 1949 sources have more than 30 net counts and 923 sources have more than 70 net counts. These two counts thresholds are based on previous X-ray spectral analysis \citep[see, e.g.,][]{lanzuisi13,lanzuisi15}: in spectra with more than $\sim$70 net counts, we can perform a fit leaving the two main fit parameters (the power-law photon index $\Gamma$ and the intrinsic absorption $N_{\rm H,z}$; see Section \ref{sec:model} for further details) free to vary, recovering uncertainties $<$30\% for the vast majority of the sources (see also Figure \ref{fig:gamma_vs_cts}, right panel). We instead choose the $\sim$30 net counts threshold as a limit for which only one of the two parameters can be constrained, with the other one fixed. In the following sections, we will describe the X-ray spectral properties of the 1855 extragalactic sources (i.e., excluding the 64 stars and the 30 sources with no redshift available) with more than 30 net counts in the 0.5-7 keV band: from now on, we will refer to this sample as the CCLS30 sample. We will also refer to the sample of 887 extragalactic sources with more than 70 net counts in the 0.5-7 keV band as the CCLS70 sample.

\subsection{\cha \leg\ optical/IR counterparts}
The optical/IR counterparts of the whole \cha \leg\ sample are described in \citet{marchesi16a}. 
1273 sources out of 1855 in CCLS30 (68.6\%) have a z$_{\rm spec}$, the remaining 582 have a z$_{\rm phot}$, obtained using the SED fitting procedure described in \citet{salvato11} and based on $\chi^2$ minimization, using the publicly available code LePhare \citep{arnouts99,ilbert06}. The spectroscopic completeness is significantly higher in CCLS70, where 732 out of 887 sources (82.5\%) have a z$_{\rm spec}$. In our analysis, we classify the sources in the CCLS30 on the basis of the optical spectroscopic classification, when available; otherwise, we use the best-fitting template derived in the SED fitting procedure adopted to estimate the photometric redshifts. The sources are divided as follows:
\begin{enumerate}
\item 696 Type 1, unobscured AGN. Broad-line AGN (BLAGN) on the basis of their spectral classification, i.e., sources with lines having FWHM$\geq$2000 km s$^{-1}$, or sources with no spectral information, and SED best-fitted by an unobscured AGN template.
\item 1111 Type 2, obscured AGN. Objects with rest-frame, absorption-corrected 2-10 luminosity L$_{2-10 keV}\geq$10$^{42}$ \lu, no broad lines in their spectra (non-BLAGN), or sources with no spectral type and SED best-fitted by an obscured AGN template (and any L$_{2-10keV}$) or a galaxy template (and L$_{2-10 keV}$$\geq$10$^{42}$ \lu).
\item 37 Galaxies. Objects with rest-frame, absorption-corrected 2-10 luminosity L$_{2-10keV}$ $<$10$^{42}$ \lu, no broad lines in their spectra (non-BLAGN), or sources with no spectral type and SED best-fitted by a galaxy template. Nine of these sources are part of a sample of 50 dwarf galaxies (i.e., having mass 10$^7$ $<$ M$_*$ $<$10$^9$), candidate Type 2 AGN that are being analyzed in a separate paper (Mezcua et al. in preparation), while other ten are part of a sample of 69 early-type galaxies analyzed in \citet{civano14}. In the remaining part of our analysis, we do not show the properties of these 37 sources.
\item 11 sources have a low-quality spectroscopic redshift, from which was possible to estimate only the $z$ value, with no information on the spectral type, and lack of SED template best-fitting information. Therefore, for these 11 sources no type information is provided.
\end{enumerate}

We point out that there is an excellent agreement between the spectral and the SED template best fitting classification, while their both available. 86\% of the BLAGN also have SED best fitted with an unobscured AGN template, and 96\% of the non-BLAGN have SED best fitted with an obscured AGN or a galaxy template. The lower agreement for BLAGN is not surprising, given that BLAGN SEDs, especially those of low-luminosity AGN, can be contaminated by stellar light \citep{luo10,elvis12,hao14}. A summary of the average redshift and X-ray properties of the three subsamples is shown in Table \ref{tab:sample}.

\begin{table*}%\caption{{\normalsize Distribution of CCLS30 sources in different classes, based on optical classification, and their redshift and X-ray average properties.}}
\centering
\scalebox{1.}{
\begin{tabular}{ccccccccc}
\hline
\hline
Class  & Number &  Fraction &  z$_{\rm min}$, z$_{\rm max}$ & f$_{spec}$ & $\langle z_{\rm spec}\rangle$, $\langle z_{\rm phot}\rangle$ & $\langle N_{\rm cts}\rangle$ & $\langle f_{\rm 2-10keV}\rangle$ & $\langle L_{\rm 2-10keV}\rangle$\\
(1) &  (2) &  (3) &  (4) &  (5) &  (6) &  (7) &  (8) &  (9)\\
\hline
Type 1 & 696 & 37.7 & 0.103--5.31 & 83.0 & 1.71, 1.92 & 217.2 & 1.7 $\times$ 10$^{-14}$ & 3.2 $\times$ 10$^{44}$\\
Type 2 & 1111 & 60.3 & 0.066--4.45 & 58.8 & 1.05, 1.58 & 102.2 & 1.4 $\times$ 10$^{-14}$ & 1.7 $\times$ 10$^{44}$\\
Galaxies & 37 & 2.0 & 0.029--0.363 & 84.6 & 0.18, 0.15 & 72.1 & 6.4 $\times$ 10$^{-15}$ & 4.0 $\times$ 10$^{41}$\\
\hline
\hline
\end{tabular}}\caption{{\normalsize Distribution of CCLS30 sources in different classes, based on optical classification, and their redshift and X-ray average properties. (1) Source class; (2) number of sources; (3) fraction of CCLS30 sample (stars and sources without redshift excluded) in each class; (4) minimum and maximum redshift; (5) fraction of spectroscopic redshift; (6) average of the spectroscopic and photometric redshifts; (7) average 0.5--7 keV net counts; (8) average 2--10 keV flux; (9) average 2--10 keV rest-frame, absorption-corrected luminosity.}}\label{tab:sample}
\end{table*}

\section{Spectra extraction}\label{sec:spec_analysis}
We first extract a spectrum in each of the fields where a source was observed, using the CIAO \citep{fruscione06} tool \texttt{specextract}. We used CIAO 4.7 and CALDB 4.6.9. The \texttt{specextract} tool creates a source and background spectrum for each input position, together with the respective response matrices, ARF and RMF. For the source spectral extraction we use a circular region with radius $r_{90}$, i.e., the radius which contains 90\% of the PSF in the 0.5-7 keV band; $r_{90}$ was computed using the CIAO tool \texttt{psfsize\_srcs}, for each source in each observation where the source has been observed (1 to 16 observations\footnote{23 \cha \leg\ fields have been observed in two or three separate observations, due to instrumental constraints.}). The most common number of observations per source is 4 (463 sources, 25\%), and 331 sources (18\%) have been observed in 8 or more fields.

To extract the background spectrum, we use event files where the detected sources have been previously removed, to avoid source contamination to the background. The background spectra have been extracted from an annular region centered on the source position and with inner radius $r_{90}$+2.5$^{\prime\prime}$ and outer radius $r_{90}$+20$^{\prime\prime}$. These radii were chosen to avoid contamination from the source emission, and to have enough counts to obtain a reliable background spectrum. As a result, the mean (median) number of background counts in the 0.5-7 keV band is 149.1 (154.7), and only 78 sources (i.e., $\sim$4\% of the CCLS30 sample) have less than 50 background counts in the 0.5-7 keV band. These 78 sources are mainly located in low-exposure pointings.

All the spectra obtained for a single source have finally been combined in a single spectrum, using the CIAO tool \texttt{combine\_ spectra}. We set the bscale method parameter, i.e., the parameter which determines how are the background counts combined, to ``counts'', because this algorithm is the suggested one to have background counts and backscale values properly weighted when the background is going to be modeled rather than subtracted\footnote{http://cxc.harvard.edu/ciao/ahelp/combine spectra.html}.

\section{Spectral fitting}\label{sec:fit}
The spectral  fitting was performed using the CIAOmodelling and fitting package SHERPA \citep{freeman01}. All the fits were performed using the Cstat statistics, which is based on the Cash statistics \citep{cash79} and is usually adopted for low-counts spectral fitting, since in principle does not require counts binning to work. The main difference between Cstat and the original Cash statistics is that the change in Cstat statistics adding or subtracting a parameter to a model ($\Delta$C) is distributed similarly to $\Delta \chi^2$. Therefore, it is possible to use the reduced Cstat, Cstat$_\nu$=Cstat/DOF, where DOF is the number of degrees of freedom of the fit, as an estimator of the fit goodness; a good  fit should have Cstat$\sim$1. It is also worth mentioning that, given that Cstat$_\nu$ is a good estimator of the fit goodness only if the fitted spectra have more than 1 count per bin, to avoid empty channels. We binned our spectra with 3 counts per bin for an easier visual inspection of the fits. Cstat does not work with background-subtracted data, being a maximum likelihood function and assuming a purely Poissonian count error. For this reason, our analysis requires a proper modelling of the background, to find the best-fit which is then included in the final model of the source+background spectrum. We describe the model we adopted to fit the \cha ACIS-I background in Appendix \ref{app:back}.

\subsection{Source modelling}\label{sec:model}
We now describe the procedure we adopted to find the best-fitting model for each source in our sample. We started from a basic model, an absorbed power-law, then we added further components, looking for a statistically significant improvement in the fit, such as $\Delta$Cstat=Cstat$_{\rm old}$-- Cstat$_{\rm new} >$ 2.71 \citep[see, e.g.,][which validated this value with extended simulations]{tozzi06,brightman14}, where Cstat$_{\rm old}$ is the Cstat value of the best fit of the original model, while Cstat$_{\rm new}$ is the Cstat value of the best fit of the model with the additional component. It is worth noticing that $\Delta$Cstat=2.71 corresponds to a fit improvement with 90\% confidence only if Cstat$_\nu\sim$1 \citep{brightman14}, which is a true statement for the majority of our fits (Figure \ref{fig:cstat_dof}). 

For all fits, we fixed the Galactic absorption to the average value observed in the direction of the COSMOS field \citep[$N_{\rm H,gal}$=2.5 $\times$ 10$^{20}$ cm$^{-2}$; ][]{kalberla05}.

\begin{enumerate}
\item For the 968 sources with 30$<$cts$<$70, we fitted an absorbed power-law with fixed photon-index $\Gamma_1$=1.9, keeping the rest frame absorbed column, $N_{\rm H,z}$, free to vary.
\item We then fitted all the 1855 sources with an absorbed power-law with $\Gamma_1$ and $N_{\rm H,z}$ free to vary. 296 of the 967 sources with $<$70 net counts have a best-fit significantly improved with respect to the fit with $\Gamma_1$=1.9.
\item For a subsample of sources, all of which are obscured AGN with $N_{\rm H,z}>$ 10$^{22}$ cm$^{-2}$,  we found an improvement to the fit adding to the model a second power-law, with $\Gamma_2$=$\Gamma_1$, no intrinsic obscuration and normalization free to vary. This second power-law models the AGN emission unabsorbed by the torus and/or a scattered component, i.e., light deflected without being absorbed by the dust and gas.
This second normalization, $norm_2$ is always signicantly smaller than the first one, $norm_1$, with the ratio $norm_2$/$norm_1$ ranging between 3 $\times$ 10$^{-2}$ and 0.15. 57 sources have best-fit significantly improved with respect to the model with a single power-law, 29 of which have fixed $\Gamma$=1.9, while the other 28 have $\Gamma$ free to vary.
\item A fraction of spectra are expected to have an excess in the single power-law fit residuals around 6--7 keV (rest-frame), this excess being due to the iron K$\alpha$ emission line at 6.4 keV. For this reason, we added to our absorbed power-law fit an emission line at 6.4 keV, modelled with a gaussian having line width $\sigma$=0.1 keV, and we refitted the spectra of all the 1855 sources in CCLS30. We freezed the redshift value of the line for those sources with a spectroscopic redshift, while we left the redshift free to vary within $z+\Delta z$, with $\Delta z$=0.5, for the photometric redshifts. More than 90\% of the sources with only a photo-z in CCLS30 have $\Delta z <$0.5, so we choose this value as a conservative threshold. We find that 130 (82) sources have best-fit significantly improved with respect to the single power-law  fit in CCLS30 (CCLS70). Moreover, 10 (6) of the sources in the $>$30 ($>$70) counts sample are best-fitted by a double power-law with emission line model. We discuss how the iron K$\alpha$ line equivalent width (EW) correlates with the best-fit photon index $\Gamma$ and $N_{\rm H,z}$ in Section \ref{sec:ew}.
\end{enumerate}

\begin{figure*}%[!h]
\begin{minipage}[b]{.5\textwidth}
  \centering
  \includegraphics[width=1.0\linewidth]{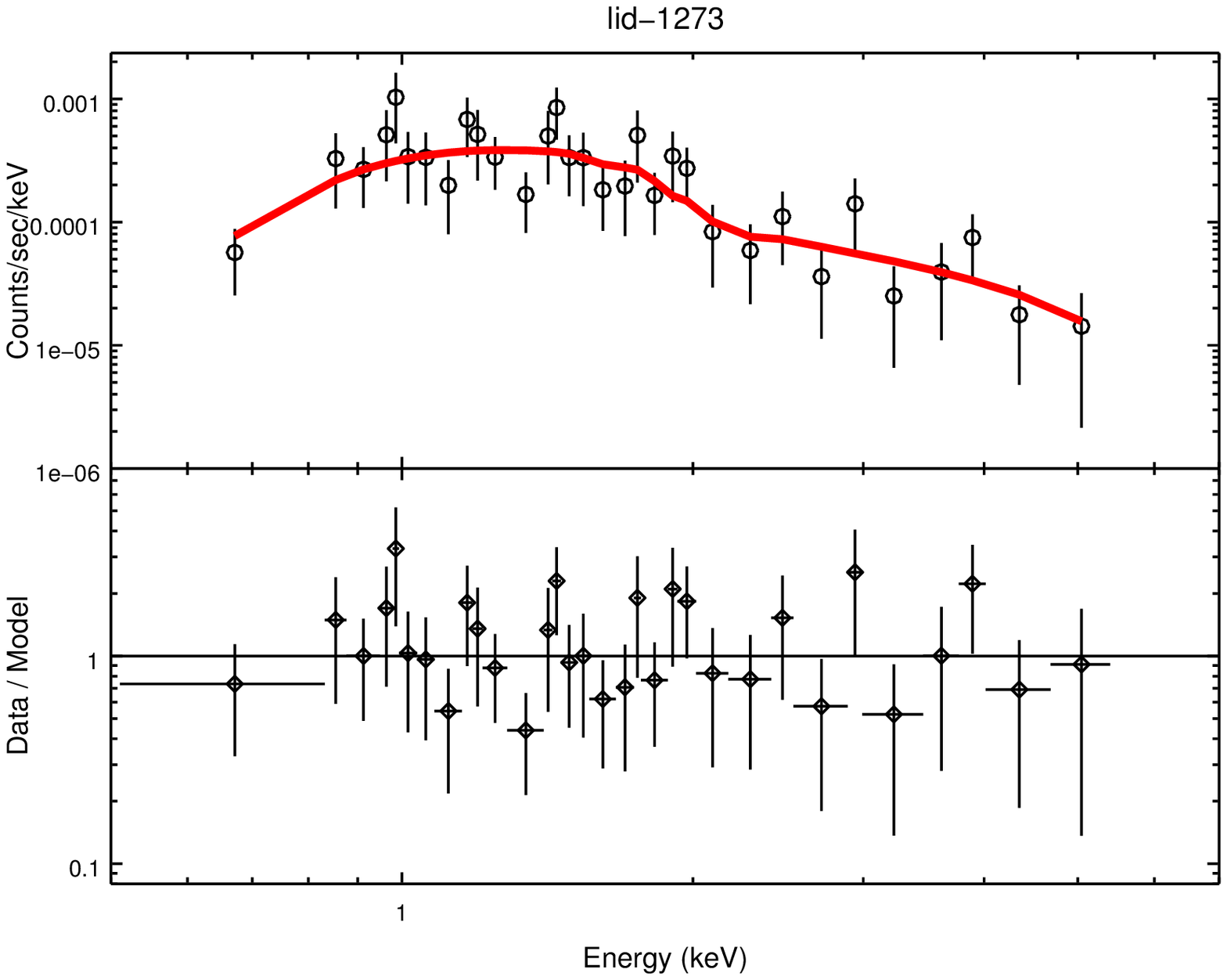}
\end{minipage}
\begin{minipage}[b]{.5\textwidth}
  \centering
  \includegraphics[width=1.00\textwidth]{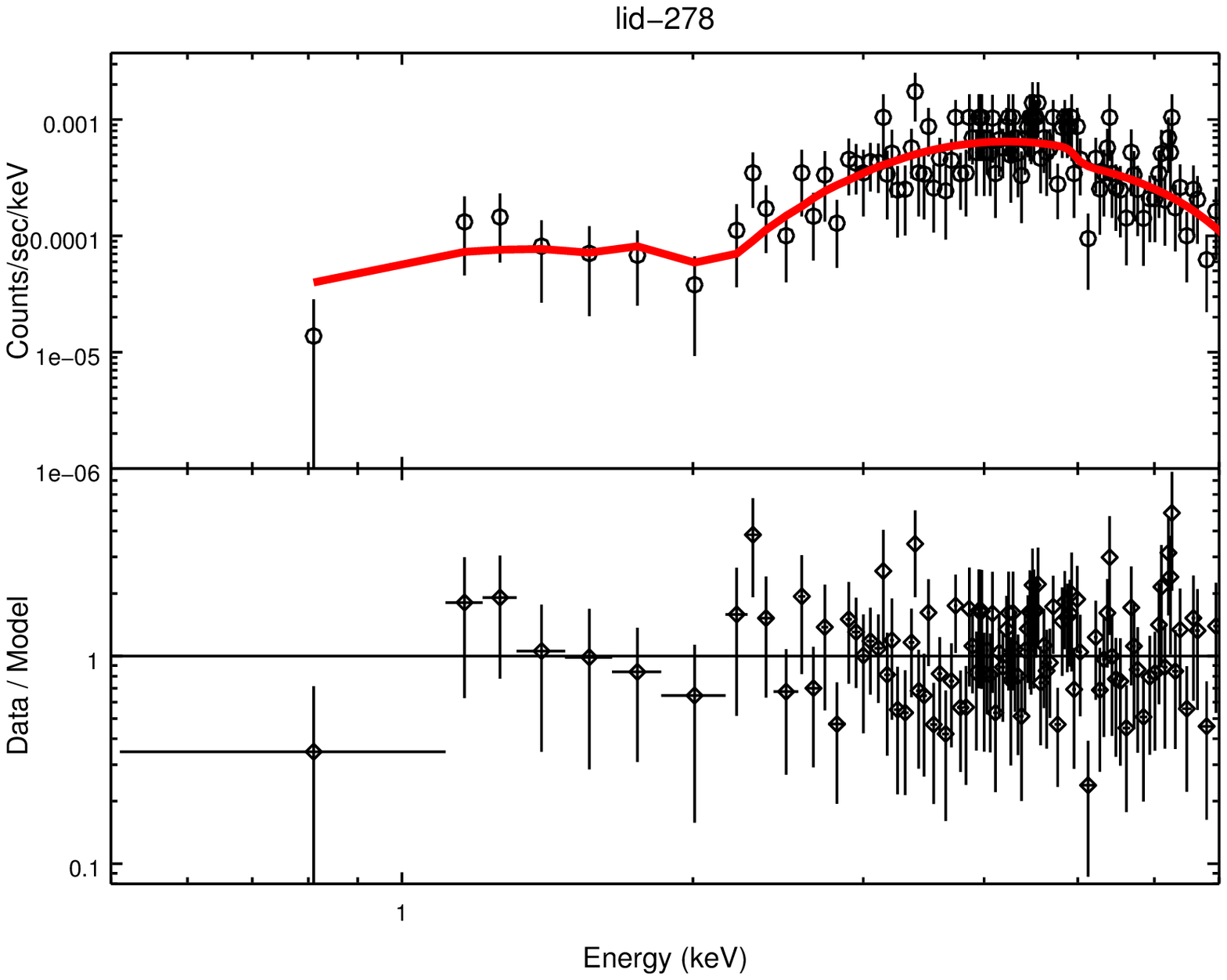}
\end{minipage}
\begin{minipage}[b]{.5\textwidth}
  \centering
  \includegraphics[width=1.0\linewidth]{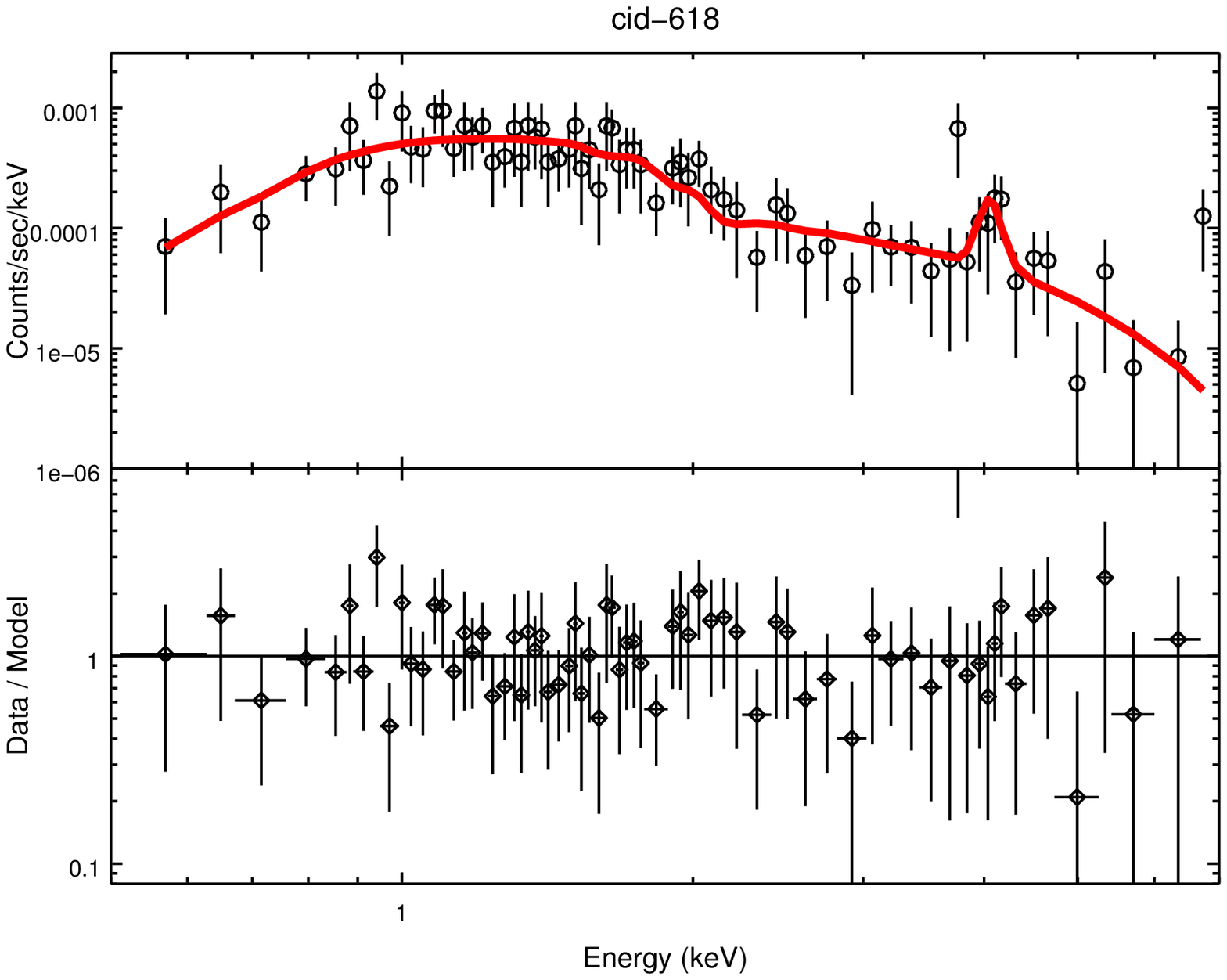}
\end{minipage}%
\begin{minipage}[b]{.5\textwidth}
  \centering
  \includegraphics[width=1.00\textwidth]{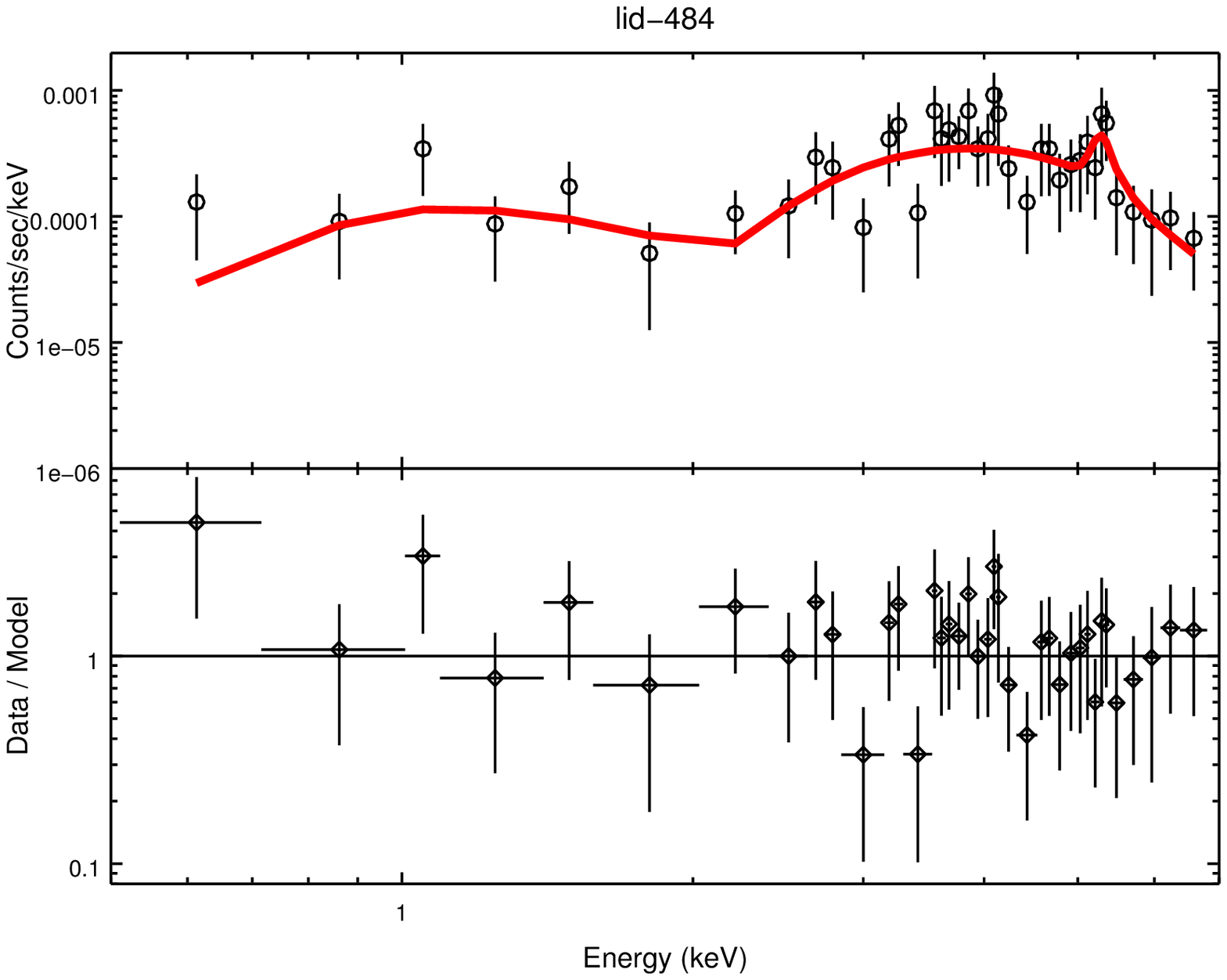}
\end{minipage}
\caption{{\normalsize Examples of different best-fits. Top left: absorbed power-law. Top right: double power-law. Bottom left: absorbed power-law with iron K$\alpha$ emission line. Bottom right: double power-law with iron K$\alpha$ emission line.}}\label{fig:fit_example}
\end{figure*}

\begin{figure*}%[!h]
\begin{minipage}[b]{.5\textwidth}
  \centering
  \includegraphics[width=1.02\linewidth]{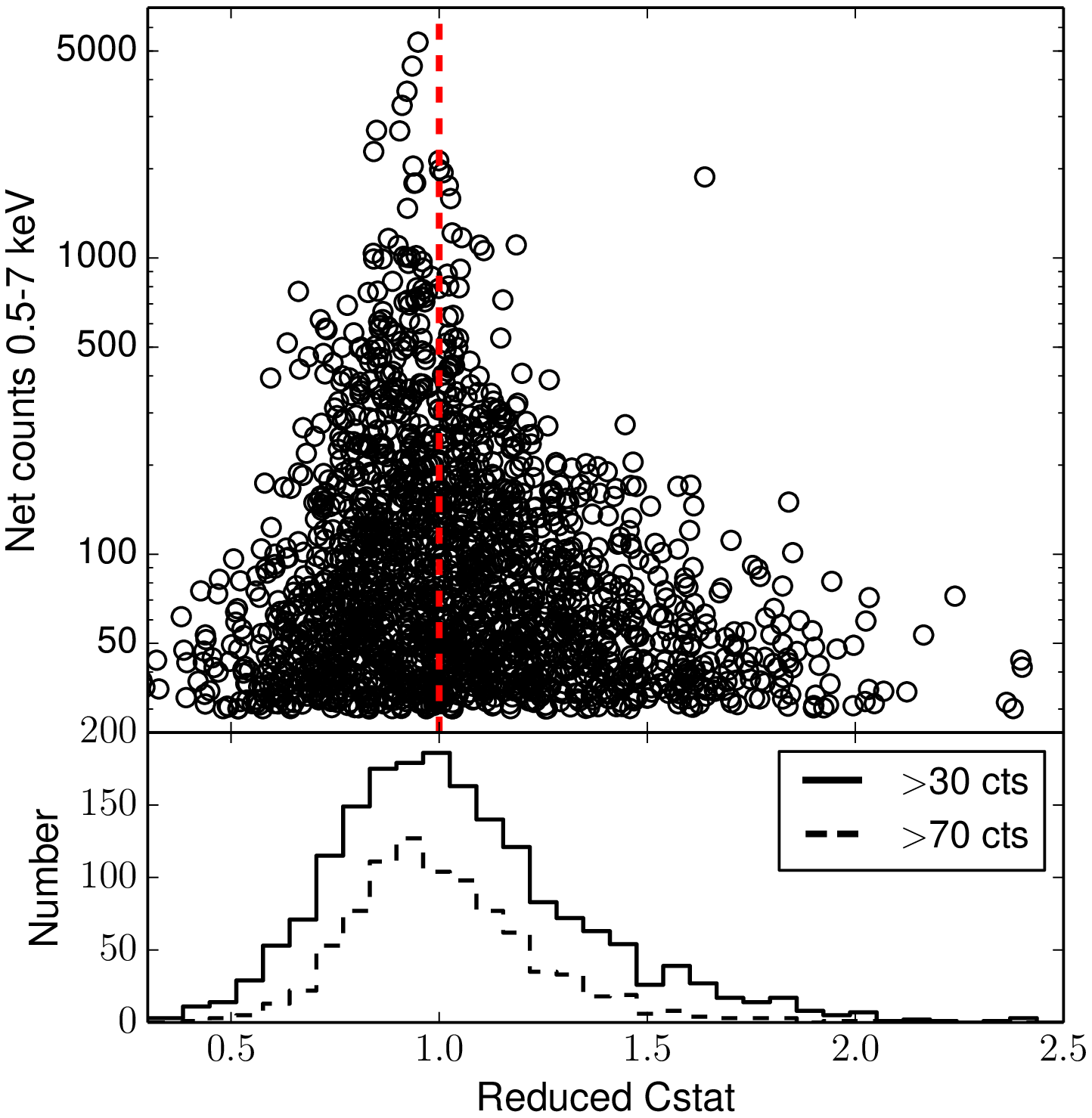}
\end{minipage}%
\begin{minipage}[b]{.5\textwidth}
  \centering
  \includegraphics[width=1.00\textwidth]{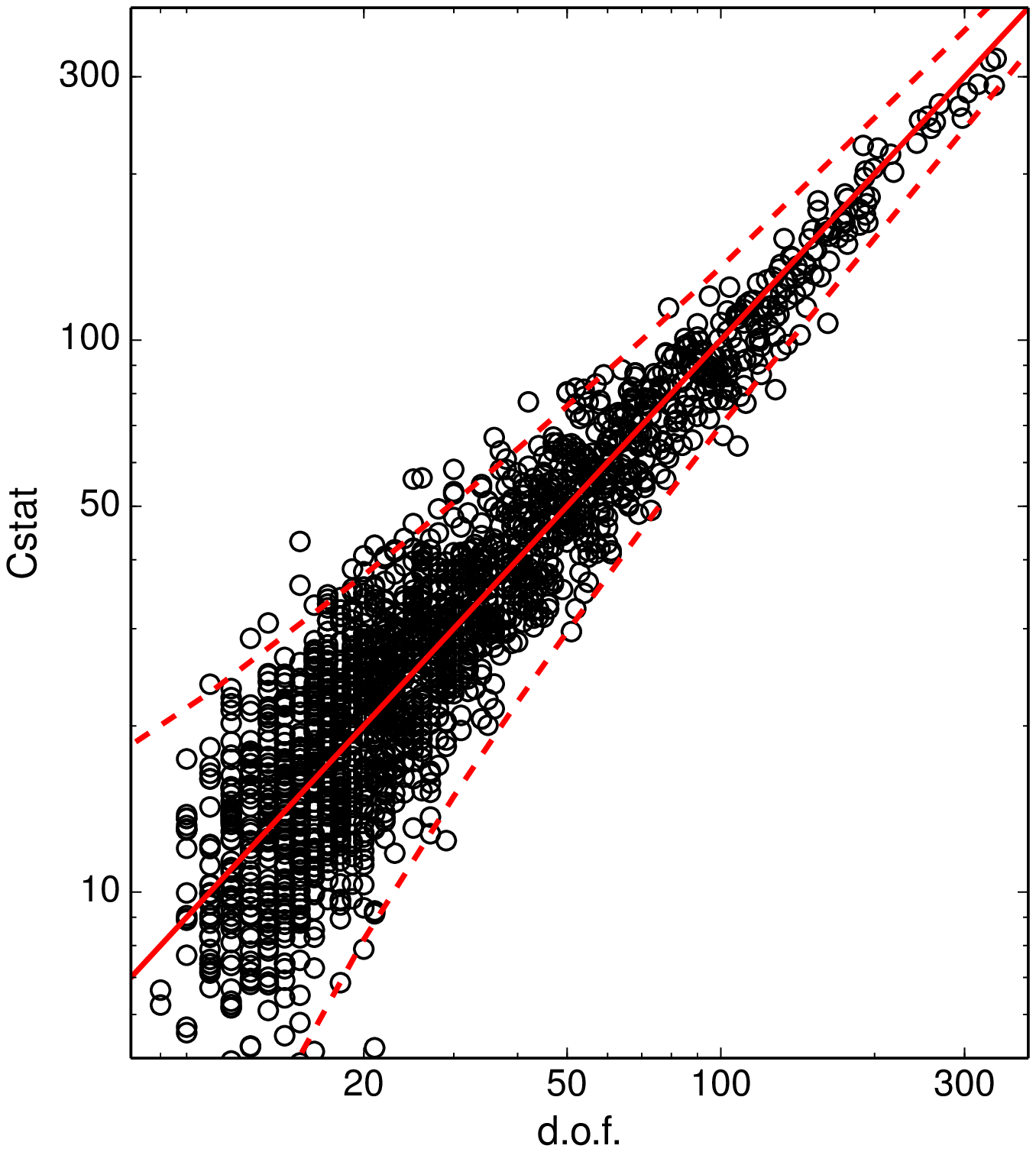}
\end{minipage}
\caption{{\normalsize Left: Reduced Cstat (Cstat$_\nu$) versus 0.5-7 keV net counts (top panel) and Cstat$_\nu$ distribution (bottom), for the CCLS30 (solid line in the bottom panel) and the CCLS70 (dashed line) samples. Right: Cstat versus degrees of freedom for each source. The red solid line represents the Cstat=DOF trend (i.e., the ideal Cstat$_\nu$=1), while the dashed lines indicate, at any DOF, the Cstat above (below) which a 1 per cent probability to find such high (low) Cstat values is expected.}}\label{fig:cstat_dof}
\end{figure*}

In Figure \ref{fig:fit_example} we show an example of each type of best-fit. In Table \ref{tab:best-fit} we report the number of sources for each class of best-fit, for the CCLS30 and the CCLS70 samples, and for the sample of sources with more than 30 and less than 70 net counts in the 0.5-7 keV band. In Table \ref{tab:best-fit_type} we show the same best-fit division, for Type 1 and Type 2 sources, and for galaxies.

\begin{table}[H]
\centering
\scalebox{1.}{
\begin{tabular}{cccc}
\hline
\hline
Fit & n$_{CCLS30}$ & n$_{30-70}$ & n$_{CCLS70}$\\
\hline
$\Gamma$=1.9 & 609 & 609 & 0\\
$\Gamma$ free &1048 & 272 & 776\\
Double PL & 57 & 35 & 22\\
Fe K$\alpha$ &131 & 48 & 83\\
2PL+Fe K$\alpha$ &10 & 4 & 6\\
\hline
Total & 1855 & 968 & 887\\
\hline
\hline
\end{tabular}}\caption{{\normalsize Number of sources for each class of best-fit, for sources with more than 30, more than 30 and less than 70, and more than 70 net counts in the 0.5-7 keV band.}}\label{tab:best-fit}
\end{table}

\begin{table*}
\centering
\scalebox{1.}{
\begin{tabular}{c|ccc|ccc|ccc}
\hline
\hline
& \multicolumn{3}{c|}{Type 1} & \multicolumn{3}{c|}{Type 2} & \multicolumn{3}{c}{Galaxies}\\
Fit & n$_{CCLS30}$ & n$_{30-70}$ & n$_{CCLS70}$ & n$_{CCLS30}$ & n$_{30-70}$ & n$_{CCLS70}$ & n$_{CCLS30}$ & n$_{30-70}$ & n$_{CCLS70}$\\
\hline
$\Gamma$=1.9 & 177 & 177 & 0 & 412 & 412 & 0 & 20 & 20 & 0\\
$\Gamma$ free & 460 & 66 & 394 & 566 & 199 & 367 & 12 & 5 & 7\\
Double PL & 11 & 6 & 5 & 45 & 28 & 17 & 1 & 1 & 0\\
Fe K$\alpha$ & 46 & 9 & 37 & 80 & 37 & 43 & 4 & 2 & 2\\
2PL+Fe K$\alpha$ & 2 & 1 & 1 & 8 & 3 & 5 & 0 & 0 & 0\\
\hline
\hline
\end{tabular}}\caption{{\normalsize Number of sources for each class of best-fit, for sources with more than 30, more than 30 and less than 70, and more than 70 net counts in the 0.5-7 keV band.}}\label{tab:best-fit_type}
\end{table*}

In Figure \ref{fig:cstat_dof} (left panel) we show the distribution of the reduced Cstat, Cstat$_\nu$=Cstat/DOF, as a function of the 0.5-7 keV net counts, for the 1855 sources in the CCLS30 sample. The distribution is peaked around Cstat$_\nu$=1, with mean (median) Cstat$_\nu$=1.05 (1.01) and standard deviation $\sigma$=0.31. The dispersion is even smaller for the CCLS70 sample, for which $\sigma$=0.23; the same subsample has mean (median) Cstat$_\nu$=1.01 (0.98).

In Figure \ref{fig:cstat_dof} (right panel) we show the distribution of Cstat versus DOF for the sources in CCLS30. The red solid line indicates the case Cstat$_\nu$=1, i.e., Cstat=DOF, while the red dashed lines show the Cstat value, at a given DOF, above (below) which there is 1\% probability to find such a high (low) value if the model is correct. More than 98\% of the sources in our sample lie within the two dashed lines, therefore suggesting that most of the fits are acceptable. Indeed there are 16 sources (0.9\% of the whole sample) below and 21 sources (1.1\%) above the dashed lines, a fraction consistent with random noise fluctuations.

\subsection{Modelling results}
\begin{figure*}%[!h]
\begin{minipage}[b]{.5\textwidth}
  \centering
  \includegraphics[width=1.0\linewidth]{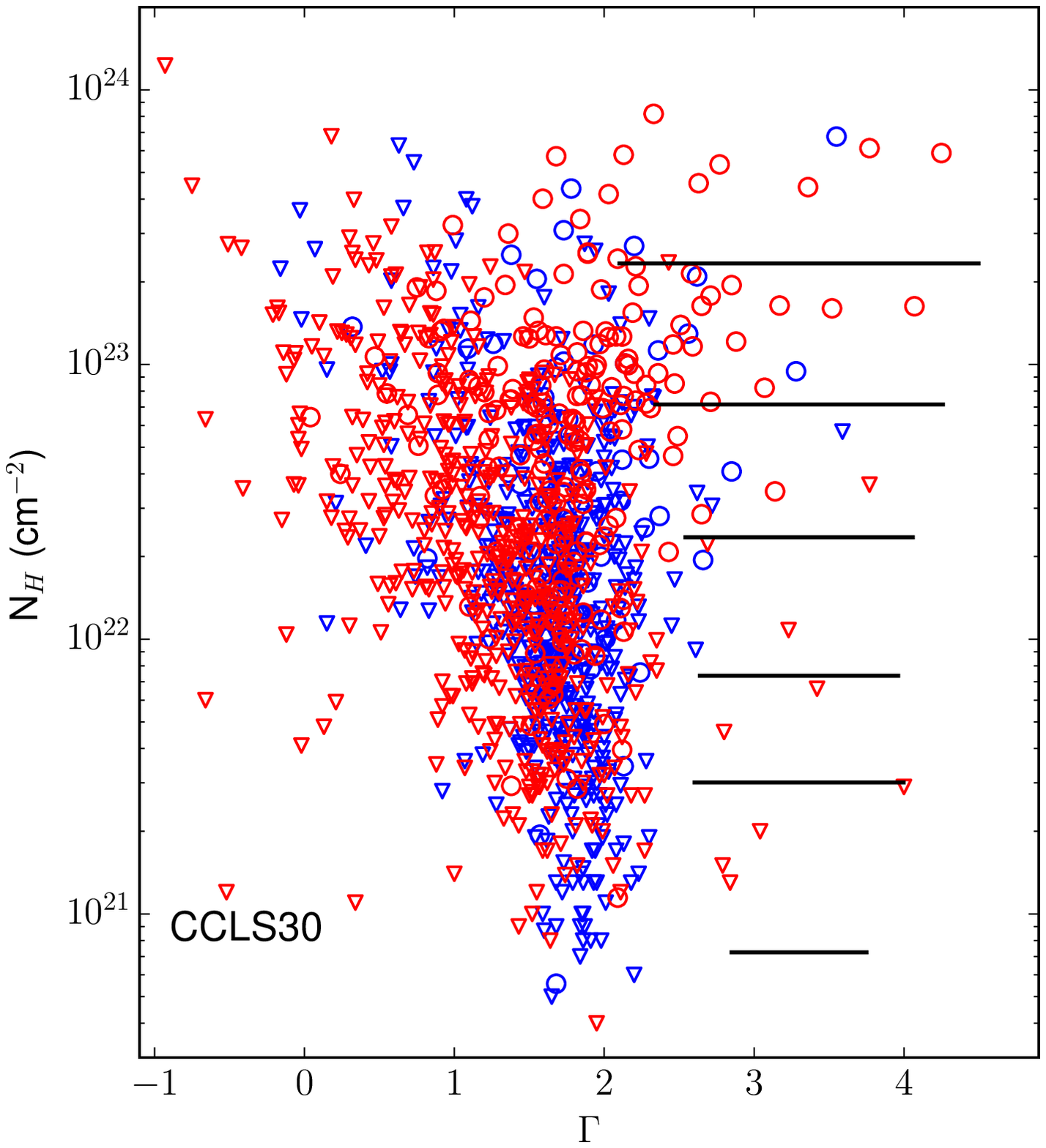}
\end{minipage}%
\begin{minipage}[b]{.5\textwidth}
  \centering
  \includegraphics[width=1.04\textwidth]{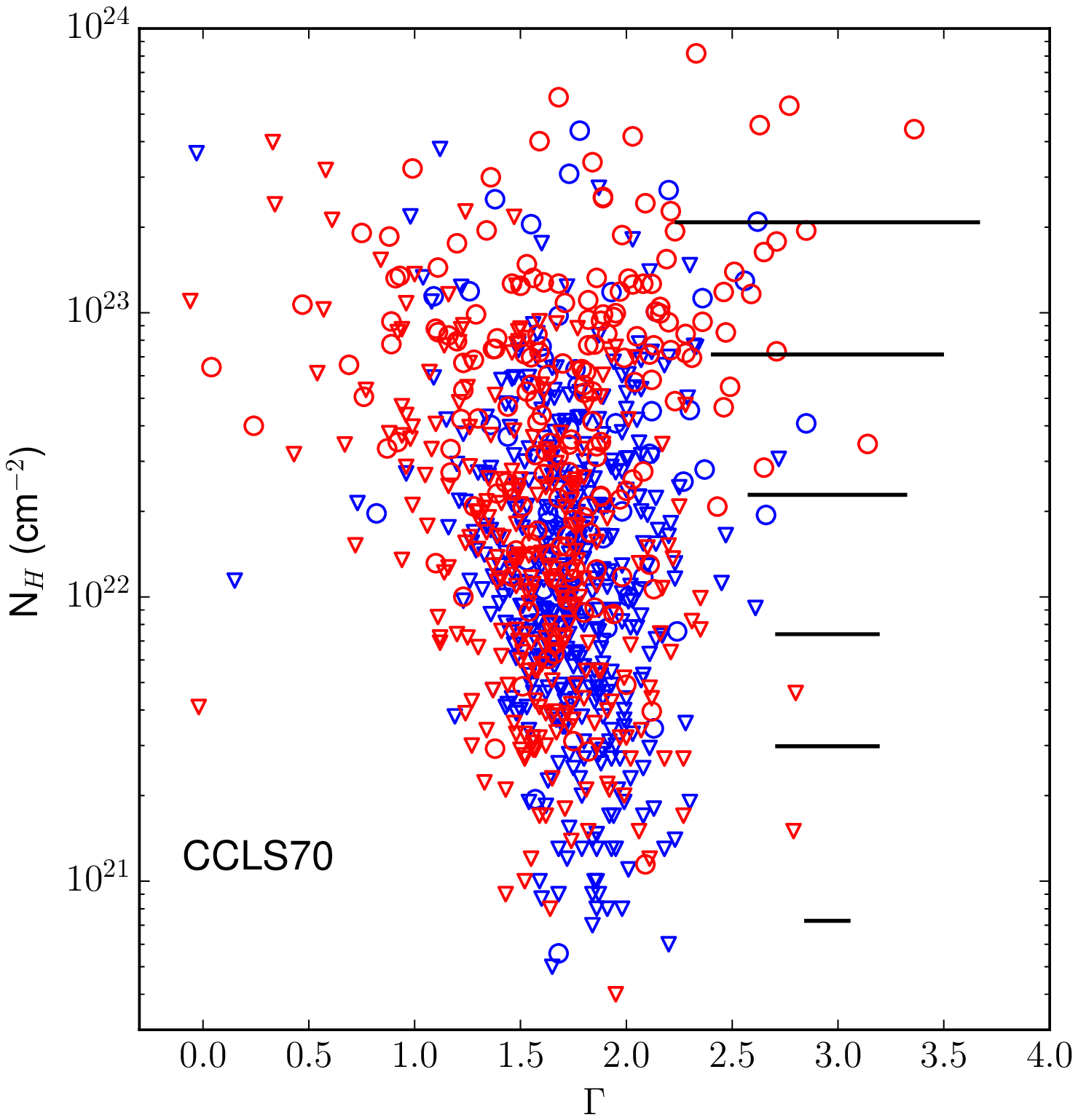}
\end{minipage}
\caption{{\normalsize $\Gamma$ versus $N_{\rm H,z}$ for Type 1 (blue) and Type 2 (red) sources in the CCLS30 (left) and CCLS70 sample (right). 90\% confidence upper limits on $N_{\rm H,z}$ are plotted as triangles. Mean errors on $\Gamma$ in different bins of $N_{\rm H,z}$ are also shown as black horizontal lines.}}\label{fig:gamma_vs_nh}
\end{figure*}

In Figure \ref{fig:gamma_vs_nh} we show the distribution of two main spectral parameters, $\Gamma$ and $N_{\rm H,z}$, for both the CCLS30 (left) and the CCLS70 (right) samples, for all the sources for which we left both parameters free to vary. It is worth noticing that the observed dispersion on $\Gamma$ is significantly smaller for those sources with $N_{\rm H,z}$ $<$10$^{22}$ cm$^{-2}$, i.e., classified as unobscured ($\sigma_{>30}$=0.47, $\sigma_{>70}$=0.31), with respect to those sources with nominal or upper limit at $N_{\rm H,z}$$>$10$^{22}$ cm$^{-2}$ ($\sigma_{>30}$=0.83, $\sigma_{>70}$=0.47). Moreover, at $N_{\rm H,z}$ $>$10$^{22}$ cm$^{-2}$ the errors on $\Gamma$ are larger, since constraining $\Gamma$ becomes more difficult for sources with larger column density. This discrepancy is mainly due to the fact that obscured sources have on average less net counts than unobscured ones, and therefore their best-fit parameters are less constrained. In CCLS30, sources with $N_{\rm H,z}$ $>$10$^{22}$ cm$^{-2}$ have mean (median) net counts in the 0.5--7 keV band $\langle cts \rangle$= 124.7 (90.2), while sources with $N_{\rm H,z}$ $<$10$^{22}$ cm$^{-2}$ have $\langle cts \rangle$= 293.4 (153.4).

\begin{figure*}%[!h]
\begin{minipage}[b]{.5\textwidth}
  \centering
  \includegraphics[width=1.02\linewidth]{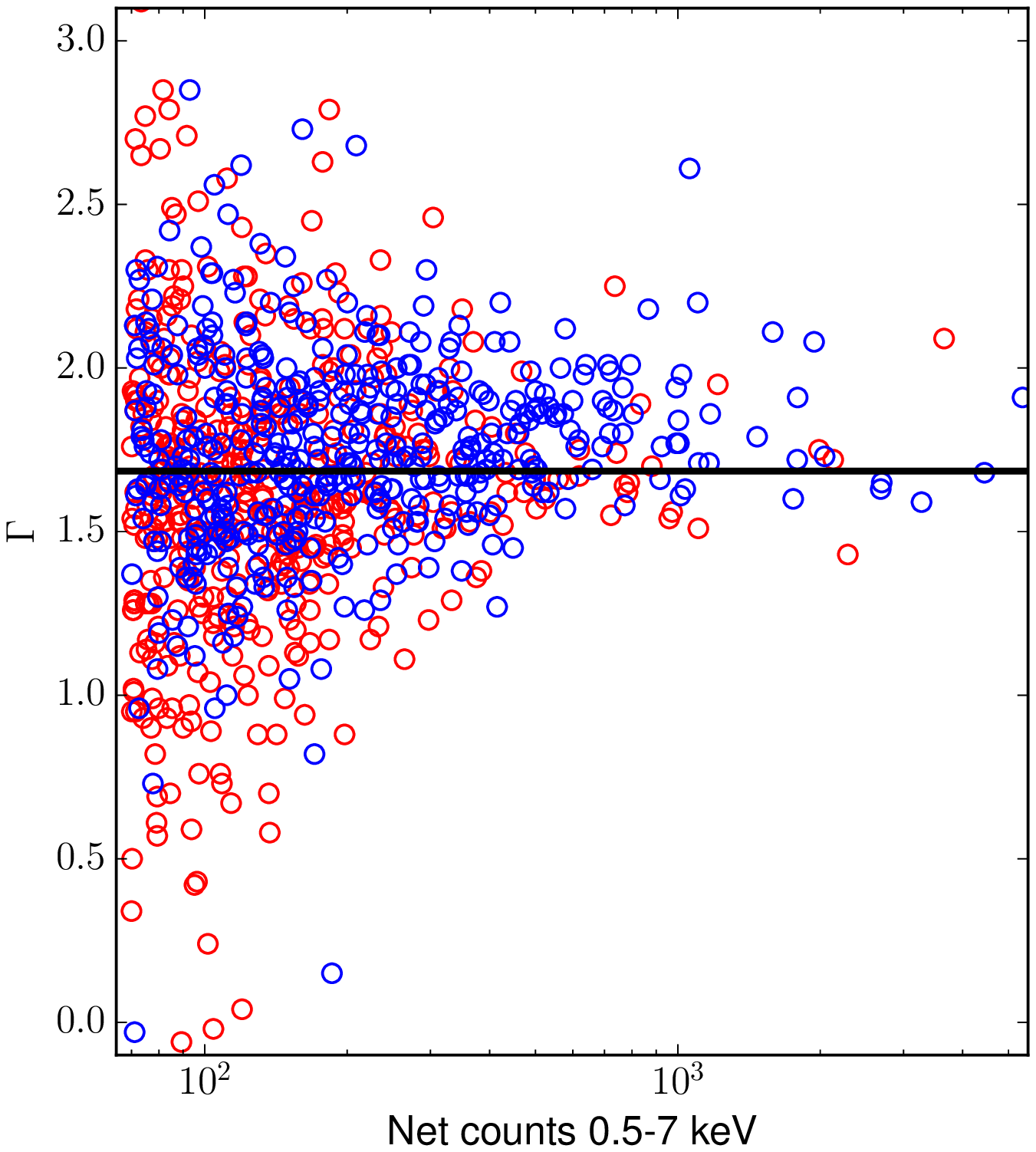}
\end{minipage}%
\begin{minipage}[b]{.5\textwidth}
  \centering
  \includegraphics[width=1.00\textwidth]{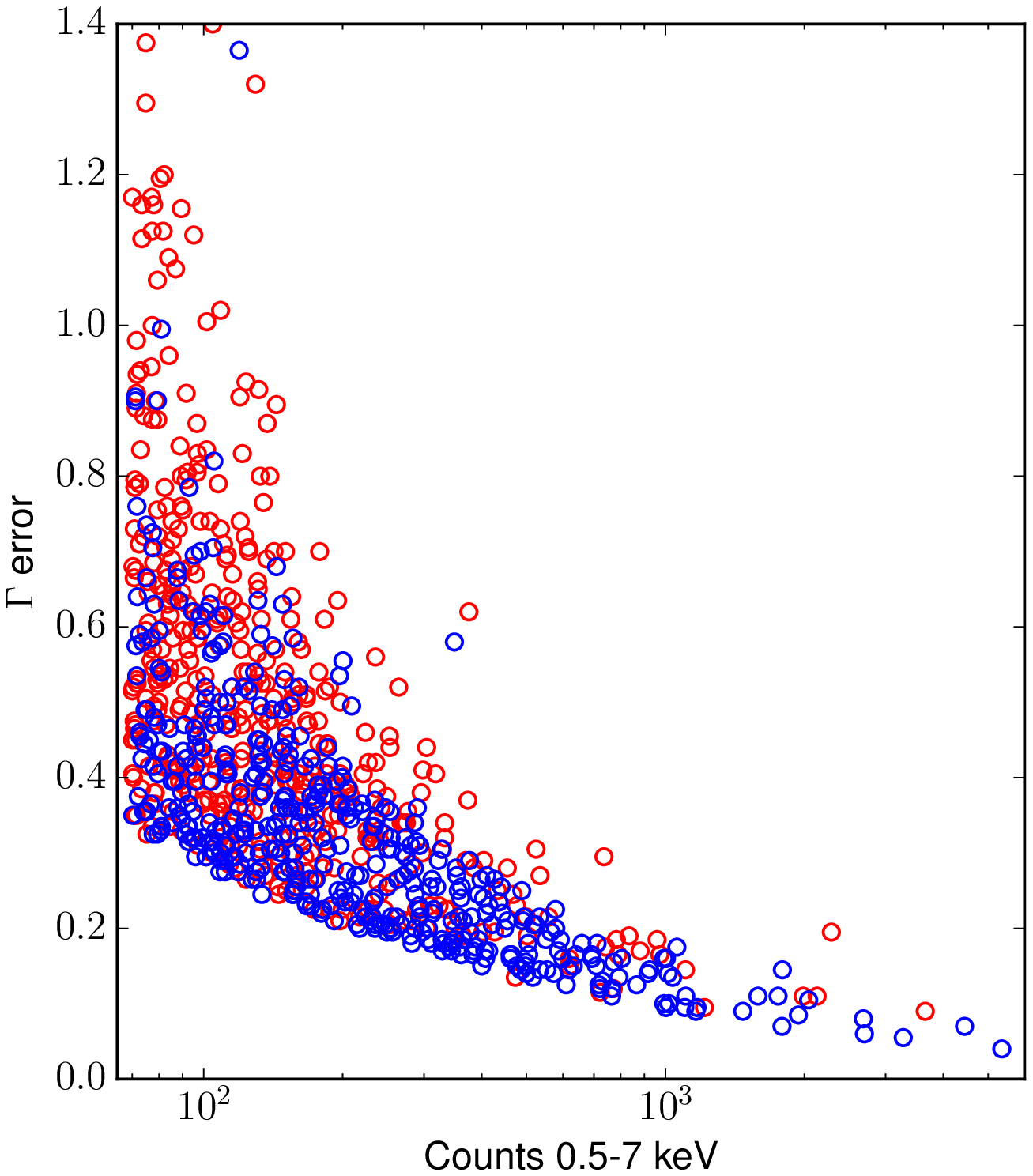}
\end{minipage}
\caption{{\normalsize $\Gamma$ versus 0.5-7 keV net counts (left) and error on $\Gamma$ versus 0.5-7 keV net counts (right), for Type 1 (blue) and Type 2 (red) sources in CCLS70. The horizontal black solid line in the left panel shows the mean photon-index $\langle \Gamma \rangle$=1.68 for the CCLS70 sample.}}\label{fig:gamma_vs_cts}
\end{figure*}

In Figure \ref{fig:gamma_vs_cts} (left) we show the distribution of the photon-index $\Gamma$ as a function of the number of 0.5-7 keV net counts for sources in CCLS70. The black solid line shows the mean $\Gamma$ value for the 877 sources in the CCLS70 sample, $\langle \Gamma \rangle$=1.68. The mean $\Gamma$ of the whole CCLS30 population, i.e., taking into account also the 609 sources with $\Gamma$=1.9 and the 345 sources with less than 70 net counts and $\Gamma$ free to vary which we do not plot in Figure 5, is $\langle \Gamma \rangle$=1.66. Sources with peculiar $\Gamma$ values (i.e., $\Gamma>$3 or $\Gamma<$1), which mainly contribute to the $\Gamma$ distribution dispersion, have for the most part less than 70 net counts and therefore their best-fit estimates are expected to be affected by larger uncertainties than those of brighter sources. A significant fraction of objects with less than 70 net counts (162 out of 967, $\sim$17\%) have flat spectra (i.e.,  $\Gamma<$1). Sources with such a photon index are candidate reflection dominated, Compton Thick (CT) AGN. However, these objects also have, on average, larger uncertainties on $\Gamma$. Therefore an extensive analysis of these objects, including a fit with more complex model than those used in this work, is required to determine how many of them are actual CT AGN, and will be performed in Lanzuisi et al. (in prep.).

The relation between number of counts and error on $\Gamma$ can be seen in Figure \ref{fig:gamma_vs_cts} (right). The fraction of sources with relative error, $\Delta$err=err$_\Gamma$/$\Gamma$, larger than 30\% is $\sim$22\% for sources with $>$30 counts, but significantly drops to $\sim$8\% ($\sim$4\%) for sources with $>$70 ($>$100) counts.

\section{Fitted parameters distribution}\label{sec:fit_distr}

\subsection{Intrinsic absorption column density, $N_{H,z}$}\label{sec:nh}
\begin{figure*}%[!h]
\begin{minipage}[b]{.5\textwidth}
  \centering
  \includegraphics[width=1.02\linewidth]{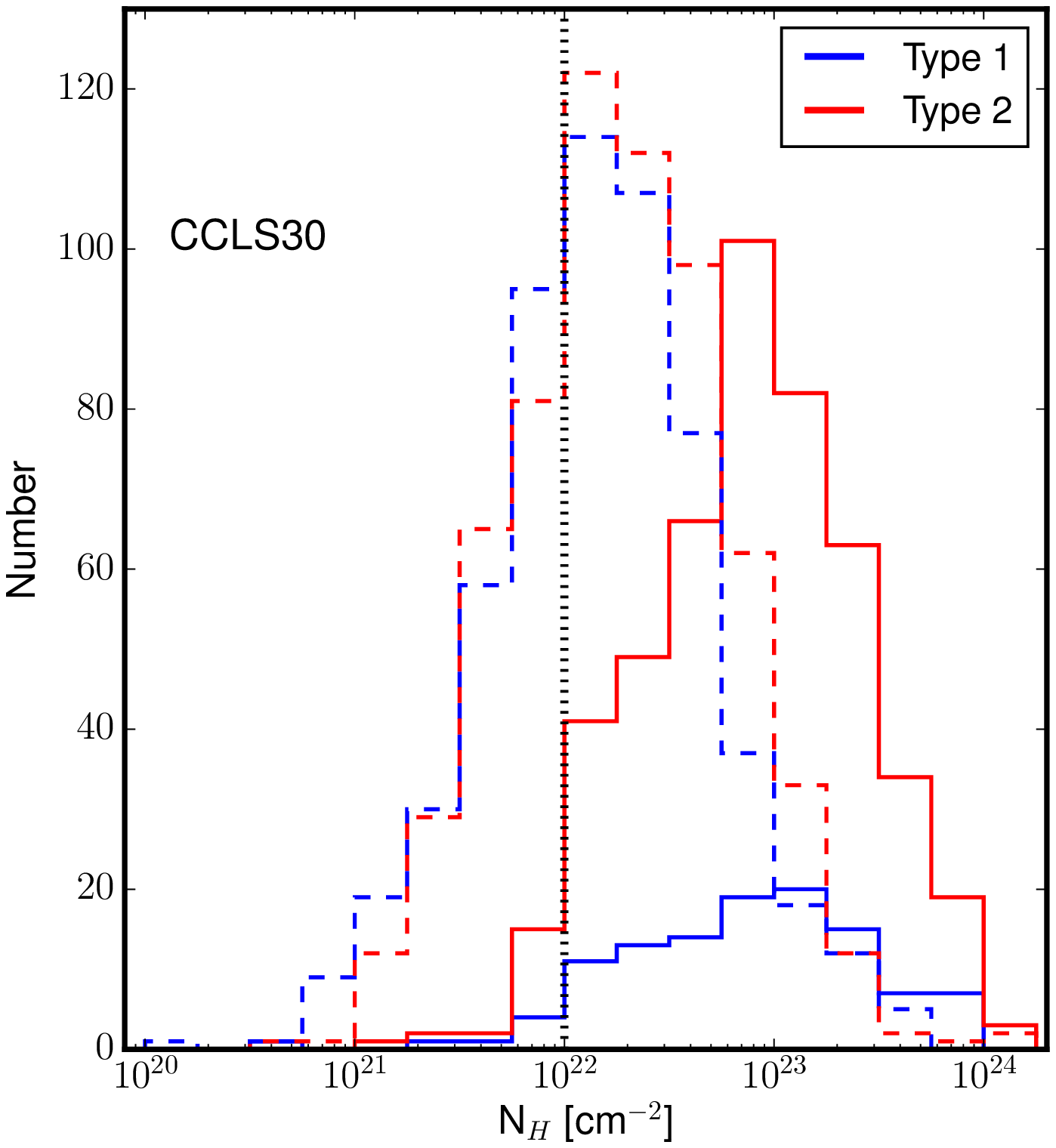}
\end{minipage}%
\begin{minipage}[b]{.5\textwidth}
  \centering
  \includegraphics[width=1.00\textwidth]{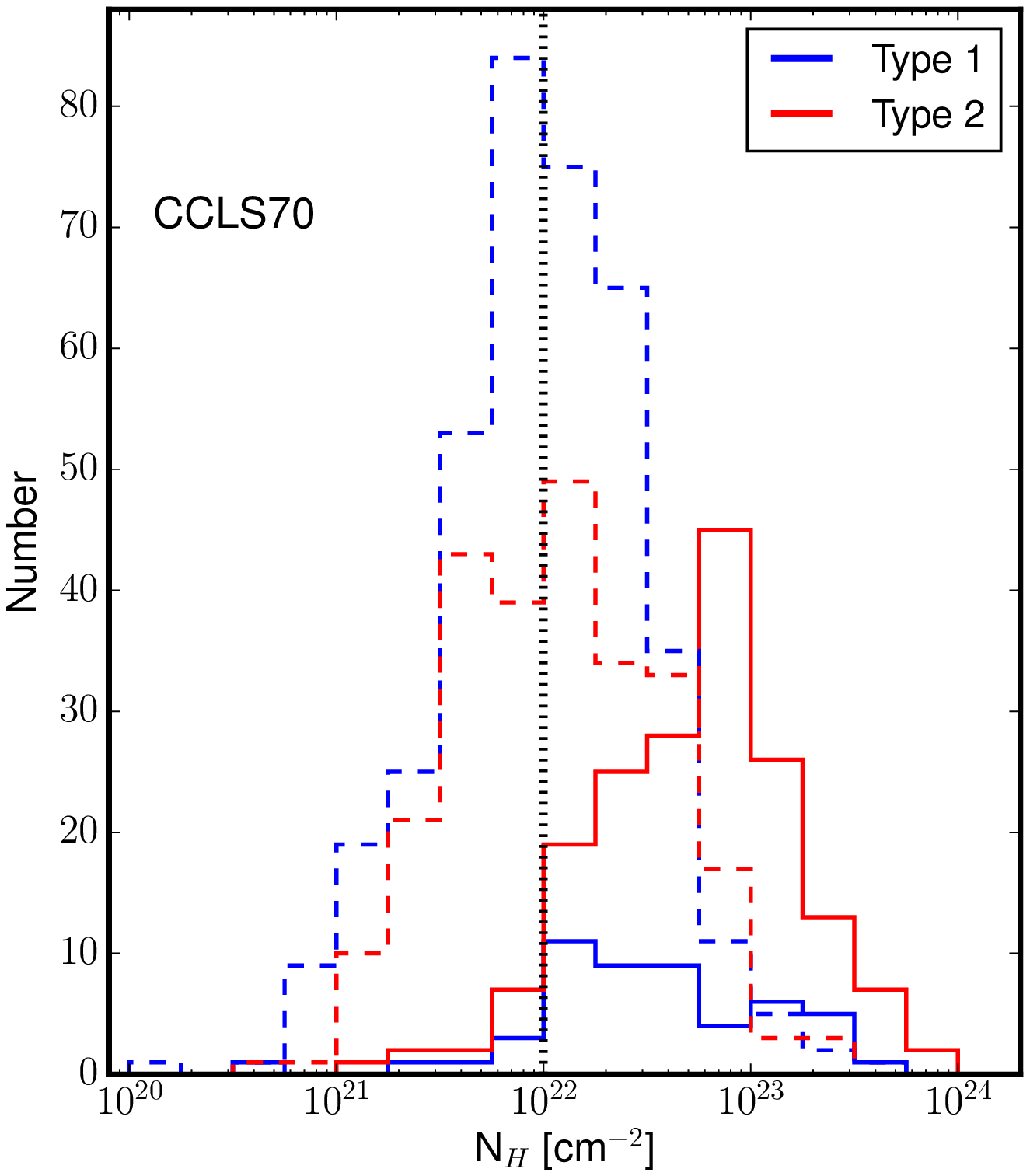}
\end{minipage}
\caption{{\normalsize $N_{H,z}$ distribution for Type 1 (blue) and Type 2 (red) AGN, for sources in CCLS30 (left) and CCLS70 (right). Nominal values are plotted with solid lines, while upper limits are plotted with dashed lines. The black dotted line at $N_{H,z}$=10$^{22}$ cm$^{-2}$ marks the conventional threshold between unobscured and obscured sources.}}\label{fig:nh_histo}
\end{figure*}

In Figure \ref{fig:nh_histo} we show the $N_{H,z}$ distribution, for both Type 1 (blue) and Type 2 (red) AGN. Nominal values are shown with solid lines, while the 90\% confidence upper limits distributions are plotted with dashed lines.

We computed the Kaplan-Meier estimators of the mean values of $N_{\rm H,z}$ for Type 1 and Type 2 sources in CCLS30 and CCLS70, using the ASURV tool, Rev 1.2 \citep{isobe90,lavalley92}, which implements the methods presented in \citet{feigelson85}, to properly take into account 90\% confidence upper limits. We report these mean values in Table \ref{tab:nh_mean}.

Type 1 AGN are significantly less obscured than Type 2 sources. In CCLS70, 45 Type 1 AGN (10.3\% of the whole Type 1 AGN population) have a 90\% confidence intrinsic absorption value $N_{\rm H,z}$$>$10$^{22}$ cm$^{-2}$ (i.e., above the threshold usually adopted to distinguish between obscured and unobscured sources), while 165 Type 2 AGN (38.3\%) have $N_{\rm H,z}$$>$10$^{22}$ cm$^{-2}$ at a 90\% confidence level, and other 35 have $N_{\rm H,z}$$>$10$^{22}$ cm$^{-2}$ within 1$\sigma$. In CCLS30 the fraction of obscured sources slightly increases in both Type 1 (106 sources, 15.2\%) and Type 2 sources (460 sources, 41.4\%). The fractions do not change significantly if we take into account only sources with spectroscopic classification, therefore ruling out a pure SED-fitting template misclassification. We summarize the number of obscured sources per AGN type in Table \ref{tab:nh_obs}.

Finally, in Table \ref{tab:nh_upper} we report the fraction of sources with only an upper limit on $N_{\rm H,z}$: as expected, the fraction of upper limits in Type 1 sources (83.8\% in CCLS30, 88.3\% in CCLS70) is significantly higher than in Type 2 sources (56.8\% in CCLS30, 58.9\% in CCLS70), at any counts threshold. We also point out that the large fraction of upper limits on $N_{\rm H,z}$ is mainly due to the low counts statistics of the majority of the objects in CCLS30, and cannot be related to intrinsically low values of $N_{\rm H,z}$. This is particularly true for Type 2 AGN, where 157 CCLS30 sources (14\%) have an upper limit on $N_{\rm H,z}$ larger than 5 $\times$ 10$^{22}$ cm$^{-2}$, i.e., well above the 10$^{22}$ cm$^{-2}$ threshold.

\begin{table}[H]
\centering
\scalebox{1.}{
\begin{tabular}{ccc}
\hline
\hline
Type & CCLS30 & CCLS70\\
\hline
Type 1 & 2.57$\pm$0.36 & 0.88$\pm$0.18\\
Type 2 & 6.53$\pm$0.40 & 4.02$\pm$0.41\\
\hline
\hline
\end{tabular}}\caption{{\normalsize $N_{\rm H,z}$ mean value for Type 1 and Type 2 AGN, for sources in CCLS30 and CCLS70, computed with the ASURV tool to take into account the 90\% confidence upper limits. All values are in units of 10$^{22}$ cm$^{-2}$.}}\label{tab:nh_mean}
\end{table}

\begin{table*}
\centering
\scalebox{1.}{
\begin{tabular}{ccccccccc}
\hline
\hline
Type & n$_{>1E22}$ & f$_{>1E22}$ & n$_{>1E22}$ & f$_{>1E22}$ & n$_{>1E22}$ & f$_{>1E22}$ & n$_{>1E22}$ & f$_{>1E22}$\\
& \multicolumn{2}{c}{CCLS30, all} & \multicolumn{2}{c}{CCLS30, spec} & \multicolumn{2}{c}{CCLS70, all} & \multicolumn{2}{c}{CCLS70, spec}\\
\hline
1 & 106 & 15.2\% & 72 & 13.5\% & 45 & 10.3\% & 41 & 10.8\%\\
2 & 460 & 41.4\% & 206 & 38.6\% & 165 & 38.3\% & 93 & 35.6\%\\
\hline
\hline
\end{tabular}}\caption{{\normalsize Number and fraction of objects with $N_{\rm H,z}>$10$^{22}$ cm$^{-2}$, for Type 1 and Type 2 AGN, for sources in CCLS30 and in CCLS70. We also computed numbers and fractions for sources with reliable spectral type only. The fraction is computed on the total number of sources of the same type.}}\label{tab:nh_obs}
\end{table*}

\begin{table}[H]
\centering
\scalebox{1.}{
\begin{tabular}{ccccc}
\hline
\hline
Type & n$_{\rm up}$/n$_{\rm tot}$ & f$_{\rm up}$ & n$_{\rm up}$/n$_{\rm tot}$ & f$_{\rm up}$\\
& \multicolumn{2}{c}{CCLS30} & \multicolumn{2}{c}{CCLS70}\\
\hline
All & 1214/1807 & 67.2\% & 640/868 & 73.7\%\\
Type 1 & 583/696 & 83.8\% & 386/437 & 88.3\%\\
Type 2 & 631/1111 & 56.8\% & 254/431 & 58.9\%\\
\hline
\hline
\end{tabular}}\caption{{\normalsize Ratio between sources with an upper limit on $N_{\rm H,z}$ and total number of sources, and fraction of sources with a 90\% confidence upper limit on $N_{\rm H,z}$, for Type 1 and Type 2 AGN, and for all sources with optical classification.}}\label{tab:nh_upper}
\end{table}

\subsection{Photon index, $\Gamma$}\label{sec:gamma}
\begin{figure}[H]
\centering
\includegraphics[width=0.5\textwidth]{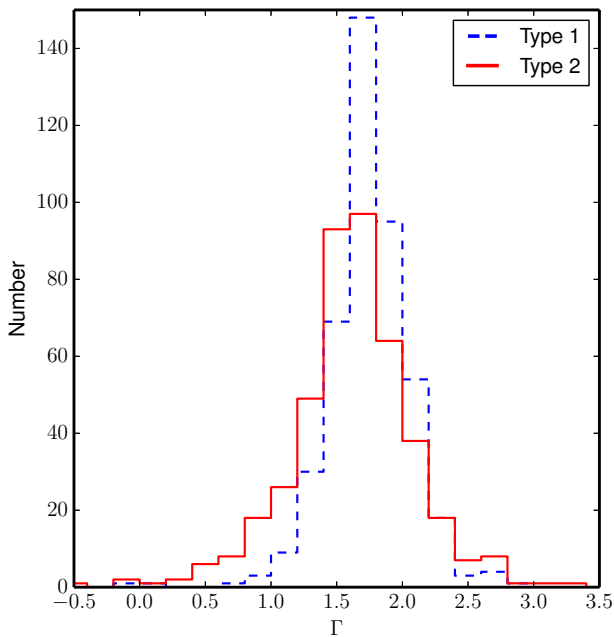}
\caption{{\normalsize Photon Index $\Gamma$ distribution for Type 1 (blue dashed line) and Type 2 (red solid line) AGN, for sources in CCLS70.}}\label{fig:histo_gamma}
\end{figure}

In Figure \ref{fig:histo_gamma} we show the distribution of $\Gamma$ for the CCLS70 sample, for Type 1 (blue dashed line) and Type 2 (red solid line) sources. We do not plot the CCLS30 histogram because for a significant fraction of sources in the 30-70 net counts sample we fixed $\Gamma$=1.9 (see Table \ref{tab:best-fit}).

The mean and $\sigma$ of the $\Gamma$ distribution for Type 1 and Type 2 AGN in CCLS70 are the following:
\begin{enumerate}
\item Type 1: the mean (median) $\Gamma$ is 1.75$\pm$0.02 (1.74), with dispersion $\sigma$=0.31.
\item Type 2: the mean (median) $\Gamma$ is 1.61$\pm$0.02 (1.62), with dispersion $\sigma$=0.47.
\end{enumerate}

The error on the mean is computed as err=$\frac{\sigma}{N}$, where $\sigma$ is the distribution dispersion and N is the number of sources in each sample.

The probability that the two distributions are drawn from the same population, on the basis of a Kolgomorov-Smirnov (KS) test, is $P$=1.9$\times$10$^{-9}$. Therefore, assuming that we are properly constraining $\Gamma$ (which might be not true for sources with high $N_{H,z}$, see Figure \ref{fig:gamma_vs_nh}), we find that Type 2 source have flatter photon index than Type 1 sources, a result already found in \citet{lanzuisi13}.

To better understand if the difference in $\Gamma$ betweenType 1 and Type 2 AGN may be caused by extreme objects, we compute the mean and $\sigma$ on $\Gamma$ of the Type 1 and Type 2 samples taking into account only those sources with 1$< \Gamma<$ 3. To this end, we exclude from our analysis very soft objects and candidate highly obscured, reflection dominated sources. In this subsample, Type 1 AGN have mean (median) photon index $\langle \Gamma \rangle$=1.77$\pm$0.01 (1.75), with dispersion $\sigma$=0.28, while Type 2 AGN have mean (median) photon index $\langle \Gamma \rangle$=1.70$\pm$0.02 (1.66), with dispersion $\sigma$=0.34. The difference between the two samples is now smaller, mainly because the Type 2 sample does not contain anymore the candidate CT-AGN, which caused both a  flattening of the average $\Gamma$ and an increasing in the dispersion. However, the KS-test still excludes that the two distributions are drawn from the same population, with probability $P$=2.2$\times$10$^{-6}$. 

Finally, we point out that a fraction of Type 2 AGN are expected to be unobscured, low Eddington ratio sources lacking of broad emission lines \citep[see, e.g., ][]{trump11,marinucci12}, and sources with low Eddington ratio are also expected to have flatter photon index \citep[see, e.g., ][]{shemmer08,risaliti09}. To verify if this is the case for CCLS70, we divide the Type 2 sources in obscured or unobscured using the $N_{\rm H,z}$=10$^{22}$ cm$^{-2}$ threshold, selecting only sources with 1$<\Gamma<$3 to reduce the number of sources where $N_{\rm H,z}$ is most likely poorly constrained. We find that this obscured Type 2 AGN subsample contains 185 sources and has $\langle \Gamma \rangle$=1.76$\pm$0.03, with dispersion $\sigma$=0.39, while the unobscured one contains 115 sources and has $\langle \Gamma \rangle$=1.71$\pm$0.03, with dispersion $\sigma$=0.28. Therefore, at least part of the observed discrepancy between Type 1 and Type 2 AGN photon index distributions may be driven by the $\Gamma$--$\lambda_{\rm edd}$ relation. However, such a result needs to be confirmed using the actual $\lambda_{\rm edd}$, since $\Gamma$ measurements may be biased in obscured objects.

\subsection{Flux and luminosity}\label{sec:flux}
We report in Table \ref{tab:flux} the mean and $\sigma$ values of the 2-10 keV flux distribution of Type 1 and Type 2 sources. The two distributions are similar, but the hypothesis that the two populations are drawn from the same parent population is rejected for CCLS30 (p-value= 6.3$\times$10$^{-4}$), even though it is worth mentioning that the KS-test does not take into account the uncertainties on the flux measurement, which can be significant for the sources with low counts statistics in CCLS30.

\begin{figure}[h]
\centering
\includegraphics[width=0.5\textwidth]{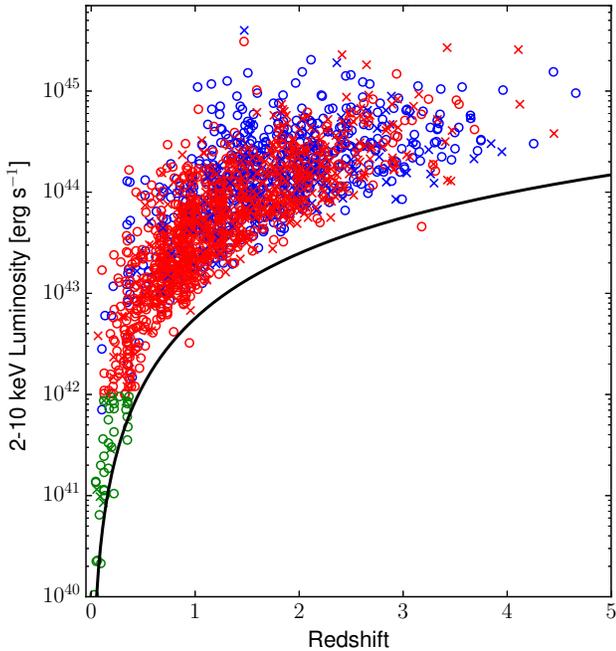}
\caption{{\normalsize 2--10 keV rest-frame, absorption-corrected luminosity as a function of redshift for the sources in CCLS30. Type 1 AGN are plotted in blue, Type 2 AGN are plotted in red, galaxies are plotted in green. Sources with spectroscopic (photometric) redshift are plotted with a circle (cross). The solid line represents the sensitivity limit of the Chandra COSMOS-Legacy survey. The four sources below the sensitivity limit have less than 40 net counts in the 0.5--7 keV band and their parameters are likely poorly constrained (e.g., $N_{H,z}$ may be underestimated and consequently the rest-frame luminosity would be underestimated too).}}\label{fig:z_vs_lx}
\end{figure}

We report the mean and $\sigma$ values of the Type 1 and Type 2 AGN 2-10 keV, absorption-corrected luminosity distributions in Table \ref{tab:lum}. Type 1 AGN are on average more luminous than Type 2 AGN, for both the net counts thresholds we adopted. The difference between the two distributions is mainly due to the fact that, as can be seen in Figure \ref{fig:z_vs_lx}, Type 1 AGN are on average at higher redshifts ($\langle z \rangle$=1.74 for CCLS30, $\langle z \rangle$=1.57 for CCLS70) than Type 2 AGN ($\langle z \rangle$=1.23 for CCLS30, $\langle z \rangle$=1.15 for CCLS70), and our sample is flux-limited, i.e., sources at higher redshifts also have higher luminosities. Nonetheless, the difference between the two distributions is also an indication of a trend with 2-10 keV luminosity of the fraction of Type 1 to Type 2 AGN: in any sample complete in both $z$ and $L_X$, the fraction of Type 2 AGN decreases for increasing luminosities, at any redshift, as already observed in several works \citep[e.g., ][]{lawrence82,ueda03,hasinger08,buchner15,marchesi16a}.

\begin{table}[H]
\centering
\scalebox{1.}{
\begin{tabular}{ccccc}
\hline
\hline
Type & $\langle CCLS30 \rangle$ & $\langle CCLS30_\sigma \rangle$ & $\langle CCLS70 \rangle$ & $\langle CCLS70_\sigma \rangle$\\
\hline
1 & --13.97 & 0.39 & --13.80 & 0.35\\
2 & --14.04 & 0.37 & --13.81 & 0.33\\
\hline
\hline
\end{tabular}}\caption{{\normalsize Logarithm of the 2-10 keV  flux distribution mean and standard deviation $\sigma$, for Type 1 and Type 2 AGN, for sources in CCLS30 and CCLS70. All values are in \flu.}}\label{tab:flux}
\end{table}

\begin{table}[H]
\centering
\scalebox{1.}{
\begin{tabular}{ccccc}
\hline
\hline
Type & $\langle CCLS30 \rangle$ & $\langle CCLS30_\sigma \rangle$ & $\langle CCLS70 \rangle$ & $\langle CCLS70_\sigma \rangle$\\
\hline
1 & 44.13 & 0.49 & 44.20 & 0.48\\
2 & 43.67 & 0.61 & 43.80 & 0.60\\
\hline
\hline
\end{tabular}}\caption{{\normalsize Logarithm of the 2-10 keV intrinsic, absorption-corrected luminosity distribution mean and standard deviation $\sigma$, for Type 1 and Type 2 AGN, for sources in CCLS30 and CCLS70. All values are in \lu.}}\label{tab:lum}
\end{table}

\subsection{Photon index dependences}\label{sec:gamma_dep}
\begin{figure*}%[!h]
\begin{minipage}[b]{.5\textwidth}
  \centering
  \includegraphics[width=1.0\linewidth]{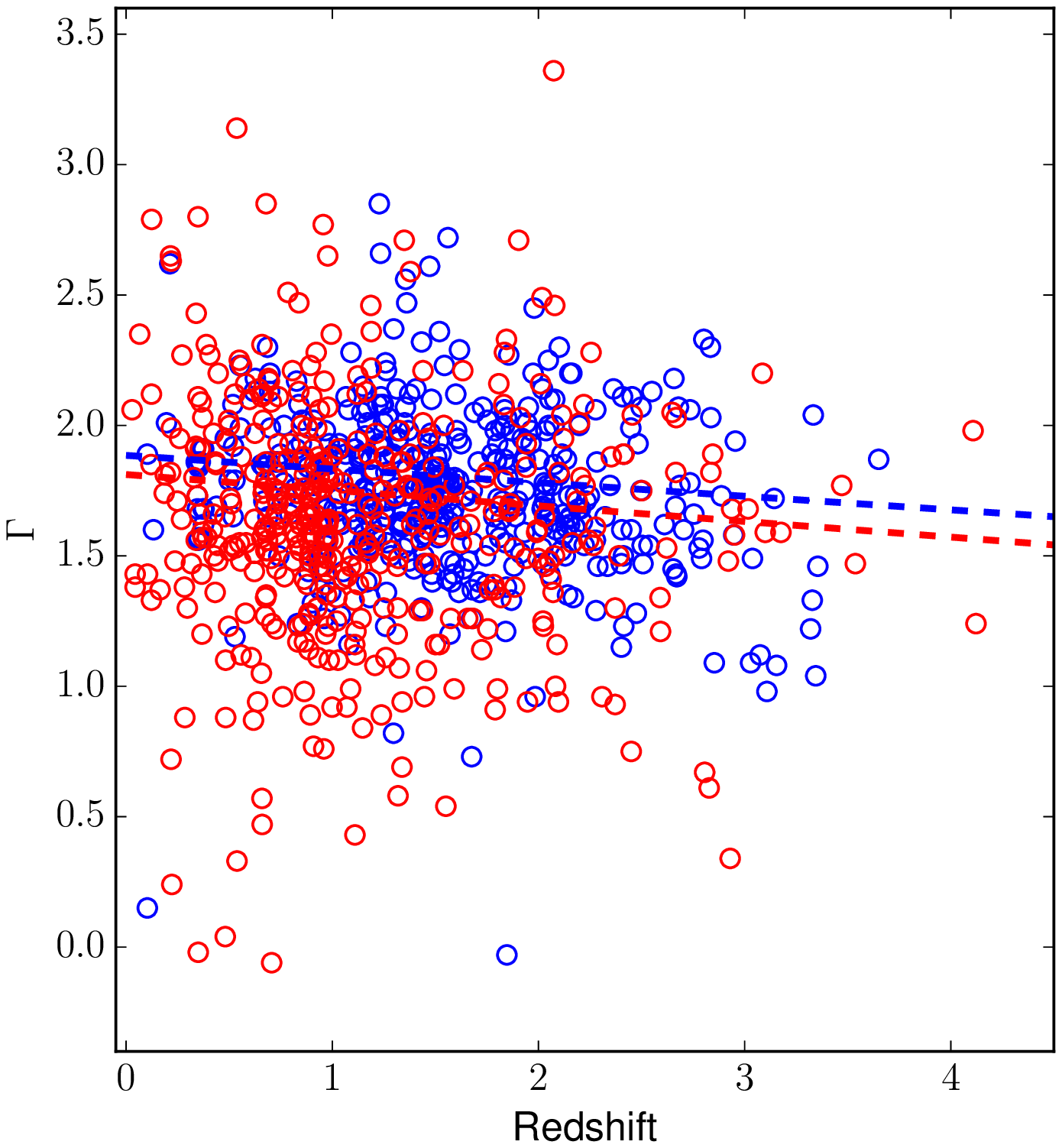}
\end{minipage}%
\begin{minipage}[b]{.5\textwidth}
  \centering
  \includegraphics[width=1.03\textwidth]{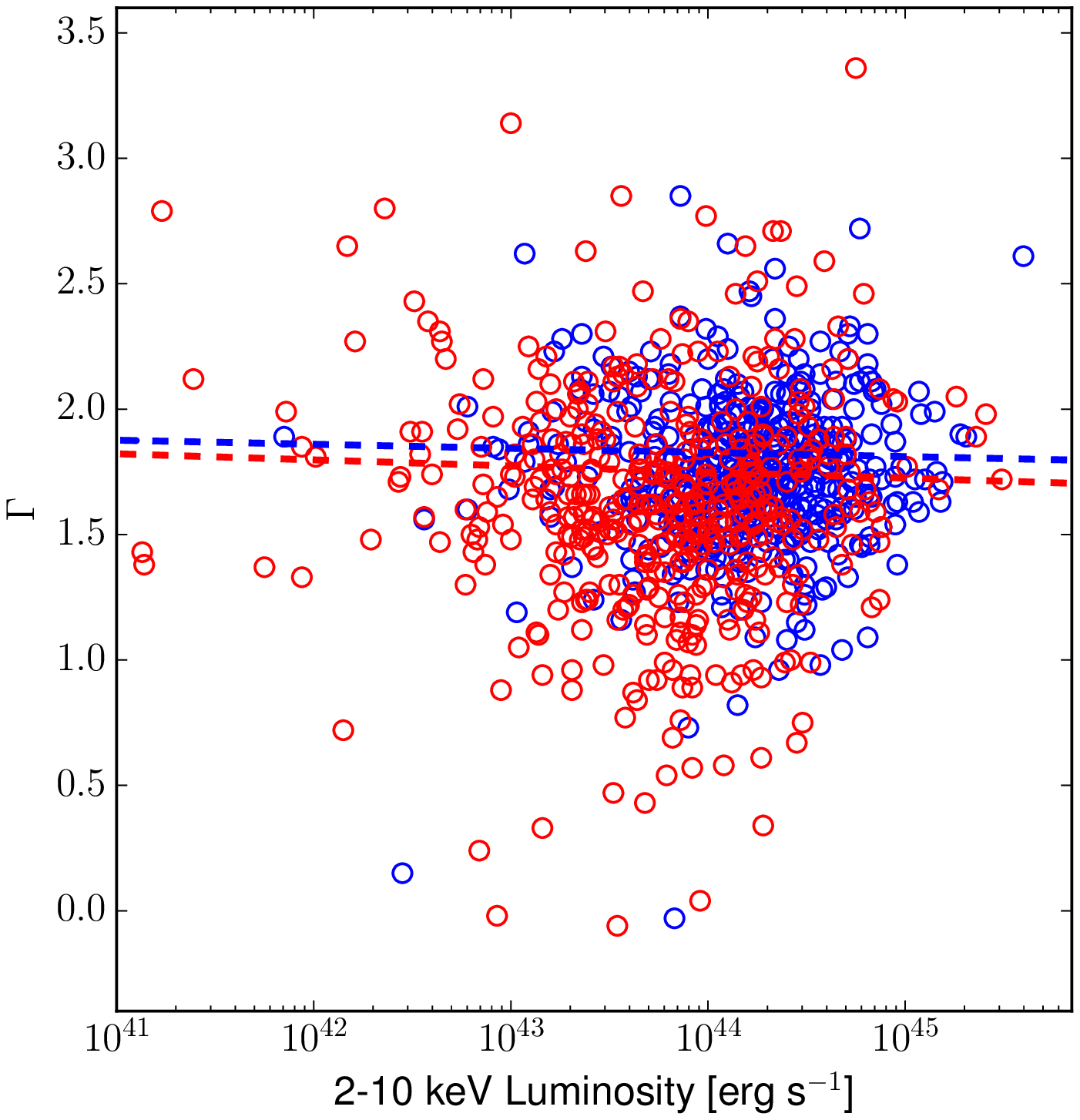}
\end{minipage}
\caption{{\normalsize $\Gamma$ as a function of redshift (left) and as a function of 2-10 keV absorption-corrected luminosity (right), for all sources in CCLS70. Blue circles are Type 1 AGN, red circles are Type 2 AGN. Linear best fits to the Type 1 and Type 2 samples are also plotted as dashed lines.}}\label{fig:gamma_z_lx}
\end{figure*}

We searched for a potential correlation between the photon index and the redshift or the X-ray luminosity. In Figure \ref{fig:gamma_z_lx}, left panel, we show the distribution of $\Gamma$ as a function of redshift, for the CCLS70 sample. 
Type 1 sources show a weak anti-correlation between z and $\Gamma$, with Spearman correlation coefficient $\rho$=--0.15 and p-value p= 2.3$\times$10$^{-3}$ for the hypothesis that the two quantities are unrelated. In Type 2 sources, instead, the correlation coefficient is smaller, $\rho$=--0.07, and the p-value p=0.12 does not allow to rule out the hypothesis of no correlation.

The results obtained for the whole CCLS70 sample do not change significantly if we exclude from the computation sources with peculiar photon index (i.e., taking into account only objects with 1$< \Gamma<$3), for both Type 1 ($\rho$=--0.14 and p-value p= 3.8$\times$10$^{-3}$) and Type 2 AGN ($\rho$=--0.08 and p-value p=0.10).

In Figure \ref{fig:gamma_z_lx}, right panel, we show the distribution of $\Gamma$ as a function of the intrinsic, absorption-corrected 2-10 keV luminosity, for the CCLS70 sample. We do not find any evidence of correlation in both Type 1 (blue, $\rho$=0.01 and p-value p=0.78) and Type 2 (red, $\rho$=0.01 and p-value p=0.90) objects, and the lack of correlation remains also in the 1$< \Gamma<$3 subsample.

\subsection{Column density dependences}\label{sec:nh_dep}
\subsubsection{Trend with redshift}
In Figure \ref{fig:nh_z_lx} (left panel) we show the distribution of $N_{\rm H,z}$ as a function of redshift for sources in CCLS30. As can be seen, $N_{\rm H,z}$ minimum value significantly increases at increasing redshifts. Such a result was already observed, by \citet{civano05}, \citet{tozzi06} and \citet{lanzuisi13} among others, and is due to the fact that moving toward high redshifts, the photoelectric absorption cut-off moves outside the limit of the observing band, 0.5 keV. Consequently, the measure of low $N_{\rm H,z}$ values becomes more difficult.

\subsubsection{Trend with 2-10 keV luminosity}
\begin{figure*}%[!h]
\begin{minipage}[b]{.5\textwidth}
  \centering
  \includegraphics[width=1.0\linewidth]{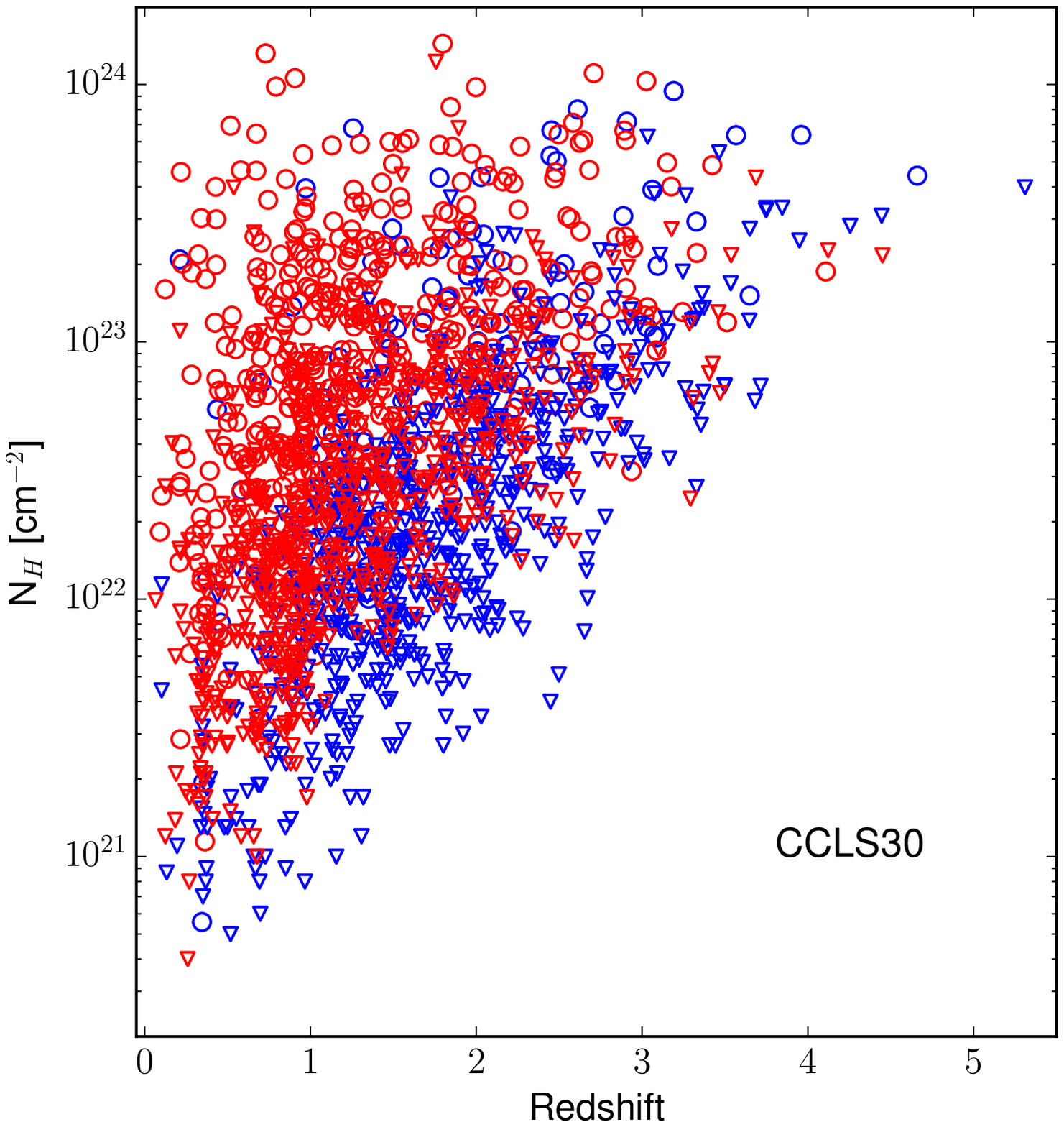}
\end{minipage}%
\begin{minipage}[b]{.5\textwidth}
  \centering
  \includegraphics[width=1.03\textwidth]{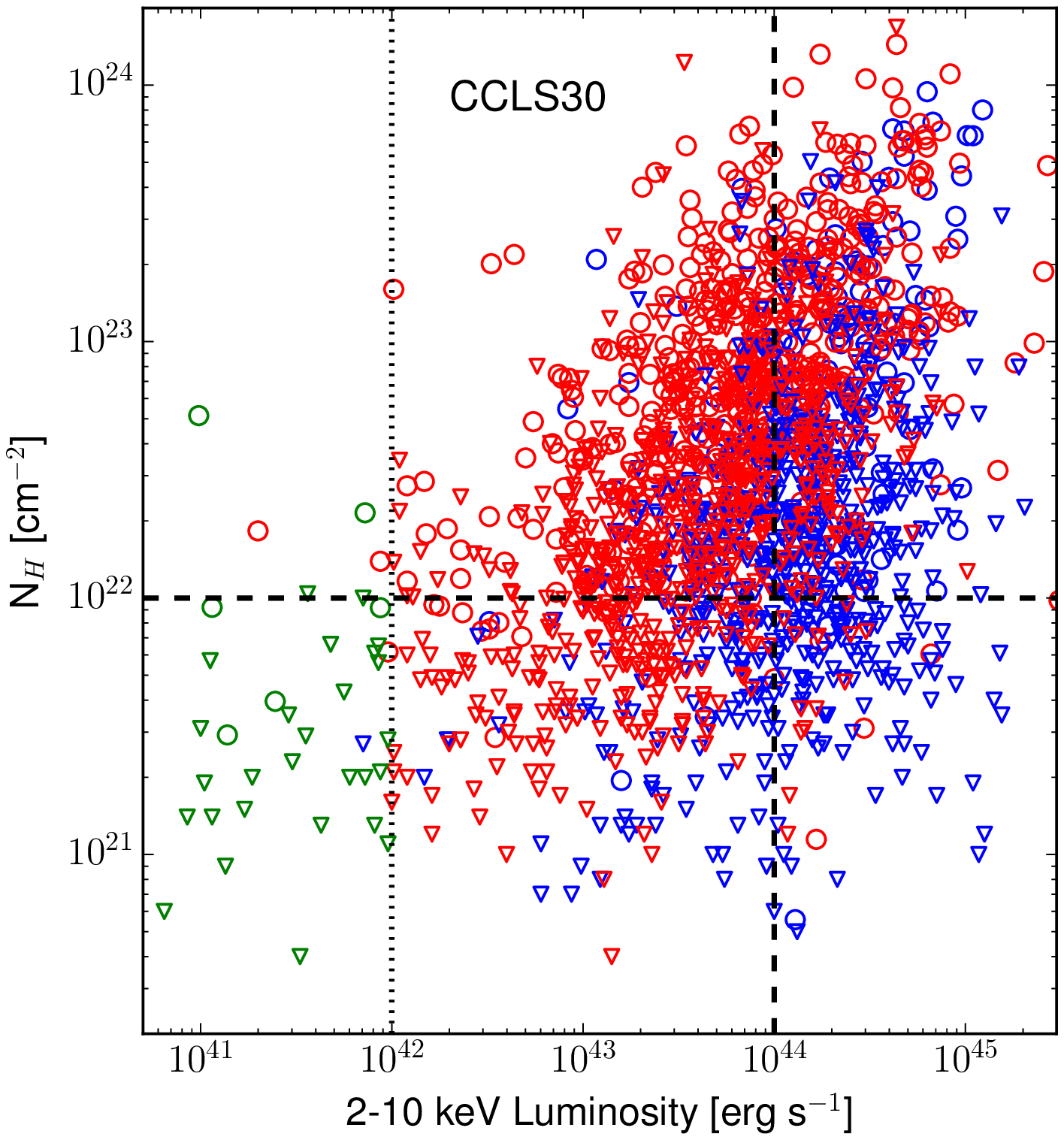}
\end{minipage}
\caption{{\normalsize \textit{Left} : $N_{\rm H,z}$ as a function of redshift, for all sources in CCLS30, for Type 1 (blue) and Type 2 (red) sources. $N_{\rm H,z}$ 90\% confidence upper limits are plotted as triangles. \textit{Right}: $N_{\rm H,z}$ as a function of 2-10 keV absorption-corrected luminosity, for all sources in CCLS30, for Type 1 (blue) and Type 2 (red) AGN, and for galaxies (green). $N_{\rm H,z}$ 90\% confidence upper limits are plotted as triangles. The horizontal black dashed line marks the threshold ($N_{\rm H,z}$=10$^{22}$ cm$^{-2}$) usually adopted to divide unobscured and obscured sources. The vertical black dotted line marks the luminosity threshold, L$_{\rm 2-10keV}$ =10$^{42}$ \lu, used to divide AGN from star-forming galaxies, while the vertical black dashed line marks the threshold usually adopted to separate quasars from Seyfert galaxies, L$_{\rm 2-10keV}$ =10$^{44}$ \lu.}}\label{fig:nh_z_lx}
\end{figure*}

In Figure \ref{fig:nh_z_lx} (right panel) we show the distribution of $N_{\rm H,z}$ as a function of 2-10 keV rest-frame absorption-corrected luminosity. As can be seen, and as already discussed in previous sections, Type 2 AGN are dominant in the region of the plot with $N_{\rm H,z}>$10$^{22}$ cm$^{-2}$ and L$_{\rm 2-10keV}<$10$^{44}$ \lu, while Type 1 sources are dominant in the region with $N_{\rm H,z}<$10$^{22}$ cm$^{-2}$ and L$_{\rm 2-10keV}>$10$^{44}$ \lu.

In CCLS30, 268 sources out of 1844 sources ($\sim$15\%) lie in the obscured quasar region, i.e., have 2--10 keV luminosity L$_{\rm 2-10keV}>$10$^{44}$ \lu\ and 90\% confidence significant intrinsic absorption $N_{\rm H,z}>$10$^{22}$ cm$^{-2}$. 83 out of these 268 sources ($\sim$31\%) are classified as Type 1 AGN, and 61 of these 83 Type 1 sources have optical spectroscopy available. The mean redshift of these 61 objects is $\langle z \rangle$=2.03. A significant fraction of obscured Type 1 AGN ($\sim$15\% of the whole population at L$_{\rm 2-10keV}>$10$^{44}$ \lu) was also found by \citet{merloni14} in XMM-COSMOS. We further discuss these obscured Type 1 AGN in the next section.

265 out of 1855 CCLS30 sources (14.3\%) lie in the L$_{\rm 2-10keV}$=[10$^{42}$-10$^{44}$] \lu, $N_{\rm H,z}<$10$^{22}$ cm$^{-2}$ area, i.e., have unobscured AGN spectral properties. 172 of these sources (64.9\%) are optically classified Type 2 AGN. The fraction is similar if we take into account only objects with a spectral type: 127 out of 203 sources (62.6\%) in the area are non-BLAGN.

In CCLS30, a total of 199 out of 1111 (17.9\%) Type 2 AGN have $N_{\rm H,z}<$10$^{22}$ cm$^{-2}$. i.e., consistent with being unobscured AGN on the basis of their X-ray spectrum. The fraction of unobscured Type 2 AGN is matter of extended debate in literature, varying from only a few percent ($<$5\%) in \citet{risaliti99}, \citet{malizia09} and \citet{davies15} to 30\% \citep{merloni14} and up to 66\% \citep{garcet07}. The first scenario suggests that optical and X-ray obscuration tend to occur at the same time, while the second points to a stronger independence between the obscuration processes in the two different energy ranges. The fraction we obtain is in an intermediate regime between the two scenarios, in good agreement with the results of \citet{panessa02}, \citet{akylas09} and \citet{koulouridis16}. We will further discuss the implications of this result in the next section.

\subsubsection{Trend with 2--10 keV luminosity of the AGN obscured fraction}
\begin{figure*}%[!h]
\begin{minipage}[b]{.5\textwidth}
  \centering
  \includegraphics[width=1.0\linewidth]{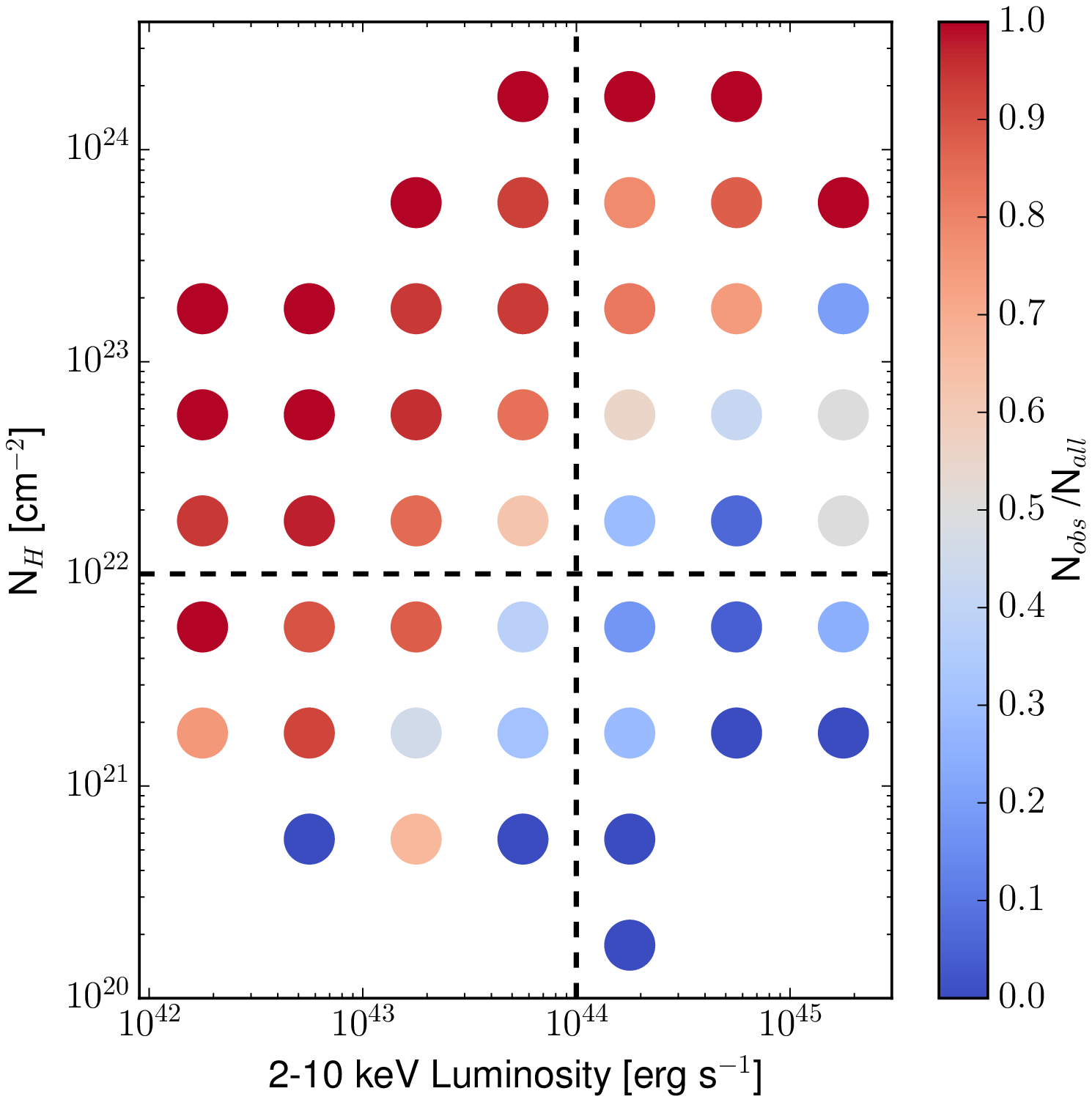}
\end{minipage}%
\begin{minipage}[b]{.5\textwidth}
  \centering
  \includegraphics[width=1.03\textwidth]{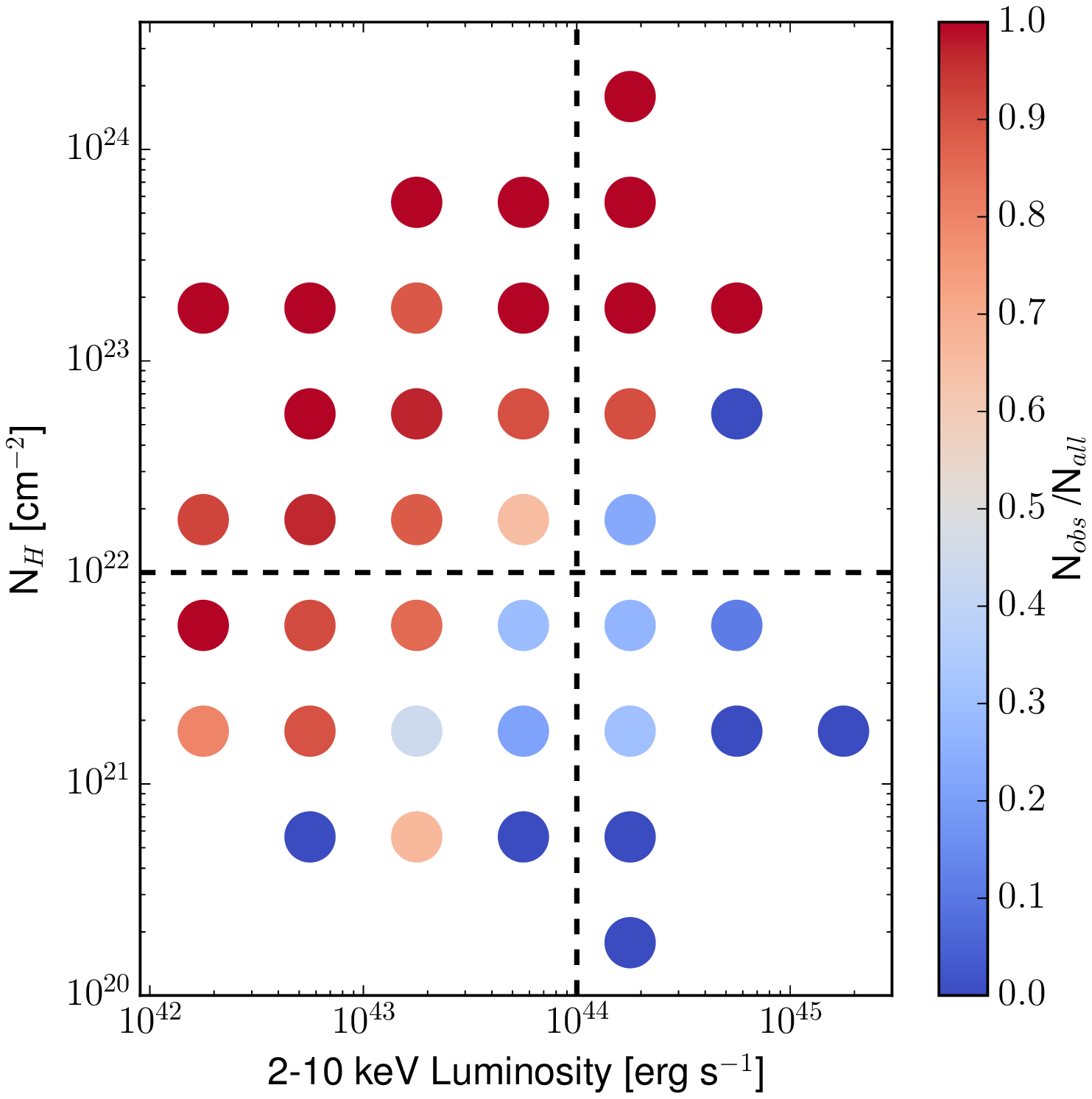}
\end{minipage}
\caption{{\normalsize $N_{\rm H,z}$ as a function of 2-10 keV absorption-corrected luminosity. The colormap shows the ratio between the number of Type 2 sources (N$_{\rm 2}$) against all sources (N$_{\rm all}$), in each bin of $N_{\rm H,z}$ and luminosity. N$_{\rm 2}$/N$_{\rm all}$=1 is plotted in red, N$_{\rm 2}$/N$_{\rm all}$=0 in blue. In the left panel we show the results for the whole sample (i.e., taking into account both the spectroscopic and the photometric type), while in the right panel only the spectroscopic type is taken into account.}}
\label{fig:ratio_obs_lx}
\end{figure*}

In Figure \ref{fig:ratio_obs_lx} we show the distribution of $N_{\rm H,z}$ as a function of 2-10 keV absorption-corrected luminosity, in bins of 0.5 dex in both $N_{\rm H,z}$ and luminosity. The color of each bin identifies the fraction of Type 2 sources $f_{\rm 2}$=N$_{\rm 2}$/N$_{\rm all}$, where N$_{\rm 2}$ is the number of Type 2 sources and N$_{\rm all}$ is the total number of sources in each bin. In the left panel we show the results for the whole sample (i.e., taking into account both the spectroscopic and the photometric type), while in the right panel only sources with optical spectroscopic classification are taken into account. It is worth noticing that there is a general good agreement between the combined classification and the one spectroscopically based, therefore all the following discussion cannot be related to a bad SED-fitting classification.

In the previous section, we mentioned that 106 out of 696 Type 1 AGN ($\sim$15\%) have $N_{\rm H,z}>$10$^{22}$ cm$^{-2}$ at a 90\% confidence level. The fraction of obscured Type 1 AGN is strongly luminosity dependent: there is only 1 obscured Type 1 AGN with log(L$_{\rm 2-10keV}$ )=[42--43] (3.2\% of the obscured sources in this luminosity range), 21 with log(L$_{\rm 2-10keV}$ )=[43--44] (8.2\%) and 83 in the quasar regime, i.e., with log(L$_{\rm 2-10keV}$ )$>$44 (31.2\%).

We expect that a fraction of optically classified classified Type 1 sources with X-ray properties consistent with those of obscured quasars to be broad absorption lines (BAL) quasars, with broad, blue-shifted absorption lines in their optical/UV spectra. Of the candidate BAL quasars with optical spectrum available, 41 out of 61 have z$>$1.5, i.e., high enough to observe the UV features (e.g., CIV at 1549 \AA) in optical spectra. However, the analysis of these objects is beyond the purpose of this work, and these aspects will be investigated in a forthcoming paper (Marchesi et al. in preparation). Moreover, a fraction of these obscured Type 1 sources may not be BAL quasars, being instead sources with dust-free gas surrounding the inner part of the nuclei, therefore causing obscuration in the X-rays and not in the optical band \citep{risaliti02,maiolino10,fiore12,merloni14}.

In the previous section, we showed that 199 ($\sim$18\%) of the 1111 CCLS30 Type 2 sources have $N_{\rm H,z}<$10$^{22}$ cm$^{-2}$, i.e., consistent with being unobscured AGN. In Figure \ref{fig:ratio_obs_lx} we can see how the distribution of these objects is strongly dependent to their X-ray luminosity, i.e., the fraction of Type 2 AGN with respect to the total number of sources in a bin is higher at log(L$_{\rm 2-10keV}$ )=[42--43] (75--90\%), and drops at L$_{\rm 2-10keV}>$10$^{44}$ \lu\ ($<$15\%), where there are only 19 unobscured Type 2 AGN (9.5\% of the whole Type 2 AGN sample). A similar result was found by \citet{merloni14} in the XMM-COSMOS survey, where $\sim$40\% of the sources with log(L$_{\rm 2-10keV}$ )=[42.75--43.25] are unobscured Type 2 AGN, while the fraction of unobscured Type 2 AGN is $<$10\% at L$_{\rm 2-10keV}<$10$^{44}$ \lu. This result is expected, since the XMM-COSMOS sample is a ``bright'' subsample of the \cha \leg\ one, since XMM-COSMOS has a flux limit $\sim$3 times higher than \cha \leg, so there is a significant overlap between the sample used by \citet{merloni14} and CCLS30. 

We point out that, according to our classification, a fraction of Type 2 sources are expected to be narrow-line Seyfert 1 galaxies \citep[NLSy1,][]{osterbrock85,goodrich89}, i.e., AGN with the properties of Seyfert 1 galaxies, but with only narrow, rather than broad, HI emission lines. These objects are by definition unobscured, although lacking of broad lines. However, NLSy1 galaxies usually have very steep photon indexes \citep[see, e.g., ][]{brandt97}, and only 16 out of 199 (8.4\%) unobscured Type 2 AGN in CCLS30 have $\Gamma>$2, and only one source has $\Gamma>$2.5. Therefore, we expect that not more than 10\% of the unobscured Type 2 AGN in CCLS30 are NLSy1 galaxies.

Besides NLSy1 galaxies, there are at least two possible reasons to explain the relatively high fraction of unobscured Type 2 AGN at low luminosities: ($i$) it is possible that a fraction of low luminosity sources is wrongly classified. A similar effect was described in \citet{oh15}, who analyzed the spectra of galaxies at $z<$0.2 in the Sloan Digital Sky Survey Data Release 7 (SDSS DR7), and found a significant fraction of previously misclassified BLAGN, i.e., sources with a stellar spectral continuum and a broad H$\alpha$ emission line. With this new classification, the number of Type 1 AGN in SDSS DR7 increased by 49\%. However, we point out that \citet{oh15} analyzed low-luminosity AGN at $z<$0.2, i.e., a sample significantly different to CCLS30, where only 39 sources out of 1855 (2.1\%) have $z<$0.2. ($ii$) AGN with low Eddington ratio, and therefore low luminosity, lack broad emission lines, even if they are intrinsically unobscured \citep[see, e.g., ][]{trump11,marinucci12}. If this is the case, the drop in BLAGN at $N_{\rm H,z}<$10$^{22}$ cm$^{-2}$ and L$_{\rm 2-10keV}<$10$^{44}$ \lu would imply an Eddington ratio threshold of $\lambda_{Edd}\sim$10$^{-3}$--10$^{-2}$, assuming average BH masses M$_{\rm BH}$=10$^8$--10$^9$ M$_\odot$. This would have strong consequences for the unification scheme. 

\subsection{Iron K$\alpha$ equivalent width dependences}\label{sec:ew}
141 CCLS30 sources are best-fitted with a model which includes an iron K$\alpha$ emission line at 6.4 keV. For each source, we compute the emission line equivalent width, EW, a measure of the line intensity, computed as follows:
\begin{equation}
EW=\int{\frac{F_l(E)-F_c(E)}{F_c(E)}dE}
\end{equation}
where $F_l$(E) is the flux of the emission line at the energy E, and $F_c$(E) is the intensity of the spectral continuum at the same energy. The mean (median) equivalent width of the 141 CCLS30 sources is EW=0.49$\pm$0.03 (0.41) keV, with dispersion $\sigma$=0.39. We also checked how much the assumption we made on the line width, which we fixed to $\sigma$=0.1, affected the mean EW value. To do so, we re-fitted the 141 spectra using $\sigma$=0, i.e., the scenario where the line shows no relativistic broadening. We find that in this case the iron K$\alpha$ mean (median) equivalent width only slightly decreases, being EW=0.44$\pm$0.03 (0.36) keV, with dispersion $\sigma$=0.34.

In Figure \ref{fig:ew} (left panel), we show the EW distribution as a function of the photon index $\Gamma$ for the 101 sources for which we compute $\Gamma$ (i.e., excluding those sources for which we fixed $\Gamma$=1.9). If we fit the data with a linear model (red dashed line), EW= a$\Gamma$ + b, we found evidence of a significant inverse correlation, with a=--0.22$\pm$0.05. 

However, a fraction of sources with prominent Iron K$\alpha$ emission line are expected to be CT AGN. A proper characterization of these objects requires fitting models more complicated than those used in this work, to properly take into account Compton scattering processes. For this reason, we are performing an extended analysis of the CCLS30 sample using more appropriate torus models and a MCMC analysis to estimate the probability of a source a certain $N_{\rm H,z}$ value (Lanzuisi et al. in preparation). As a preliminary result, we find that 12 out of the 141 CCLS30 sources with significant iron K$\alpha$ emission line have a significant probability to be CT AGN or heavily obscured sources (i.e., with Log($N_{\rm H,z}$)$>$23.5). We plot these sources as black stars in Figure \ref{fig:ew}, left panel. If we re-fit the sample without these sources, which are likely to have a wrongly computed $\Gamma$ using our basic models, the correlation between EW and $\Gamma$ disappears (a=0.10$\pm$0.08; red solid line in Figure \ref{fig:ew}). This result is due to the fact that heavily obscured AGN are poorly fitted by basic models like those used in this work, which try to mimic the flat spectra of obscured sources with a flat photon index and no obscuration.

In Figure \ref{fig:ew} (central panel) we show the EW distribution as a function of $N_{\rm H,z}$. 99 sources (7\%) have only an upper limit on $N_{\rm H,z}$, while the remaining 30\% have a $N_{\rm H,z}$ value significant at a 90\% confidence level. We do not find any significant correlation between EW and $N_{\rm H,z}$. We remark that a similar result is not unexpected, since a trend between EW and $N_{\rm H,z}$ is observed only at $N_{\rm H,z}>$10$^{23}$ cm$^{-2}$ \citep[see, e.g., ][]{makishima86}, and only 12\% of the sources analyzed in this section have intrinsic absorption values above this threshold. Finally, candidate CT AGN (plotted with full markers) have on average low $N_{\rm H,z}$ values, a further indication that a standard spectral fitting procedure does not work properly with these extreme sources.

\subsubsection{The X-ray Baldwin effect}
We select the 33 Type 1 AGN in CCLS30 best-fitted with a model containing an Iron K$\alpha$ line to check for the presence of the so-called ``X-ray Baldwin'' effect, i.e., the existence of an anti-correlation between the Iron K$\alpha$ line EW and the AGN 2-10 keV luminosity \citep{iwasawa93}.

\begin{figure*}%[!h]
\begin{minipage}[b]{.33\textwidth}
  \centering
  \includegraphics[width=1.06\linewidth]{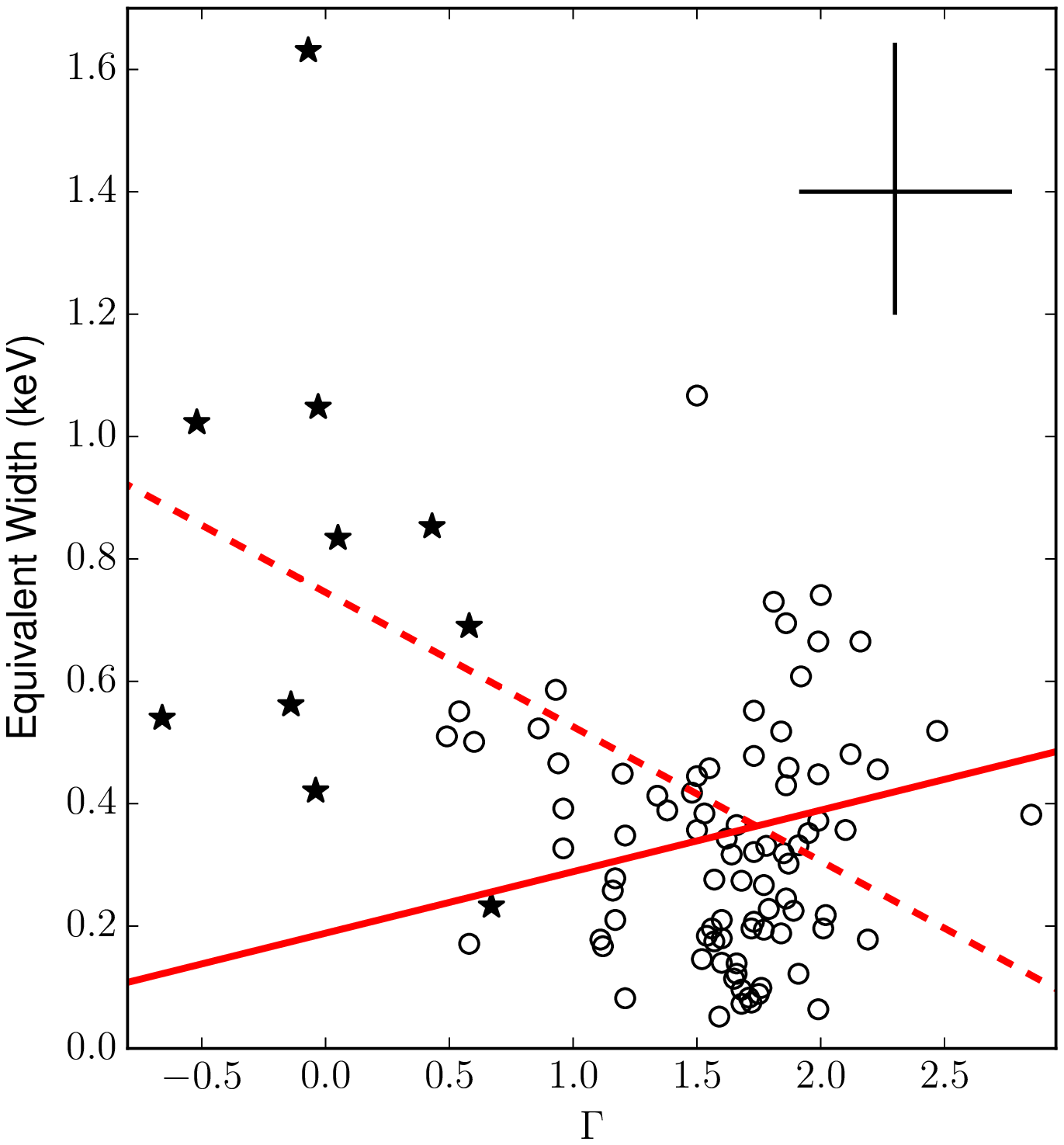}
\end{minipage}%
\begin{minipage}[b]{.33\textwidth}
  \centering
  \includegraphics[width=1.04\textwidth]{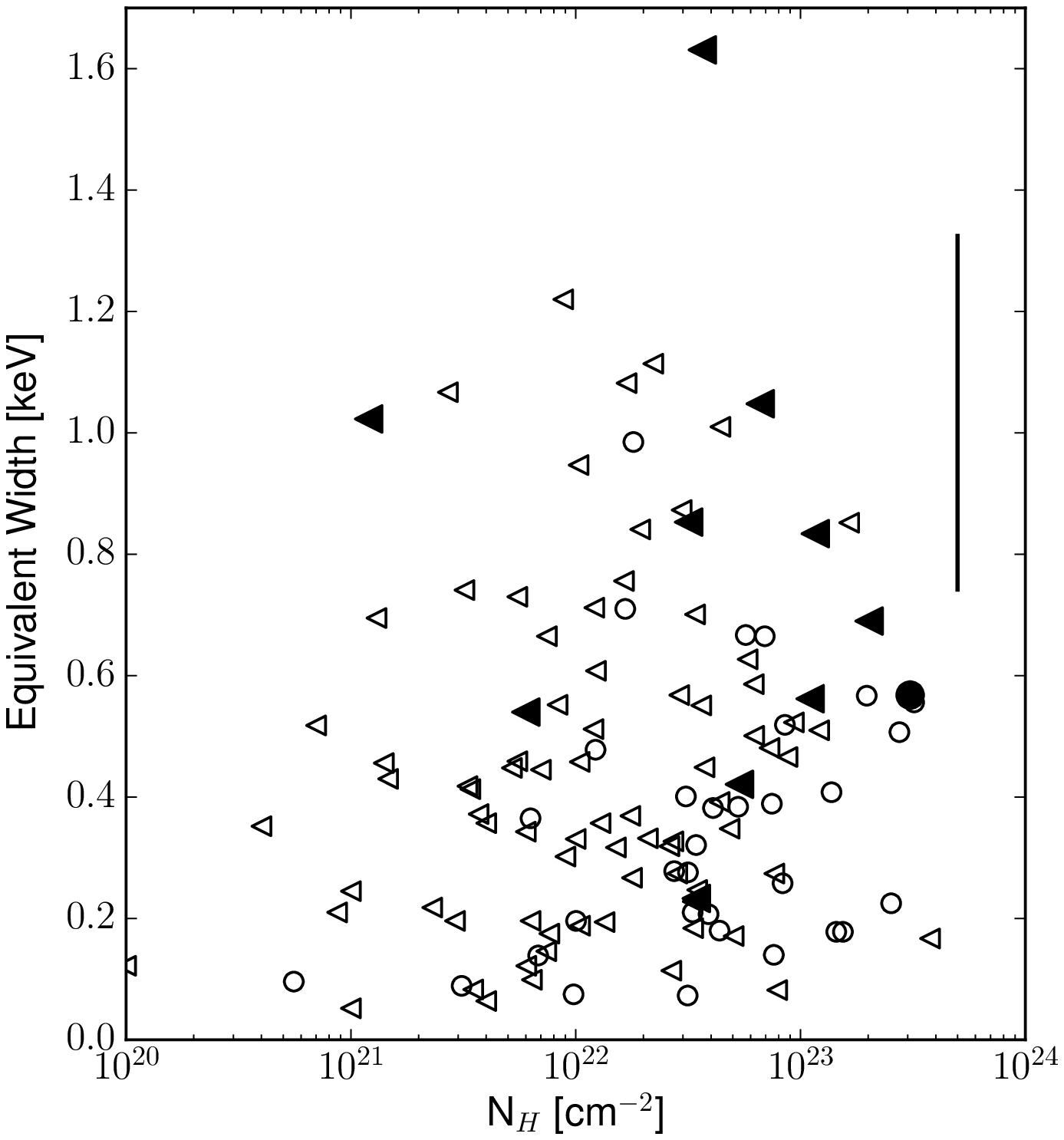}
\end{minipage}
\begin{minipage}[b]{.33\textwidth}
  \centering
  \includegraphics[width=1.02\textwidth]{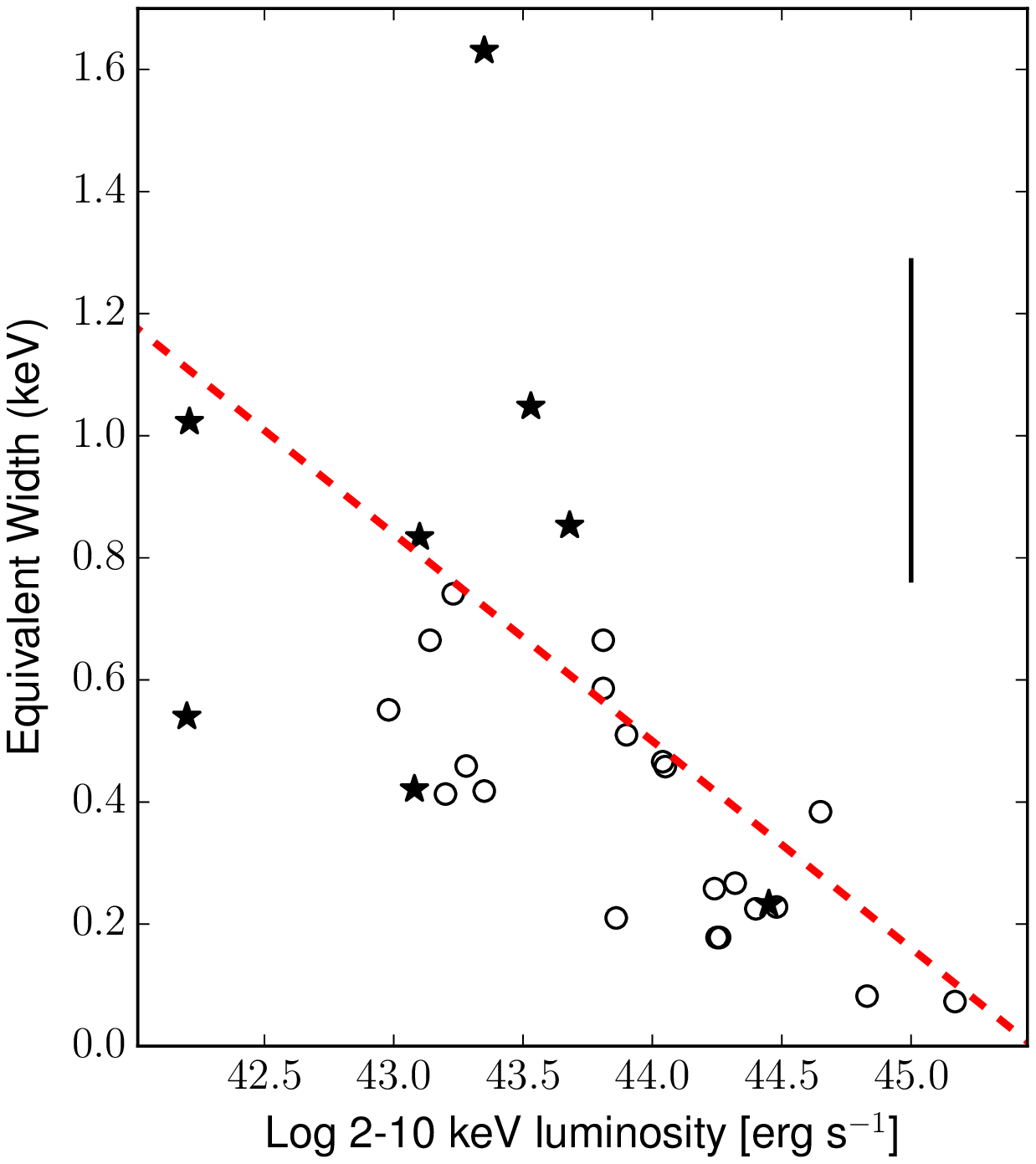}
\end{minipage}
\caption{{\normalsize Iron K$\alpha$ equivalent width (EW, in keV) as a function of best-fit photon index ($\Gamma$, left), intrinsic absorption ($N_{\rm H,z}$, center) and 2-10 keV luminosity (right). In the left and right panels, the linear best fit to the data is plotted as a red dashed line; candidate highly obscured AGN on the basis of the fit with complex torus models are plotted as stars. The linear best fit to the data without the candidate CT AGN is plotted in the left panel as a red solid line. In the central panel, nominal $N_{\rm H,z}$ values are plotted as circles, upper limits are plotted as triangles; candidate highly obscured AGN on the basis of the fit with the complex torus model are plotted with full markers. Mean errors on EW and $\Gamma$ (left) and on EW (center and right) are also plotted.}}
\label{fig:ew}
\end{figure*}

The existence of this anti-correlation is confirmed by our data, as can be seen in Figure \ref{fig:ew} (right panel); the Spearman correlation coefficient is $\rho$=--0.72, with p-value p= 9.9$\times$10$^{-6}$, for the hypothesis that the two quantities are unrelated. The best-fit to our data is expressed by the relation EW(K$\alpha$)$\propto$ $L_X$(2--10 keV)$^{-0.34\pm0.07}$, in fair agreement with the result obtained by \citet{iwasawa93}, which measured a trend expressed by the relation EW(K$\alpha$)$\propto$ $L_X$(2--10 keV)$^{-0.20\pm0.03}$.

\subsection{Host galaxy mass and star formation rate dependences for Type 2 AGN}\label{sec:sfr_mgal}
Suh et al. (in prep.) computed host galaxy properties such as mass (M$_*$) and star formation rate (SFR) for Type 2 AGN in \cha \leg. These quantities have been computed using SED-fitting techniques: the host galaxy properties are derived using a 3-component SED-fitting decomposition method, which combines a nuclear dust torus model \citep{silva04}, a galaxy model \citep{bruzual03} and starburst templates \citep{chary01,dale02}. The SFR is then estimated by combining the contributions from UV and total host-galaxy IR luminosity computed with the SED fitting ($L_{8-1000\mu m}$).

We first study the trend of the photon index $\Gamma$ as a function of M$_*$ and SFR for Type 2 objects in CCLS70; we also perform a separate analysis on sources spectroscopically classified as non-BLAGN sources and on sources with only SED template best-fitting classification. We performed a Spearman correlation test for each of the subsamples we described: we report the results in Table \ref{tab:sfr_mgal_gamma}. In the case of M$_*$, we find no evidence of correlation in the SED template best-fitting subsample and in the whole Type 2 sample, and a weak evidence of anti-correlation in the spectroscopic sample, although the hypothesis that the two quantities are uncorrelated cannot be ruled out (p-value=0.06). We obtain a similar result while correlating $\Gamma$ and SFR: in this case, the hypothesis that $\Gamma$ and SFR are uncorrelated in the spectroscopic sample is ruled out with 97\% confidence. The SED template best-fitting and the whole Type 2 sample do not show evidence of correlation.

\begin{table}[H]
\centering
\scalebox{1.}{
\begin{tabular}{ccccc}
\hline
\hline
Sample & $\rho_{M_*}$ & p-value$_{M_*}$ & $\rho_{SFR}$ & p-value$_{SFR}$\\
\hline
Spec & --0.12 & 0.06 & --0.14 & 0.03\\
SED & 0.03 & 0.73 & 0.04 & 0.66\\
All & --0.06 & 0.24 & --0.07 & 0.16\\
\hline
\hline
\end{tabular}}\caption{{\normalsize Spearman correlation coefficient $\rho$ and p-value for the photon-index $\Gamma$ in relation to M$_*$ or SFR, for sources with spectroscopic classification, sources with only best-fit SED template and for the whole CCLS70 Type 2 sample.}}\label{tab:sfr_mgal_gamma}
\end{table}

\begin{figure*}%[!h]
\begin{minipage}[b]{.5\textwidth}
  \centering
  \includegraphics[width=1.0\linewidth]{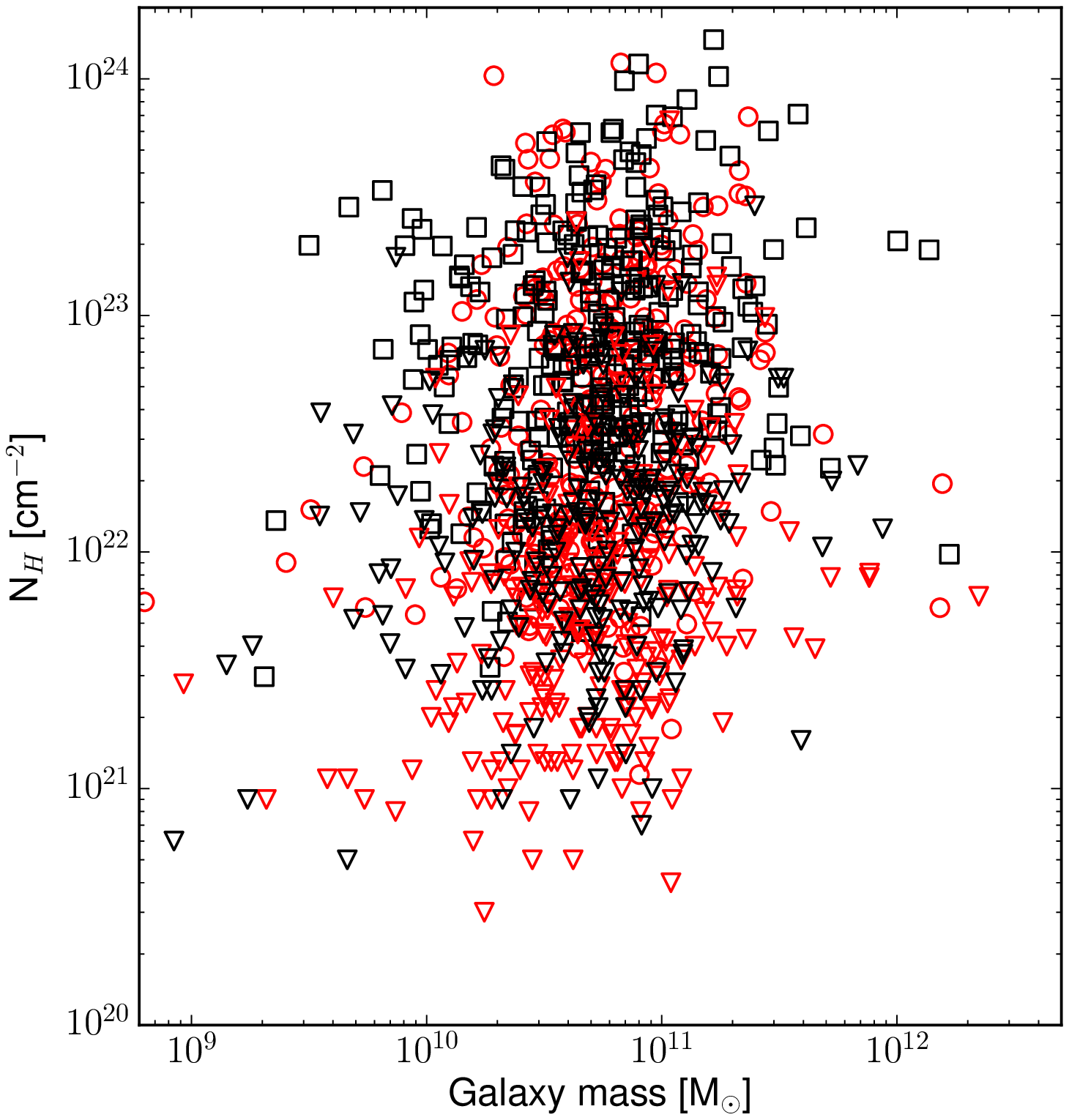}
\end{minipage}%
\begin{minipage}[b]{.5\textwidth}
  \centering
  \includegraphics[width=1.0\textwidth]{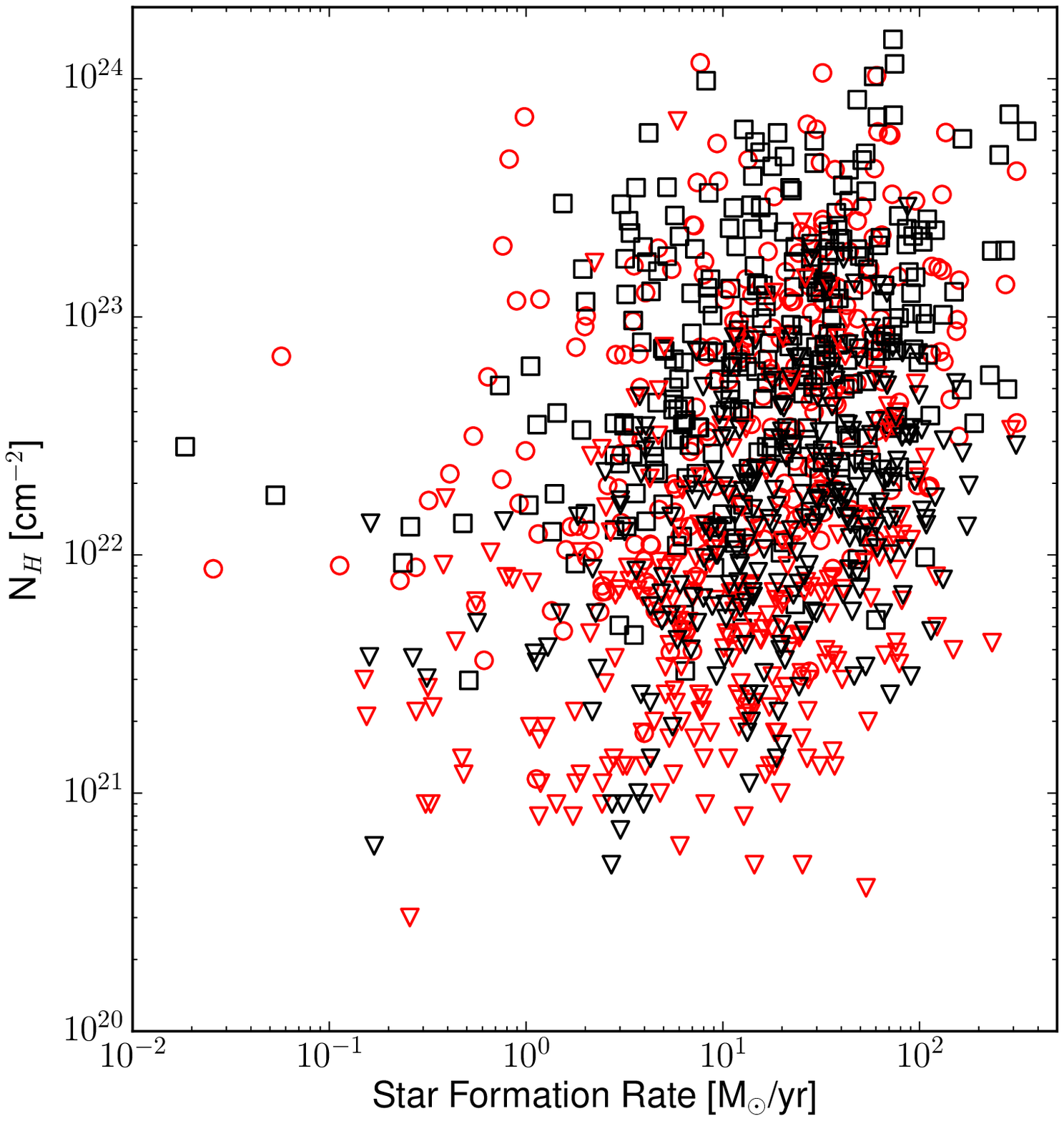}
\end{minipage}
\caption{{\normalsize Intrinsic absorption $N_{\rm H,z}$ as a function of host galaxy mass (left) and star formation rate (right), for all Type 2 AGN in CCLS30. Sources spectroscopically classified are plotted as red circles, while sources with only SED template best-fitting information are plotted as black squares. Upper limits on $N_{\rm H,z}$ are plotted as triangles.}}
\label{fig:sfr_mgal}
\end{figure*}

In Figure \ref{fig:sfr_mgal} we show the distribution of the intrinsic absorption $N_{\rm H,z}$ as a function of M$_*$ (left) and SFR (right), for the 1011 Type 2 objects in CCLS30. Sources spectroscopically classified as Type 2 AGN are plotted as red circles, while sources with only SED template best-fitting classification are plotted with black squares. Upper limits on $N_{\rm H,z}$ are plotted with triangles. We studied the existence of a correlation between $N_{\rm H,z}$ and M$_*$ or SFR computing the Spearman correlation coefficient using the ASURV tool, Rev 1.2 \citep{isobe90,lavalley92}, which implements the methods presented in \citet{isobe86}, to properly take into account the 565 sources having only a 90\% confidence upper limit on $N_{\rm H,z}$. We report the results of the fit in Table 10: we find a significant correlation between $N_{\rm H,z}$ and both M$_*$ and SFR, i.e., the objects with higher $N_{\rm H,z}$ values are also those with higher M$_*$ and SFR. We obtain the same result fitting separately only the 533 sources with spectral type and the 478 with SED template best-fitting type. We point out that SFR and M$_*$ are correlated, i.e., more massive galaxies have higher SFR, therefore the correlation with $N_{\rm H,z}$ can be intrinsic only for one of the two parameters, more likely SFR. 
Finally, we performed a partial correlation analysis to understand how much of the observed correlation between $N_{\rm H,z}$ and M$_*$ (or SFR) is driven by a redshift selection effect, i.e., if the observed correlation is due to the fact that at high redshifts we observe only sources with significant $N_{\rm H,z}$ values (as shown in Figure \ref{fig:nh_z_lx}, left panel), and with high values of M$_*$ and SFR. To do so, we compute the partial Spearman correlation coefficient between $N_{\rm H,z}$ and M$_*$ (or SFR), conditioned by the distance $\dot{z}$, using the equation:
\begin{equation}
\rho(a,b,\dot{c})=\frac{\rho_{ab}-\rho_{ac}\rho_{b\dot{c}}}{\sqrt{(1-\rho_{a\dot{c}}^2)(1-\rho_{b\dot{o}}^2)}}
\end{equation}

\citep{conover80}. The partial correlation coefficient we obtain are $\rho(a,b,\dot{c})_{M_*}$=0.06 and $\rho(a,b,\dot{c})_{SFR}$=0.11. Following Equation 6 of \citet{macklin82}, these values correspond to confidence levels of $\sigma_{M_*}$=1.58 and $\sigma_{M_*}$=2.68. Therefore, the observed relation between M$_*$ and $N_{\rm H,z}$ seems to be mainly driven by a redshift selection effect, while the relation between SFR and $N_{\rm H,z}$ remains significant, although at $<$3$\sigma$, even taking into account the redshift contribution.

\begin{table}[H]
\centering
\scalebox{1.}{
\begin{tabular}{ccccc}
\hline
\hline
Sample & $\rho_{M_*}$ & p-value$_{M_*}$ & $\rho_{SFR}$ & p-value$_{SFR}$\\
\hline
Spec & 0.17 & 3 $\times$ 10$^{-4}$ & 0.29 & 0\\
SED & 0.13 & 3 $\times$ 10$^{-3}$ & 0.19 & 0\\
All & 0.14 & 0 & 0.24 & 0\\
\hline
\hline
\end{tabular}}\caption{{\normalsize Spearman correlation coefficient $\rho$ and p-value for the intrinsic absorption $N_{\rm H,z}$ in relation to M$_*$ or SFR, for sources with spectroscopic classification, sources with only best-fit SED template and for the whole CCLS70 Type 2 sample.}}\label{tab:sfr_mgal_nh}
\end{table}

\section{High-redshift sample}\label{sec:high-z}
In this section, we summarize the results of the spectral fitting of the 20 Chandra COSMOS-Legacy sources at
$z\geq$3 in CCLS70. 15 of these sources have z$_{\rm spec}$, the remaining 5 have z$_{\rm phot}$. An extended analysis of the \cha \leg\ $z\geq$3 sample, which contains 174 sources, is reported in \citet{marchesi16b}.

We first fitted our spectra with the best-fit model obtained with the procedure described in Section \ref{sec:model}. For 19 out of 20 sources the best-fit is an absorbed power-law model, while for cid\_83 a Fe K$\alpha$ emission line is also required. For each source, we estimate the spectral slope $\Gamma$ and the intrinsic absorption $N_{\rm H,z}$. We report the results of this fit in Table \ref{tab:z_gt3}. The mean (median) photon index is $\Gamma$=1.50$\pm$0.08 (1.49), with dispersion $\sigma$=0.35.

15 out of the 20 sources only have an upper limit on $N_{\rm H,z}$, while the other 5 have a significant $N_{\rm H,z}$ value at a 90\% confidence level.

We then repeated the fit, this time using the \texttt{pexrav} model, which takes into account the presence of a reflection component caused by cold material close to the black hole accretion disk. We do so because at $z>$3 the reflection component, which contributes to the spectral emission at energies greater than 30 keV in the rest-frame, is observed in the 2--10 keV band. Therefore, a lack of the reflection component in the fit produces an artificial flattening in the photon index estimate. We fix the reflection parameter to 1: since is $R$=$\Omega$/2$\pi$, where $\Omega$ is the solid angle of the cold material visible from the hot corona, $R$=1 is the case where the reflection is caused by an in infinite slab illuminated by the isotropic corona emission. The results of this second fit are reported in Table \ref{tab:z_gt3}: the presence of a reflection component implies a general steepening of the spectral slope, which now has mean (median) value $\Gamma$=1.65$\pm$0.07 (1.65), with dispersion $\sigma$=0.32.

12 out of the 20 sources have only an upper limit on $N_{\rm H,z}$ fitting the spectra with the pexrav model. Two sources, lid\_1577 and cid\_507, have $N_{\rm H,z}>$10$^{23}$ cm$^{-2}$ even if the 90\% confidence error is taken into account.

\begin{figure*}%[!h]
\begin{minipage}[b]{.5\textwidth}
  \centering
  \includegraphics[width=1.0\linewidth]{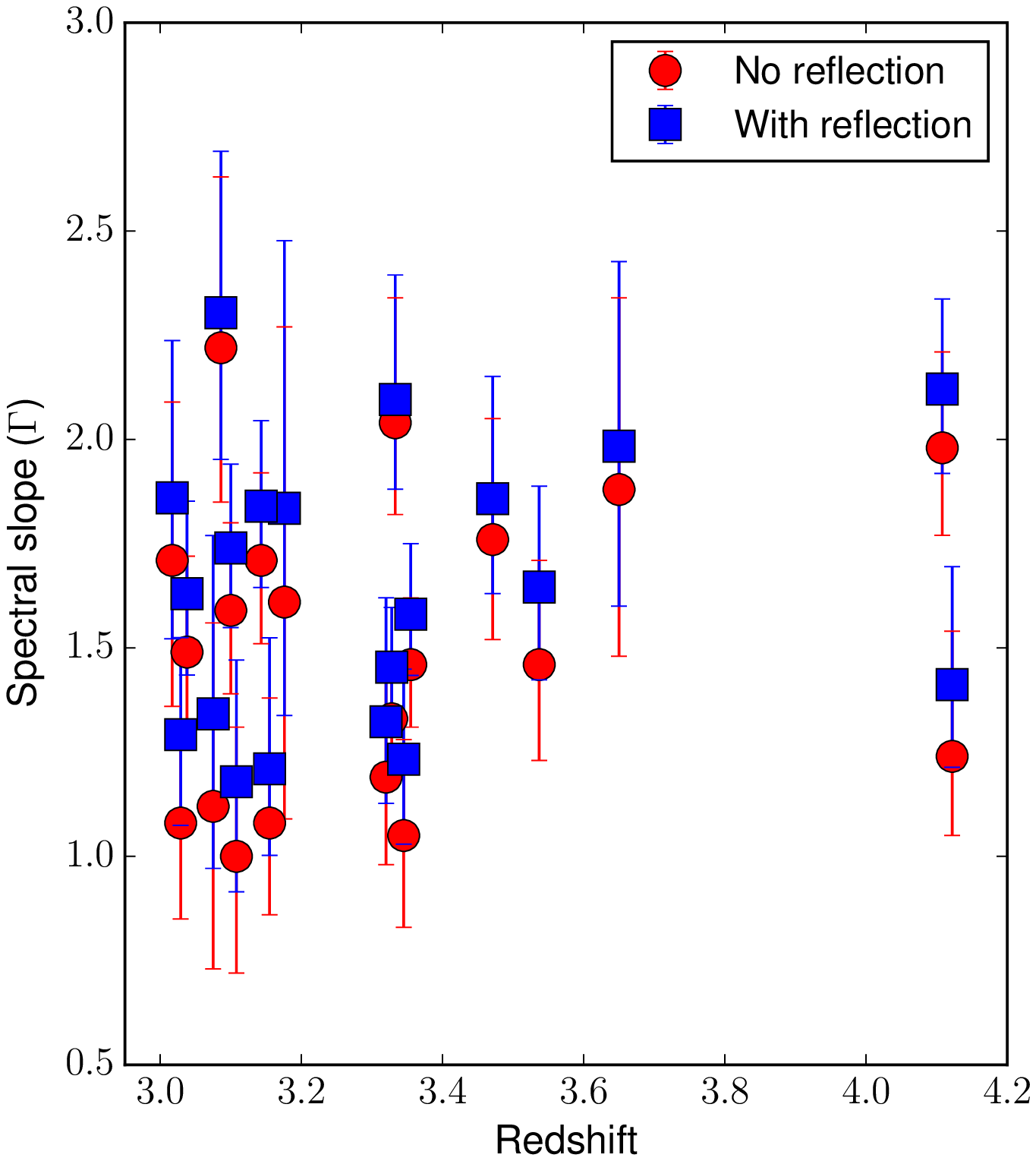}
\end{minipage}%
\begin{minipage}[b]{.5\textwidth}
  \centering
  \includegraphics[width=1.03\textwidth]{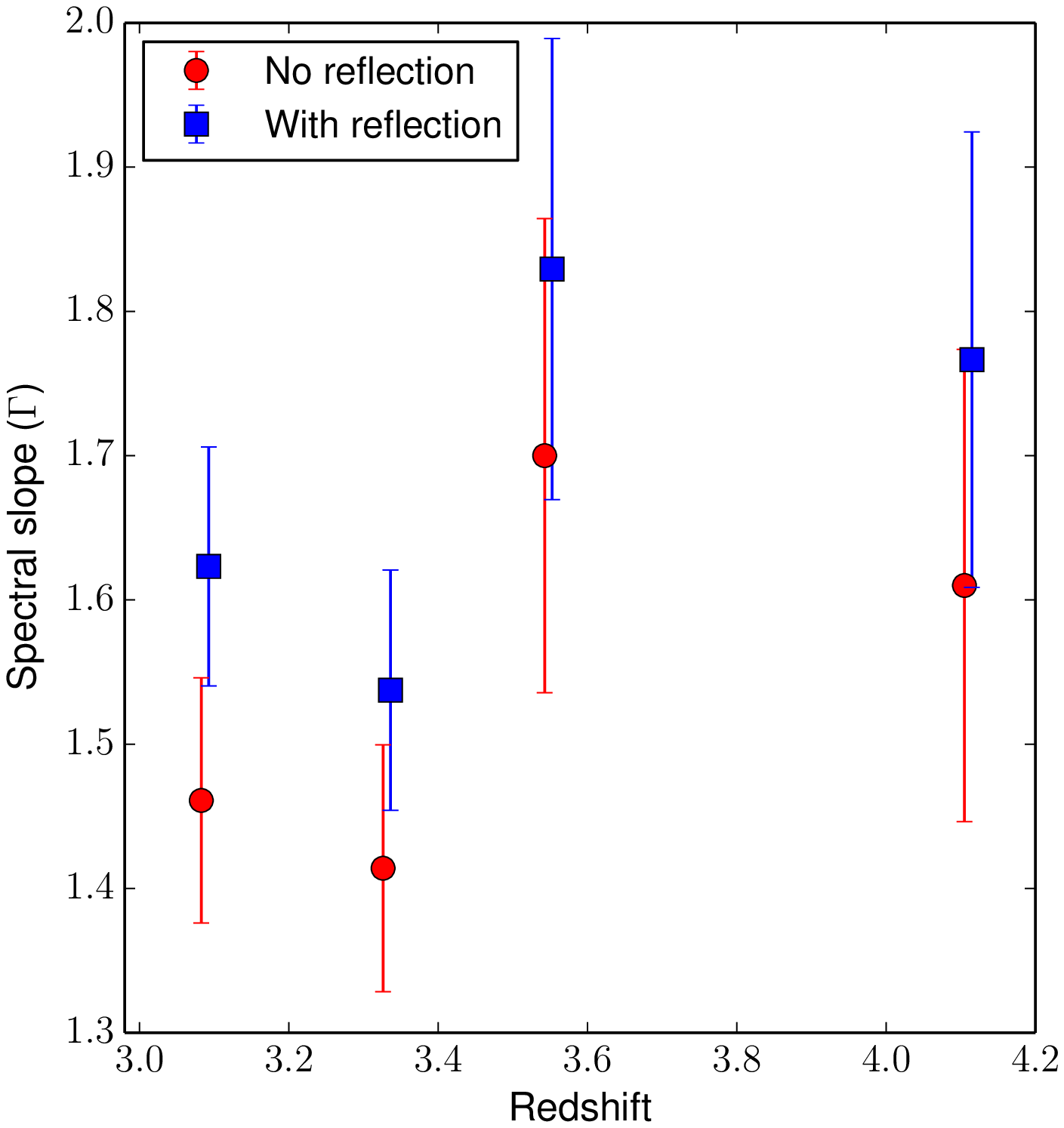}
\end{minipage}
\caption{{\normalsize \textit{Left}: Evolution with redshift of the photon index $\Gamma$, without (red circles) or with (blue squares) the contribution of a reflection component, for the 20 \cha \leg\ sources with z$\geq$3 and more than 70 net counts in the 0.5-7 keV band. \textit{Right}: same as in the left panel, but binned in four different bins of redshift. Note that the y-axes ranges are different in the two panels.}}
\label{fig:high-z}
\end{figure*}

In Figure \ref{fig:high-z} we show the evolution with redshift of the X-ray spectral slope $\Gamma$, both without (red circles) and with (blue squares) the contribution of a reflection component: the photon index distribution has a large spread and no clear trend with redshift is observed.

\begingroup
\renewcommand*{\arraystretch}{1.5}
\begin{table*}
\centering
\scalebox{1.}{
\begin{tabular}{ccccc|ccc|ccc}
\hline
\hline
& & & & & \multicolumn{3}{c|}{Power-law} & \multicolumn{3}{c}{\texttt{Pexrav}}\\
\hline
  id  & $z$ & spec & type & cts& $\Gamma$ & $N_{\rm H,z}$ & Cstat/DOF & $\Gamma$ & $N_{\rm H,z}$ & Cstat/DOF\\
\hline
  cid\_407  & 3.471  & N & 2 & 70 & 1.77$_{-0.39}^{+0.51}$ & $<$6.33 & 21.0/23 & 1.86$_{-0.37}^{+0.50}$ & $<$6.68 & 20.8/23\\
  cid\_1118 & 3.65   & Y & 1 &  75 & 1.87$_{-0.65}^{+0.82}$ & $<$27.51& 12.5/29 & 1.98$_{-0.61}^{+0.76}$ & $<$27.11& 12.5/29\\
  lid\_1577 & 3.176  & Y & 2 &  77 & 1.59$_{-0.81}^{+1.19}$ & 40.14$_{-31.23}^{+62.35}$ & 29.9/23 & 1.84$_{-0.81}^{+1.10}$ & 45.35$_{-32.02}^{+62.17}$ & 30.0/23\\
  cid\_472  & 3.155  & Y & 1 &   79 & 1.08$_{-0.37}^{+0.57}$ & $<$10.95 & 48.5/30 & 1.21$_{-0.34}^{+0.54}$ & $<$10.17 & 47.6/30\\
  lid\_1519 & 3.32   & Y & 1 &  80 & 1.22$_{-0.34}^{+0.53}$ & $<$12.33 & 17.8/25 & 1.32$_{-0.32}^{+0.52}$ & $<$12.40 & 17.7/25\\
  cid\_325  & 3.086  & Y & 2 &  86 & 2.20$_{-0.59}^{+0.69}$ & 9.26$_{-8.38 }^{+11.67}$ & 31.1/27 & 2.30$_{-0.56}^{+0.67}$ & 10.02$_{-8.00 }^{+12.22}$ & 30.7/27\\
  cid\_83   & 3.075  & N & 1 & 95 & 1.12$_{-0.62}^{+0.77}$ & $<$37.69 & 34.2/31 & 1.34$_{-0.60}^{+0.74}$ & 14.61$_{-14.55 }^{+25.36}$ & 33.9/31\\
  cid\_113  & 3.333  & Y & 1 &  97 & 2.04$_{-0.36}^{+0.55}$ & $<$5.46  & 36.3/33 & 2.10$_{-0.35}^{+0.52}$ & $<$5.38  & 35.9/33\\
  lid\_1865 & 4.122  & N & 2 & 107& 1.24$_{-0.30}^{+ 0.52}$ & $<$22.70 &  27.8/33 & 1.41$_{-0.32}^{+ 0.48}$ & $<$18.42 &  28.7/33\\
  cid\_529  & 3.017  & Y & 2 & 107& 1.68$_{-0.56}^{+0.65}$ & 12.66$_{-10.33}^{+15.74}$ & 29.4/36 & 1.86$_{-0.55}^{+0.66}$ & 14.48$_{-10.34}^{+16.69}$ & 30.5/36\\
  lid\_721  & 3.108  & Y & 1 & 111& 0.98$_{-0.42}^{+0.52}$ & $<$21.82 & 63.1/37 & 1.18$_{-0.42}^{+0.50}$ & $<$23.65 & 62.2/37 \\
  cid\_64   & 3.328  & Y & 1 & 134& 1.33$_{-0.24}^{+0.25}$ & $<$2.73  & 60.4/46 & 1.45$_{-0.23}^{+0.29}$ & $<$2.18  & 59.0/46\\
  lid\_476  & 3.038  & Y & 1 & 142& 1.49$_{-0.28}^{+0.39}$ & $<$7.76  & 50.4/44 & 1.63$_{-0.31}^{+0.38}$ & $<$9.72  & 50.4/44\\
  lid\_205  & 3.355  & Y & 1 & 149& 1.46$_{-0.25}^{+0.30}$ & $<$4.76  & 42.2/47 & 1.58$_{-0.24}^{+0.29}$ & $<$4.70  & 41.4/47\\
  cid\_413  & 3.345  & Y & 1 & 151& 1.04$_{-0.34}^{+0.38}$ & $<$13.31 & 33.0/46 & 1.23$_{-0.32}^{+0.38}$ & $<$14.92 & 33.0/46\\
  cid\_75   & 3.029  & Y & 1 & 176& 1.09$_{-0.37}^{+0.41}$ & 11.42$_{-8.76 }^{+12.06}$ & 62.2/61 & 1.29$_{-0.35}^{+0.40}$ & 13.66$_{-5.95 }^{+7.20}$ & 62.4/61\\
  cid\_1269 & 3.537  & Y & 2 & 185& 1.47$_{-0.38}^{+0.43}$ & $<$21.74 & 41.0/61 & 1.65$_{-0.35}^{+0.42}$ & 9.53$_{-8.63}^{+12.76}$ & 40.5/61\\
  cid\_558  & 3.1    & N & 2 & 195& 1.59$_{-0.32}^{+0.34}$ & $<$9.34  & 54.7/57 & 1.74$_{-0.31}^{+0.34}$ & 4.71$_{-4.15}^{+5.94}$  & 56.2/57\\
  lid\_460  & 3.143  & Y & 1 & 204& 1.72$_{-0.33}^{+0.37}$ & $<$12.36 & 62.7/59 & 1.84$_{-0.31}^{+0.36}$ & $<$12.35 & 61.4/59\\
  cid\_507  & 4.108  & N & 2 & 242& 1.98$_{-0.34}^{+0.38}$ & 18.76$_{-10.55}^{+13.49}$ & 66.2/67 & 2.12$_{-0.32}^{+0.38}$ & 19.82$_{-9.60}^{+13.82}$ & 67.5/67\\
\hline
\hline
\end{tabular}}\caption{{\normalsize Summary of the X-ray spectral fitting with an absorbed power-law for the 20 \cha \leg\ sources with $z\geq$3 and more than 70 net counts in the 0.5-7 keV band. ``spec'' indicates if the redshift is spectroscopic (Y) or photometric (N).}}\label{tab:z_gt3}
\end{table*}
\endgroup

\subsection{Stacking of low statistics spectra}
To complete our analysis of the $z\geq$3 sample, we perform a fit stacking the spectra of the 154 sources with less than 70 net counts in the 0.5-7 keV band. Each spectrum was corrected for background, detector response and Galactic absorption in the same way as done in \citet{iwasawa12}, and rebinned into 1 keV intervals in the 3-23 keV band in the galaxy rest frame. The spectral stacking is then a straight sum of these individual spectra. As for the sources in CCLS70, we first fit the data with an absorbed power-law and then with a \texttt{pexrav} model with reflection parameter $R$=1. We report the results of the fits in Table \ref{tab:stack_zgt3}, for the whole sample and for different subsamples bsased on the optical classification of the sources. The stacked spectrum obtained combining all the 154 sources contains 3583 net counts in the 3-23 keV rest frame band, has $\Gamma$=1.44$^{+0.16}_{-0.14}$ and $N_{\rm H,z}$=5.22$^{+4.41}_{-3.23}$ $\times$10$^{22}$ cm$^{-2}$ while fitting the data with an absorbed power-law, and $\Gamma$=1.63$^{+0.16}_{-0.13}$ and $N_{\rm H,z}$=7.02$^{+4.57}_{-3.32}$ $\times$10$^{22}$ cm$^{-2}$ while fitting with a \texttt{pexrav} model. The indication of a flattening in the spectral slope, with respect to a typical value $\Gamma$=1.9, is confirmed even if we stack separately the spectra on the basis of their optical classification, taking into account both the classification in Type 1 and Type 2 AGN and the presence or the absence of a spectroscopic redshift.

To study how much the low counts population affects the stacking results, we re-fitted the data removing from the stacking those sources with less than 10 counts in the 3-23 keV rest-frame band: the difference with respect to the fit to the whole sample is completely negligible. 

Finally, we repeat the fit after removing from the stacking also those sources having more than 45 counts in the 3-23 keV rest-frame band, to estimate how much the brightest sources effect the fitting. We find a good agreement, at a 90\% confidence level, between the spectral parameters $\Gamma$ and $N_{\rm H,z}$ computed with and without the faintest and brightest sources in the sample.

\begingroup
\renewcommand*{\arraystretch}{1.5}
\begin{table*}
\centering
\scalebox{1.}{
\begin{tabular}{cccc|ccc|ccc}
\hline
\hline
Type & $n_{src}$ & cts & $\langle z \rangle$ & $\chi^2$/d.o.f. & $\Gamma$ & $N_{\rm H,z}$ & $\chi^2$/d.o.f. & $\Gamma$ & $N_{\rm H,z}$\\
 & & & & \multicolumn{3}{c|}{Power-law} & \multicolumn{3}{c}{\texttt{Pexrav}}\\
\hline
All & 154 & 3586 & 3.58 & 22.3/17 & 1.44$^{+0.16}_{-0.14}$ & 5.22$^{+4.41}_{-3.23}$ & 27.3/17 & 1.63$^{+0.16}_{-0.13}$ & 7.02$^{+4.57}_{-3.32}$\\
1 spec, all & 40 & 1341 & 3.52 & 12.7/17 & 1.50$^{+0.24}_{-0.23}$ & $<$5.90 & 10.6/17 & 1.70$^{+0.24}_{-0.21}$ & 4.29$^{+5.76}_{-4.23}$\\
1 spec, 10-45 cts & 31 & 899 & 3.56 & 21.5/17 & 1.68$^{+0.35}_{-0.31}$ & $<$12.02 & 19.6/17 & 1.83$^{+0.31}_{-0.29}$ & $<$12.83\\
1 phot, all & 33 & 699 & 3.55 & 16.6/17 & 1.48$^{+0.35}_{-0.29}$ & $<$11.01 & 18.0/17 & 1.67$^{+0.33}_{-0.27}$ & $<$16.74\\
1 phot, 10-45 cts & 25 & 564 & 3.34 & 16.9/17 & 1.53$^{+0.33}_{-0.24}$ & 4.48$^{+3.99}_{-2.90}$ & 18.0/17 & 1.69$^{+0.31}_{-0.27}$ & 5.66$^{+7.46}_{-5.64}$\\
2 spec, all & 22 & 518 & 3.52 & 15.4/17 & 1.50$^{+0.43}_{-0.27}$ & 8.67$^{+12.29}_{-7.16}$ & 16.8/17 & 1.70$^{+0.42}_{-0.27}$ & 10.78$^{+13.04}_{-4.25}$\\
2 spec, 10-45 cts & 18 & 385 & 3.54 & 17.2/17 & 1.85$^{+0.62}_{-0.31}$ & 13.99$^{+19.90}_{-10.47}$ & 18.3/17 & 2.05$^{+0.59}_{-0.31}$ & 16.53$^{+21.04}_{-5.42}$\\
2 phot all & 59 & 1028 & 3.66 & 29.4/17 & 1.36$^{+0.38}_{-0.27}$ & 9.05$^{+13.67}_{-4.32}$ & 33.0/17 & 1.56$^{+0.38}_{-0.27}$ & 11.46$^{+15.02}_{-4.21}$\\
2 phot, 10-45 cts & 39 & 763 & 3.70 & 22.0/17 & 1.19$^{+0.24}_{-0.18}$ & $<$5.93 & 24.1/17 & 1.33$^{+0.29}_{-0.16}$ & $<$7.67\\
\hline
\hline
\end{tabular}}\caption{{\normalsize Spectral parameters obtained fitting the stacked spectra of AGN at $z\geq$3 and with less than 70 net counts in the 0.5-7 keV band. $N_{\rm H,z}$ is in units of 10$^{22}$ cm$^{-2}$. ``spec'' and ``phot'' indicate sources with spectroscopic redshift available and with only photometric redshift, respectively. ``10-45 cts'' indicates that we stacked only source having between 10 and 45 net counts in the 3-23 keV rest-frame band. The source net counts are computed in the 3-23 keV rest-frame band.}}\label{tab:stack_zgt3}
\end{table*}
\endgroup

\section{Conclusions}\label{sec:concl}
We analyzed the X-ray spectra of the 1855 \cha \leg\ extragalactic sources with more than 30 net counts in the 0.5-7 keV band (CCLS30). 1273 out of 1855 sources ($\sim$69\%) have a spectroscopic redshift, while the remaining 582 have a photometric redshift.

90\% of the sources are well fitted with a basic power-law model,  while the remaining 10\% showed a statistically significant improvement while adding to the basic model further components, such as an iron K$\alpha$ line at 6.4 keV and/or a second power-law. The source spectra have been fitted together with the background spectra, which we reproduced with a complex multi-component model, described in Appendix \ref{app:back}.

We now summarize the main results we obtained.
\begin{enumerate}
\item 37.7\% of the CCLS30 sources are classified as Type 1 AGN and 60.3\% are classified as Type 2 AGN, on the basis of either their spectroscopic or their SED template best-fitting classification. Finally, 2.0\% sources are classified as low-redshift, passive galaxies.
\item The majority of sources in CCLS30 (67.2\%) have only a 90\% confidence upper limit on $N_{\rm H,z}$ (see Figure \ref{fig:nh_histo}, left panel). Type 2 AGN are also significantly more obscured than Type 1 AGN. 41.4\% of Type 2 AGN are obscured (i.e., with $N_{\rm H,z} >$10$^{22}$ cm$^{-2}$ at a 90\% confidence level) in CCLS30, while only 15.2\% of Type 1 AGN are obscured.
\item The majority of sources in CCL30 with less than 70 net counts (672 out 968, 69.4\%) have been fitted with a typical AGN photon index $\Gamma$=1.9, to better constrain $N_{\rm H,z}$. In CCLS70 (Figure 7), the mean $\Gamma$ of Type 1 sources is $\langle \Gamma \rangle$=1.75$\pm$0.02 (1.74), with dispersion $\sigma$=0.31, while Type 2 sources have on average flatter photon indexes, with $\langle \Gamma \rangle$=1.61$\pm$0.02 (1.62), with dispersion $\sigma$=0.47. While this difference may be intrinsic, it is worth noticing that Type 1 and Type 2 AGN have significantly closer $\langle \Gamma \rangle$ values if we take into account only sources with 1$<\Gamma<$3, i.e., we remove from the computation very soft objects and candidate highly obscured, reflection dominated sources, for which a flat $\Gamma$ is trying to fit an obscured spectrum. Using this subsample of sources, Type 1 AGN have mean photon index $\langle \Gamma \rangle$=1.77$\pm$0.01, with $\sigma$=0.28, while Type 2 AGN have $\langle \Gamma \rangle$=1.70$\pm$0.02, with $\sigma$=0.34. 
\item Type 1 AGN have on average higher 2--10 keV intrinsic, absorption-corrected luminosity values than Type 2 AGN. The median luminosity of Type 1 AGN is 2.3 $\times$ 10$^{44}$ \lu, and $\sim$64\% of Type 1 AGN have L$_{\rm X}$ $>$10$^{44}$ \lu, i.e., in the quasar regime. The majority ($\sim$71\%) of Type 2 sources have instead LX below the quasar threshold and are mainly Seyfert galaxies. However, this is not an intrinsic difference, but is caused by having a flux-limited sample, i.e., sources at higher redshifts also have higher luminosities, and by the fact that Type 1 AGN are on average at higher redshifts ($\langle z \rangle$=1.74 for CCLS30) than Type 2 AGN ($\langle z \rangle$=1.23). Nonetheless, the difference between the two distributions may also suggest a trend with 2-10 keV luminosity of the fraction of Type 1 to Type 2 AGN, Type 2 AGN being more numerous at lower luminosities.
\item We studied the distribution of $N_{\rm H,z}$ as a function of 2-10 keV rest-frame absorption-corrected luminosity (Figure \ref{fig:nh_z_lx}, right panel). In CCLS30, 268 sources out of 1844 ($\sim$15\%) lie in the obscured quasar region (i.e.,  L$_{\rm 2-10keV}$ =10$^{44}$ \lu\ and $N_{\rm H,z}>$10$^{22}$ cm$^{-2}$). 83 out of these 268 sources ($\sim$31\%) are optically classified Type 1 AGN, and a fraction of these 83 sources are expected to be BAL quasars.
\item A significant fraction of optical Type 2 sources lie in the L$_{\rm 2-10keV}$ =[10$^{42}$--10$^{44}$] \lu, $N_{\rm H,z}<$10$^{22}$ cm$^{-2}$ area, i.e., these sources have unobscured AGN X-ray properties. In CCLS30, 172 Type 2 AGN lie in the unobscured AGN area (15.5\% of the whole Type 2 AGN population), while the fraction slightly increases (24.0\%) if we take into account only objects with a spectral type. The fraction of unobscured Type 2 AGN strongly decreases with increasing 2--10 keV luminosity (Figure \ref{fig:ratio_obs_lx}) and can be explained by a misclassification of low-luminosity BLAGN and/or by the lack of broad emission lines in intrinsically unobscured, low accretion AGN.
\item For the 141 CCLS30 sources best-fitted with a model which includes an iron K$\alpha$ emission line at 6.4 keV, we computed the emission line equivalent width, EW. The mean (median) equivalent width of the 141 CCLS30 sources is EW=0.49$\pm$0.38 (0.39) keV. Fitting the EW distribution as a function of $\Gamma$ with a linear model (EW= $a\Gamma+b$), we find a significant anti-correlation, with a=--0.22$\pm$0.05 (Figure \ref{fig:ew}, left panel). However the correlation between EW and $\Gamma$ disappears if we exclude from the sample the 12 candidate CT AGN we obtained on the basis of more complex fitting models, which include proper characterization of the dusty torus surrounding the SMBH. This implies that basic models may fail in fitting heavily obscured sources, fitting the flat spectra of these objects with flat, unphysical photon indexes and no $N_{\rm H,z}$.
\item We searched for a correlation between $\Gamma$ and $N_{\rm H,z}$ and the host galaxy mass (M$_*$) and star formation rate (SFR) for Type 2 AGN. M$_*$ and SFR have been computed using SED  fitting techniques (Suh et al. in preparation). We do not find any significant correlation between $\Gamma$ and either M$_*$ or SFR. We instead find a significant correlation between $N_{\rm H,z}$ and both M$_*$ and SFR: objects with higher $N_{\rm H,z}$ values also have higher M$_*$ and SFR (Figure \ref{fig:sfr_mgal}). However, a partial correlation test showed that these relations are partially driven by a redshift selection effect.
\item We studied the properties of the 20 CCLS70 sources at z$\geq$3. The mean (median) slope of these 20 sources is $\Gamma$=1.50$\pm$0.08 (1.49). 15 out of the 20 sources have only an upper limit on $N_{\rm H,z}$. Two sources have $N_{\rm H,z}>$ 10$^{23}$ cm$^{-2}$. We also repeated the fitting using the \texttt{pexrav} model, to take into account the potential presence of a reflection component close to the black hole accretion disk. The reflection component affects the spectrum at energies greater than 30 keV in the rest-frame, i.e., inside the observed 0.5-7 keV energy range we use in our fitting. The presence of a reflection component implies a general steepening of $\Gamma$, with $\langle \Gamma \rangle$=1.65$\pm$0.07. The evidence for a photon-index flatter than the standard value in our $z\geq$3 sample is confirmed by a fit of the stacked spectra of the 154 sources not in CCLS70, which have $\Gamma$=1.44$^{+0.16}_{-0.14}$, while fitting the data with a power-law and $\Gamma$=1.63$^{+0.16}_{-0.13}$ adding a reflection component to the model.
\end{enumerate}

This work is the first step towards large sample of sources at high redshift (z$>$0.5) and medium-low luminosities (L$_{\rm 2-10keV}$ =[10$^{42}$--10$^{43}$] \lu), where X-ray spectral properties can be constrained. The next generation of X-ray missions, such as Athena+ \citep{nandra13} or X-ray Surveyor \citep{vikhlinin12}, will provide even larger samples of such sources, significantly reducing selection biases and allowing statistical studies for column densities and SFR, searching for correlation between the central engine and the galaxy properties. Moreover, the improved statistics will allow to detect iron emission lines in the high-z universe \citep{georgakakis13} and even measure redshifts from X-ray features \citep{comastri04}.

\section{Acknowledgments}
We thank the anonymous referee for the suggestions which helped improving this paper. This research has made use of data obtained from the \cha Data Archive and software provided by the \cha X-ray Center (CXC) in the CIAO application package.

This work was supported in part by NASA Chandra grant number GO3-14150C and also GO3-14150B  (F.C., V.A., M.E.); PRIN-INAF 2014 "Windy Black Holes combing galaxy evolution" (A.C., M.B., G.L. and C.V.); the FP7 Career Integration Grant ``eEASy'': ``Supermassive black holes through cosmic time: from current surveys to eROSITA-Euclid Synergies" (CIG 321913; M.B. and G.L.); the Spanish MINECO under grant AYA2013-47447-C3-2-P and MDM-2014-0369 of ICCUB (Unidad de Excelencia ``Mar\'ia de Maeztu''; K.I.); UNAM-DGAPA Grant PAPIIT IN104216 and  CONACyT Grant Cient\'ifica B\'asica \#179662 (T.M.); the Swiss National Science Foundation Grant PP00P2\_138979/1 (K.S.); the Center of Excellence in Astrophysics and Associated Technologies (PFB 06), by the FONDECYT regular grant 1120061 and by the CONICYT Anillo project ACT1101 (E.T.).

\bibliographystyle{aa}
\bibliography{xray_spectral_fitting_arxiv}

\begin{appendix}
\section{Background modelling}\label{app:back}
\begin{figure*}%[!h]
  \centering
  \includegraphics[width=0.75\linewidth]{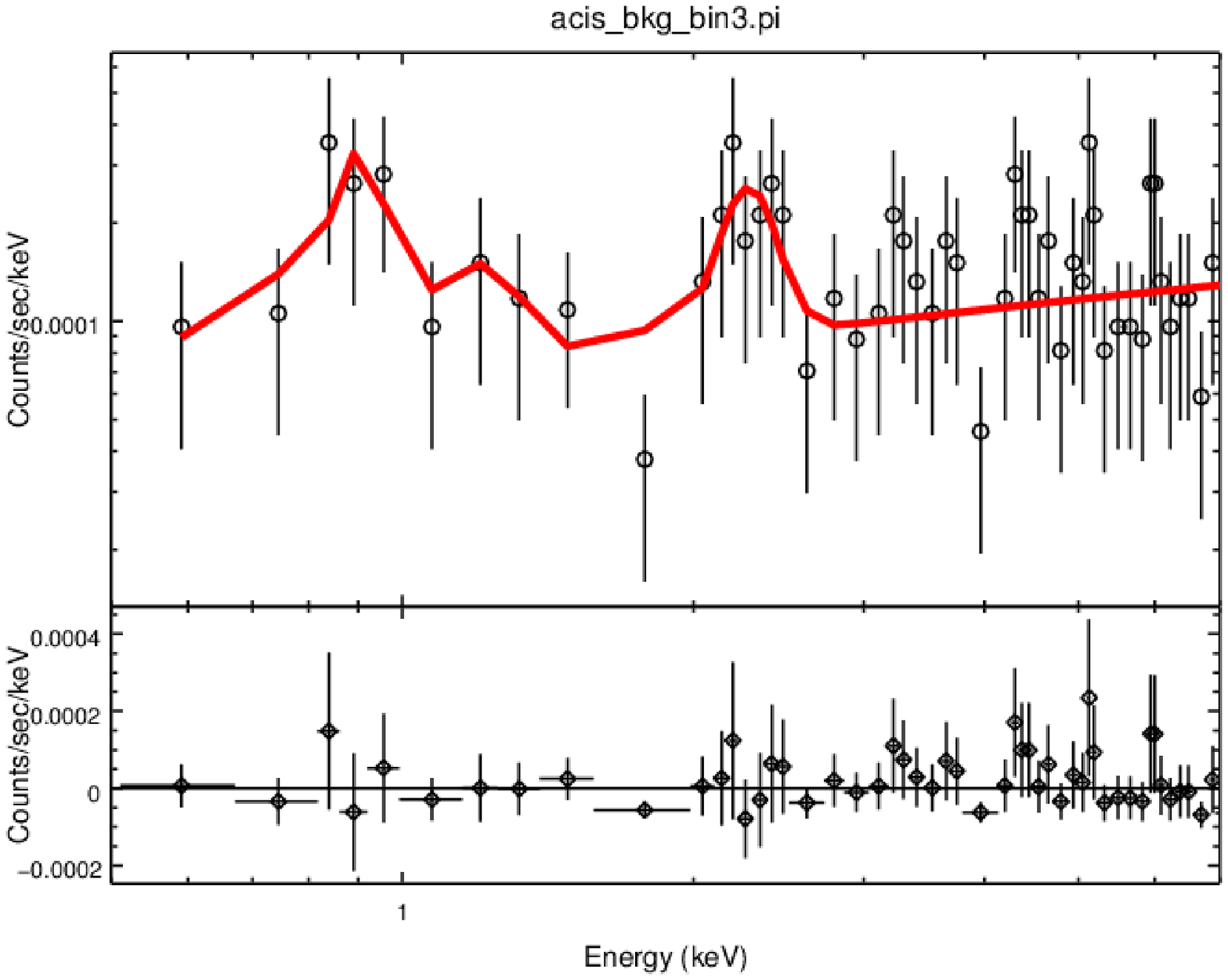}
  \centering
  \includegraphics[width=0.75\textwidth]{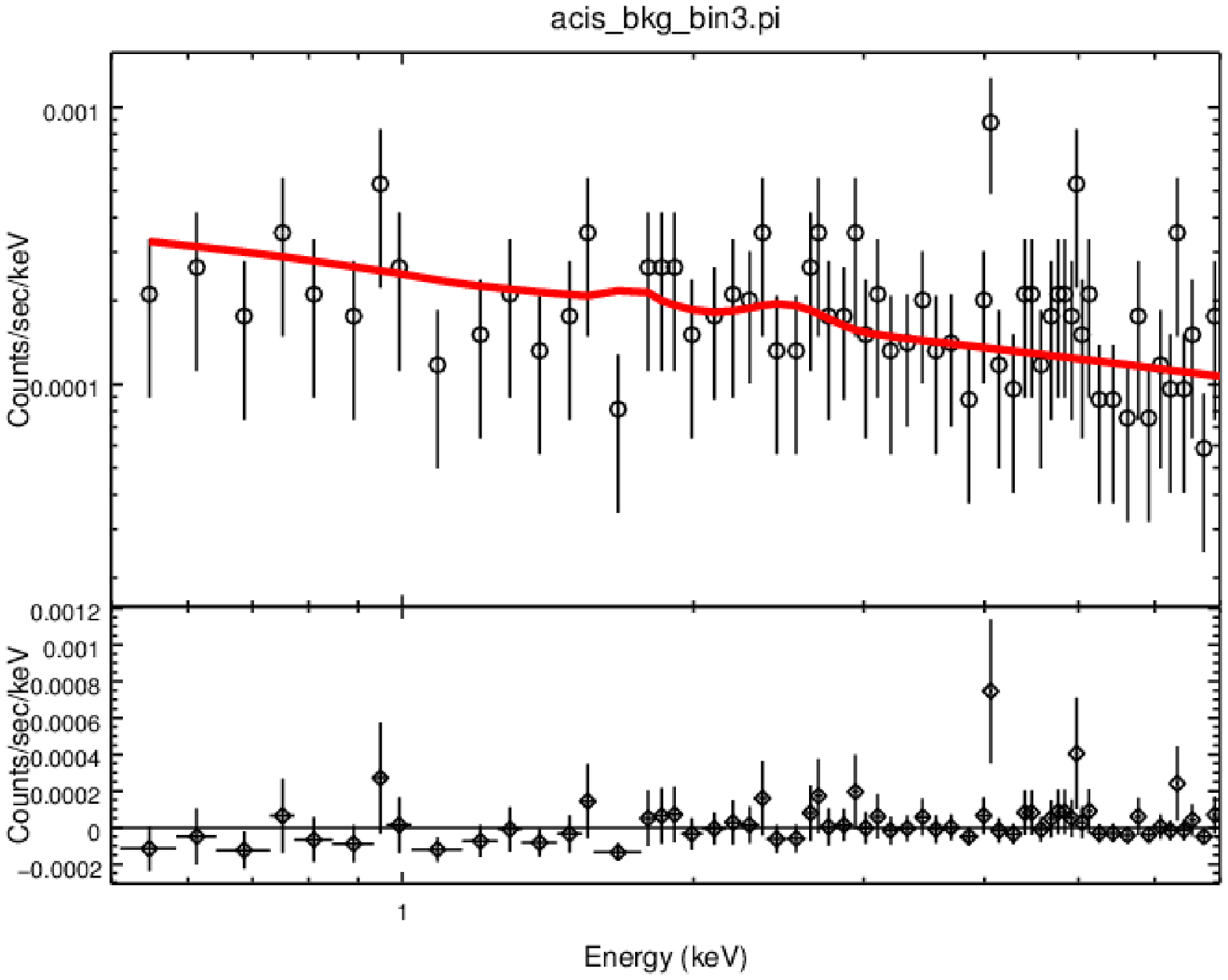}
\caption{{\normalsize Example of two background spectral fittings, for sources cid\_29 (top) and cid\_33 (bottom).}}\label{fig:bkg_example}
\end{figure*}

In Figure \ref{fig:bkg_example} we show an example of \cha ACIS-I background in the 0.5-7 keV band for two sources in our sample. As can be seen, the background is rather flat and shows different emission features. The background is formed by different components, both astrophysical and instrumental, which we fit with different models:
\begin{enumerate}
\item The cosmic background, which we fit with an absorbed \texttt{apec} component, i.e., a thermal plasma emission model, mainly caused by the local ``super-bubble''. The temperature $kT$ of this thermal component is left free to vary between 0 and 2 keV.
\item The instrumental particle background, which we fit with a power-law and two emission lines: one between 1 and 1.8 keV, to fit the Al and Si emission features, and one between 1.8 and 2.5 keV, to fit the Au emission feature. Since these are instrumental components, we did not convolve this part of model through the ARF.
\end{enumerate}

Once we find the background best-fit, we freeze all the background fit parameters and we then re-fit the combined source and background spectrum.

\section{Comparison with \xmm}
The COSMOS field has been observed with \xmm\ to a flux limit of 7 $\times$ 10$^{-16}$ \flu\ in the 0.5-2 keV band \citep{hasinger07}. The point-source catalog development procedure is described in \citet{cappelluti07}, while the spectral analysis of the X-ray bright sample is reported in \citet{mainieri07} and \citet{lanzuisi15}. The spectral extraction procedure for XMM-COSMOS sources has been the same used for CCLS30, i.e., the spectra where extracted from each observation and the combined, weighting the response matrices ARF and RMF by the contribution of each spectrum. The spectral fitting was performed as in CCLS30, using the Cstat minimization technique and modelling rather than subtracting the background component. The fits were performed in the 0.5-7 keV band to be fully consistent with the \cha procedure, even if in principle \xmm\ is sensitive up to $\sim$10 keV.

1010 CCLS30 sources ($\sim$52\%) have a counterpart with more than 30 net counts in the 0.5-7 keV band in the XMM-COSMOS catalog. In the following sections we study how $\Gamma$ and $N_{H,z}$ computed with \cha and \xmm\ correlate.

\subsection{Photon index}
720 out of 1010 sources have less than 150 net counts in the 0.5--7 keV band in XMM-COSMOS. For these sources, we fix $\Gamma$=1.9, due to lack of good statistics. 190 of these sources have $\Gamma$=1.9 also in CCLS30.

\begin{figure*}%[!h]
\begin{minipage}[b]{.5\textwidth}
  \centering
  \includegraphics[width=1.\linewidth]{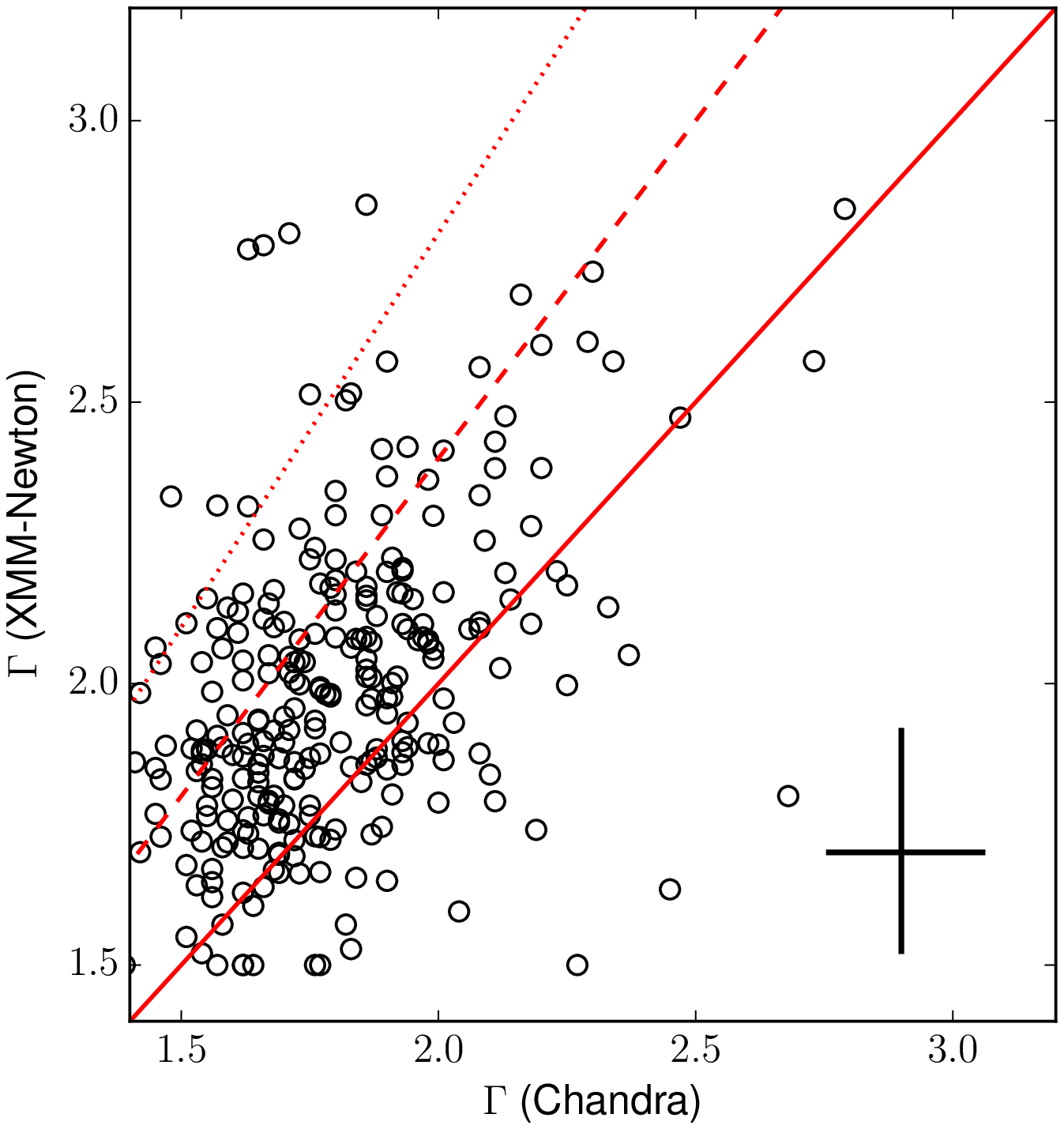}
\end{minipage}%
\begin{minipage}[b]{.5\textwidth}
  \centering
  \includegraphics[width=1.04\textwidth]{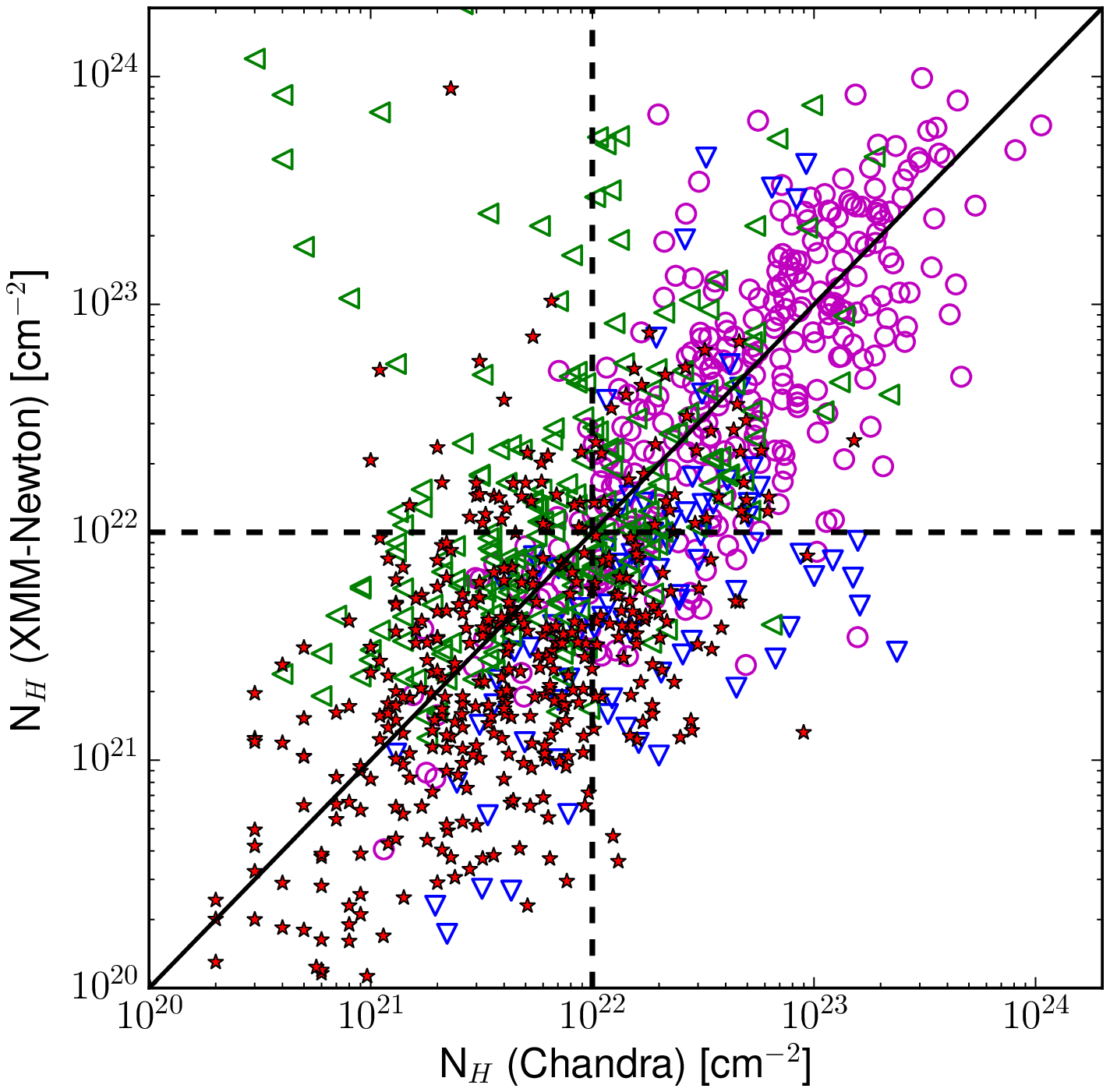}
\end{minipage}
\caption{{\normalsize \textit{Left} : photon index $\Gamma$ obtained fitting the \xmm\ spectra as a function of $\Gamma$  computed for CCLS70, for the 287 sources with $\Gamma$  left free to vary in the spectral fitting. The 1:1 relation is plotted with a red solid line; the $y = x \times$ 1.2 ($y = x \times$ 1.4) relation is plotted with a red dashed (dotted) line. \textit{Right}: intrinsic absorption $N_{H,z}$ obtained fitting the \xmm\ spectra as a function of $N_{H,z}$ computed for CCLS30, for the 1010 sources in CCLS30 with XMM-COSMOS counterparts. Sources with $N_{H,z}$ significant value significant at 1$\sigma$ are plotted as magenta circles, sources with upper limit in \xmm\ and significant detection in \cha are plotted as blue downwards triangles, sources with upper limit in \cha and significant detection in \xmm\ are plotted as green leftwards triangles, and sources with upper limits on $N_{H,z}$ in both samples are shown as red stars. The 1:1 relation is plotted as a black solid line, and the standard threshold adopted to divide unobscured from obscured sources ($N_{H,z}$=10$^{22}$ cm$^{-2}$) as black dashed lines.}}\label{fig:xmm_cha}
\end{figure*}

For the remaining 290 sources with $\Gamma$ free to vary in XMM-COSMOS, we show in Figure \ref{fig:xmm_cha} (left panel) the trend of the photon index from the \xmm\ best-fit $\Gamma_{\rm XMM}$ as a function of the photon index in CCLS30, $\Gamma_{\rm Cha}$. In the same figure, we plot the 1:1 relation with a red solid line, and the $y = x \times$ 1.2 ($y = x \times$ 1.4) relation with a red dashed (dotted) line. As can be seen, $\Gamma_{\rm XMM}$ is on average steeper than $\Gamma_{\rm Cha}$: in fact, the mean and standard deviation of the XMM-COSMOS sample are $\langle \Gamma_{\rm XMM} \rangle$= 1.98 and  $\sigma_{\rm XMM}$= 0.29, while the mean and standard deviation for the CCLS30 counterparts are $\langle \Gamma_{\rm Cha} \rangle$= 1.76 and $\sigma_{\rm Cha}$ = 0.28.

A similar result was already found in \citet{lanzuisi13}, and it is independent from the number of net counts in both the spectra. Moreover, \citet{lanzuisi13} showed that the same discrepancy between $\Gamma_{\rm XMM}$ and $\Gamma_{\rm Cha}$ is found using the XMM-COSMOS photon indexes computed by \citet{mainieri07}, which used a different fitting technique with respect to the one used in our work, based on background subtraction and $\chi^2$ minimization.

\subsection{Column density}
In Figure 17 (right panel) we show the intrinsic absorption $N_{\rm H,z}$ obtained fitting the \xmm\ spectra as a function of $N_{\rm H,z}$ computed for CCLS30, for the 1010 sources in CCLS30 with XMM-COSMOS counterparts. Sources with 1$\sigma$ significant $N_{\rm H,z}$ are plotted as magenta circles, sources with upper limit in \xmm\ but not in \cha are plotted as blue downwards triangles, sources with upper limit in \cha but not in \xmm\ are plotted as green leftwards triangles. Finally, upper limits on $N_{\rm H,z}$ in both samples are shown as red stars. We also plot the 1:1 relation as a black solid line and the standard threshold adopted to divide unobscured from obscured sources ($N_{\rm H,z}$=10$^{22}$ cm$^{-2}$) as black dashed lines.

304 sources out of 1010 (30\%) have a $N_{\rm H,z}$ value significant at 1$\sigma$ in both \cha and \xmm. For these sources, there is a general good agreement in the $N_{\rm H,z}$ estimates, at any range of values. 417 sources (41\%) have instead an upper limit on $N_{\rm H,z}$ in both \cha and \xmm. 

The fraction of sources with a significant $N_{\rm H,z}$ value in \xmm\ and an upper limit in \cha (205, 20\%) is higher than the fraction of sources with significant value in \cha and upper limit in \xmm\ (88, 8\%). Moreover, 108 sources are classified as obscured in \xmm\ while have only an upper limit on $N_{\rm H,z}$ in \cha, and 59 sources are obscured in \cha and have an upper limit in \xmm. In the whole sample, 351 sources (35\%) are obscured according to the \xmm\ spectral information, in reasonable agreement with the \cha results, where 329 sources (33\%) have $N_{\rm H,z} >$10$^{22}$ cm$^{-2}$.

In conclusion, there is a general good agreement between the measures on $N_{\rm H,z}$ obtained with \cha and \xmm. While 15--20\% of the sources have significantly different $N_{\rm H,z}$ values from the two different telescopes, this discrepancy can be explained by instrumental effects, as observed in the photon index measurements, or by a lack of statistics.

\end{appendix}

\end{document}